\documentclass[final, 11pt,times,pdf]{elsarticle}

\usepackage{hyperref}
\usepackage{multirow}
\usepackage{amsmath}
\usepackage{nicefrac}
\usepackage{subfigure}
\usepackage{rotating}
\usepackage{float}
\usepackage{booktabs,caption}
\usepackage[export]{adjustbox}
\usepackage{lmodern}
\usepackage[a4paper,margin=2cm]{geometry}
\usepackage{pdfrender,xcolor}
\usepackage{euscript}
\usepackage{subcaption}
\usepackage{array}
\usepackage{makecell}
\usepackage{filecontents}
\usepackage{graphicx}
\usepackage{soul}

\makeatletter
\newcommand*{\rom}[1]{\expandafter\@slowromancap\romannumeral #1@}
\makeatother

\hypersetup{
	colorlinks   = true,
	urlcolor     = blue, 
	linkcolor    = blue, 
	citecolor   = red 
}

\setcitestyle{square}

\definecolor{burntorange}{rgb}{0.8, 0.33, 0.0}
\definecolor{darkorange}{rgb}{1.0, 0.55, 0.0}
\definecolor{brightmaroon}{rgb}{0.76, 0.13, 0.28}
\definecolor{darkgreen}{rgb}{0.01, 0.42, 0.27}

\newcommand{\revised}[1]{#1}

\begin{document}

	\makeatletter
	\def\ps@pprintTitle{%
		\let\@oddhead\@empty
		\let\@evenhead\@empty
		\let\@oddfoot\@empty
		\let\@evenfoot\@oddfoot
	}
	\makeatother

	\begin{frontmatter}
		
		\title{Physics Informed Neural Network-based Computational Method for \\ Accelerating Time-Periodic Unsteady \revised{CFD Simulations}}

		\author[mymainaddress]{Lakshya Chaplot}
		\author[mymainaddress]{Harshita Agarwal}
		\author[mymainaddress]{Atul Sharma\corref{mycorrespondingauthor}}
		\cortext[mycorrespondingauthor]{Corresponding author}
		\ead{atulsharma@iitb.ac.in}

		\address[mymainaddress]{Department of Mechanical Engineering, Indian Institute of Technology Bombay, India}
		
		\begin{abstract}
			Presently, there is a steady state approach in Computational fluid dynamics (CFD) to obtain a steady solution directly from the steady state governing equations. Whereas, for obtaining a time-periodic flow solution, the present unsteady governing equations-based CFD approach starts from an initial condition and requires a large computational time---during the initial non-periodic transient phase---before reaching the periodic state.  For obtaining the periodic flow directly, without transient simulations that may not be of interest, our objective is to propose a Physics Informed Neural Network (PINN)-based periodic CFD approach. The motivation is a substantial reduction in computational time by a meshless PINN-based \emph{periodic} CFD solver as compared to the present mesh-based \emph{transient-to-periodic} solver. \revised{Proof-of-concept, for the periodic CFD approach, is demonstrated here for 2D periodic heat diffusion and fluid flow problems.} The proposed PINN-based periodic solver primarily focuses on the time-periodic state,  optimizing the neural network model's trainable parameters to precisely fit a smaller time window (one time-period) rather than the temporal domain starting from the initial condition. After presenting a verification study, effect of the PINN-related various hyperparameters---such as the number of collocation points, neural network architecture, and point spacing for numerical differentiation---on computational time and accuracy are presented. Our results demonstrate that the PINN-based periodic solver takes substantially less computational time to achieve almost same accuracy as that obtained by the traditional transient-to-periodic solver. 			
			\\
			
		\end{abstract}	

	\end{frontmatter}

	\section{Introduction}
	
	Ubiquity of time-periodic unsteady fluid flow and heat transfer problems is evident across a wide range of natural and engineering applications, making their understanding and accurate simulation critically important. These problems are characterized by an initial transient phase which subsequently develops into a temporally periodic state, giving rise to repetitive patterns over time. A typical example of the periodic flow, from classical fluid mechanics, is the von-Karman vortex shedding caused by flow separation over bluff bodies \cite{Colonius2011}.  In biomedical engineering, pulsatile flow of blood through arteries is another crucial example \cite{Morab2024}. This type of flow is particularly relevant in the study of cardiovascular diseases, where understanding the flow patterns can help in diagnosing and assessing severity of conditions like stenosis \cite{Mehrabi2012}. Engineering applications further illustrate the diversity of periodic flow problems. In turbomachinery, the relative motion between the rotor and stator creates rotating flows that are inherently periodic in nature \cite{McMullen2001}. In water-hammer events, oscillatory pressure fields are known to induce periodic flow behaviour in water pipeline systems \cite{Cao2022}. Similar phenomenon is encountered in systems driven by a pulsating thermal boundary condition, like the solar energy harvester where the heat flux continuously varies throughout the day \cite{Xue2021}. Owing to their widespread occurrence in various domains, several studies have been conducted on time-periodic unsteady flow and heat transfer problems. \\

\begin{figure}
		\hbox{\hspace{3em}
			\includegraphics[width=150mm,scale=5]{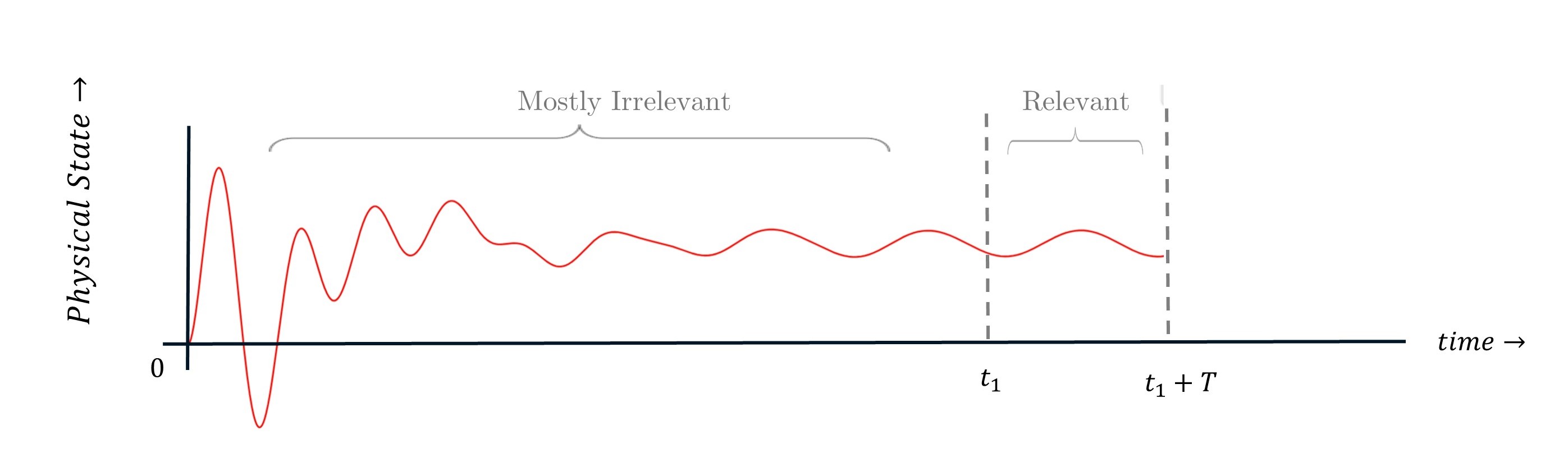}}
		\caption{Illustration of a transient solution reaching time-periodic unsteady state }
		\label{fig:Periodic state}
	\end{figure}

	One of the earliest strategies for addressing time-periodic unsteady problems dates back to 1995, with the computational method rooted in the philosophy of time marching \cite{Anderson1995}. In this finite difference-based method, the temporal derivatives are discretized using a Taylor series expansion of a specific order, and the flow properties are computed at successive time steps until they attain temporal periodicity. Such a straightforward time marching philosophy, however, has certain limitations. First, for an explicit method \cite{Sharma2021}, the strict stability criterion necessitates a very small time-step size that leads to high computational costs and substantial storage requirements. Second, for the periodic solutions, a significant portion of the total computational time is spent on simulating the initial non-periodic transient phase, which, while necessary, is mostly not needed for engineering applications; as depicted schematically in Figure \ref{fig:Periodic state}. Traditional finite difference method (FDM) and finite volume method (FVM) solvers, employing time marching techniques, offer no alternative but to initiate simulations from the initial condition, which entails solving through the entire transient phase to reach the desired periodic state. One potential solution to this issue is to initialize time-periodic unsteady simulations with the results of the steady-state equations-based simulation. However, as reported by Pfaller et al. \cite{Pfaller2021}, in the context of cardiovascular CFD, such simulations, even when initialized by a steady-state solution, often still require several cardiac cycles to reach a reasonable periodic state. A study by Johnson et al. \cite{Johnson1993} on uniperiodic flow past two cylinders reveals similar insights, where a large number of transient cycles elapse before the flow attains temporal periodicity at a Strouhal number of 0.222. Due to the challenges associated with simulating periodic flows, considerable work has been dedicated to expediting the attainment of periodic states in different partial differential equation (PDE) systems, as presented below.\\

	Carte et al. \cite{Carte1995} explored laminar wake behind a rectangular cylinder in a 2D channel, using a method that combines finite volume spatial discretization with Fourier series for time-periodic variables. Their discretization transformed the Navier-Stokes equations into a system of equations with real and imaginary parts which were solved separately, simplifying the problem. However, this method requires careful handling of real and imaginary components and does not address potential issues with discretization accuracy or the inclusion of sufficient harmonics. In the works of Hall et al. and Ekici and Hall  \cite{Hall2012, Ekici2012}, a harmonic balance technique was used to simulate unsteady flow in 2D cascades with harmonically vibrating airfoils. This method computes aerodynamic forces by representing time-dependent variables as harmonics of the fundamental frequency. However, the complexity of the resulting equations increases with the number of harmonics, requiring careful analytical handling of Fourier coefficients to accurately resolve spatial derivatives and achieve accurate simulation results. In 2007, Lubke et al. \cite{R2007} introduced a cascadic multilevel method to accelerate cyclic steady state (CSS) calculations in ModiCon-SMB processes. This technique involves three levels of resolution to enhance computational efficiency. However, while it reduces computational burden, it does not significantly shorten the transition from the initial transient state to the periodic state, and the authors concluded it to be less effective for speeding up the periodic calculations. Zhang et al. \cite{Zhang2011} developed a high-accuracy temporal-spatial pseudospectral (TSP) method for time-periodic unsteady fluid flow and heat transfer. This method uses detailed pseudospectral discretization, in both spatial and temporal dimensions, to achieve precise solutions for PDEs. However, it is highly complex and specific to the PDE being addressed, with no comparative analysis of computational time. Despite its accuracy, the method was slow to converge, taking 8 hours on a 3GHz CPU, which raises concerns about its practical applicability due to high computational demands. More recently, in 2021, Richter \cite{Richter2021} proposed an averaging scheme to efficiently approximate time-periodic flow problems, aiming to shorten the lengthy transition phase to a periodic state in Navier-Stokes solutions. However, this method, which depends on domain size, viscosity, and discretization parameters, has certain limitations: it is complex and specific to Navier-Stokes equations, lacks comparative analysis of computational time and accuracy, and only converges for the linear symmetric Stokes operator; limiting its general applicability. \revised{For the above discussed computational methods for solving periodic problems, a brief summary is presented in Table \ref{Table1}.}
	In contrast, a Physics Informed Neural Network (PINN)-based method offers a promising alternative for solving linear as well as non-linear PDEs---circumventing the need of the extensive analytical and algebraic treatment discussed in the aforementioned works. \\

	Raissi et al. \cite{Raissi2019} first introduced Physics-Informed Neural Networks (PINNs) for solving forward and inverse problems, such as the viscous Burgers, Poisson, and Schrödinger PDEs. For using neural networks in PINNs, their motivation is the well-known capability of neural networks as universal function approximators, as highlighted in the study by Dissanayake and Phan-Thien \cite{Dissanayake1994}. The core philosophy behind PINNs is to incorporate the governing PDEs into the training loss function. During training, the PINN aims to minimize the residual loss of the differential equations based on the model's output, which corresponds to the dependent variable of the PDE system. This physics-informed loss function ensures that the PINN model adheres to the PDEs, as well as the prescribed initial conditions (ICs) and boundary conditions (BCs), even without any labeled data. This approach makes PINNs a new class of \emph{meshless} PDE solvers that operate on the principle of neural network \emph{optimization}. Recently, PINNs have been applied to periodically fully developed flow in cases with repeating geometries, as demonstrated by Shah and Anand \cite{Shah2024}. However, their study reported $L_2$-norm of errors of $O(10^{-1})$ as compared to reference data and further, they found that the computational cost was significantly higher than that of a conventional FVM solver in ANSYS. This raises questions about the suitability of using the meshless PINNs over traditional mesh-based methods for periodic flows. To improve the accuracy and robustness of PINNs for more challenging problems, several extensions to the original formulation by Raissi et al. \cite{Raissi2019} have been proposed over the past few years. 
	These developments include new optimization algorithms for adaptive training \cite{McClenny2023, Wang2022, Wang2021}, where techniques on adaptively selecting samples of training data \cite{Nabian2021}, novel network architecture \cite{Jagtap2022}, new types of activation functions \cite{Jagtap2020}, and sequential learning strategies \cite{Mattey2022}. \\

	\begin{table}[h!] \small
		\caption{Summary of literature survey on  studies for time-periodic unsteady flow calculation through different computational methods}
		\centering
		\begin{tabular}{||c c c c||} 
			\hline
			\textbf{Author} & \textbf{Solver} & \textbf{Periodic Problem} & \textbf{Shortcomings} \\ [0.5ex] 
			\hline\hline
			Carte et al. \cite{Carte1995} & \makecell{FVM spatial discretization \\ and Fourier series for time- \\ periodic variables}  & \makecell{2D laminar flow \\ over square \\ cylinder} & \makecell{Insufficient harmonics \\ \& discretization accuracy} \\ \hline 
			Hall et al. \cite{Hall2012} & \makecell{Harmonic balance technique} & \makecell{2D cascade of \\ airfoils} & \makecell{Complex analysis of Fourier \\ coefficients to resolve \\ spatial derivatives} \\ \hline 
			Lubke et al. \cite{R2007} & \makecell{Cascadic multilevel method} & \makecell{ModiCon-SMB \\ processes} & \makecell{Ineffective for expediting \\  periodic computation} \\ \hline 
			Zhang et al. \cite{Zhang2011} & \makecell{Temporal-spatial \\ pseudospectral method } & \makecell{2D natural \\ convection in an \\ enclosure} & \makecell{High computational cost} \\ \hline 
			Richter \cite{Richter2021} & \makecell{Averaging scheme for \\ the approximation of \\ time-periodic flows} & \makecell{2D linear \\ Stokes  equations} & \makecell{Convergent only for \\ linear symmetric \\ Stokes operator} \\ [1ex] 
			\hline
		\end{tabular}
		\label{Table1}
	\end{table}

	The literature survey indicates that, to the best of our knowledge, there is no comprehensive demonstration of \emph{computationally efficient} PINN-based methods for expediting time-periodic unsteady problems, nor a thorough comparison of their efficacy against traditional diffusion and flow solvers. Therefore, based on the motivation and inadequacies discussed above, the present study has three objectives. The first objective is to present a \emph{PINN-based meshless CFD method,} for accelerating the attainment of temporally periodic states in simulations of simple and complex geometry unsteady \revised{2D} heat diffusion \revised{and fluid flow} problems. The second objective is to investigate computational performance (time and error norm) of our PINN-based solver under the effect of various model hyperparameters – such as the feedforward neural network architecture, number of collocation and boundary points, and the point spacing for numerical differentiation. The third and final objective is to compare computational performance of the present meshless PINN-based periodic diffusion \revised{and fluid flow}  solver with that of a FVM-based unsteady-to-periodic CFD solver on a Cartesian and curvilinear mesh. Such a comparison is intended to highlight the practical applicability, versatility, and limitations of our PINN-based periodic solver presented here for time-periodic unsteady heat diffusion \revised{and fluid flow} problems, involving simple and complex geometry.\\
	
	For the above objectives, Section \ref{Mathematical Formulation} presents detailed mathematical formulation and numerical methodology of the proposed PINN-based CFD model and its individual building blocks for the heat diffusion PDE. Thereafter, Section \ref{Computational Setup} presents the problem-specific computational setup and implementation details for two simple-geometry and one complex-geometry problems for heat diffusion, and one simple-geometry problem for fluid flow. \revised{Further, Section \ref{Performance Study} presents the results involving verification of the present PINN-based periodic solver for unsteady heat diffusion and fluid flow problems, and  performance study with respect to various coarse-grid FVM-based solvers.	
		Finally, conclusions
		are presented in \ref{Conclusion} for the proposed approach.}
	The discussed methodology for the PINN-based solver is implemented in \emph{Python} and the source codes are publicly made available on GitHub - \url{https://github.com/Lakshya-Chaplot/PINN-Periodic-Heat-Diffusion}.
	\\

	\section{Mathematical Formulation and Numerical Methodology} \label{Mathematical Formulation}
	
\begin{figure}[h!]
		\hbox{\hspace{1em}
			\includegraphics[width=160mm,scale=5]{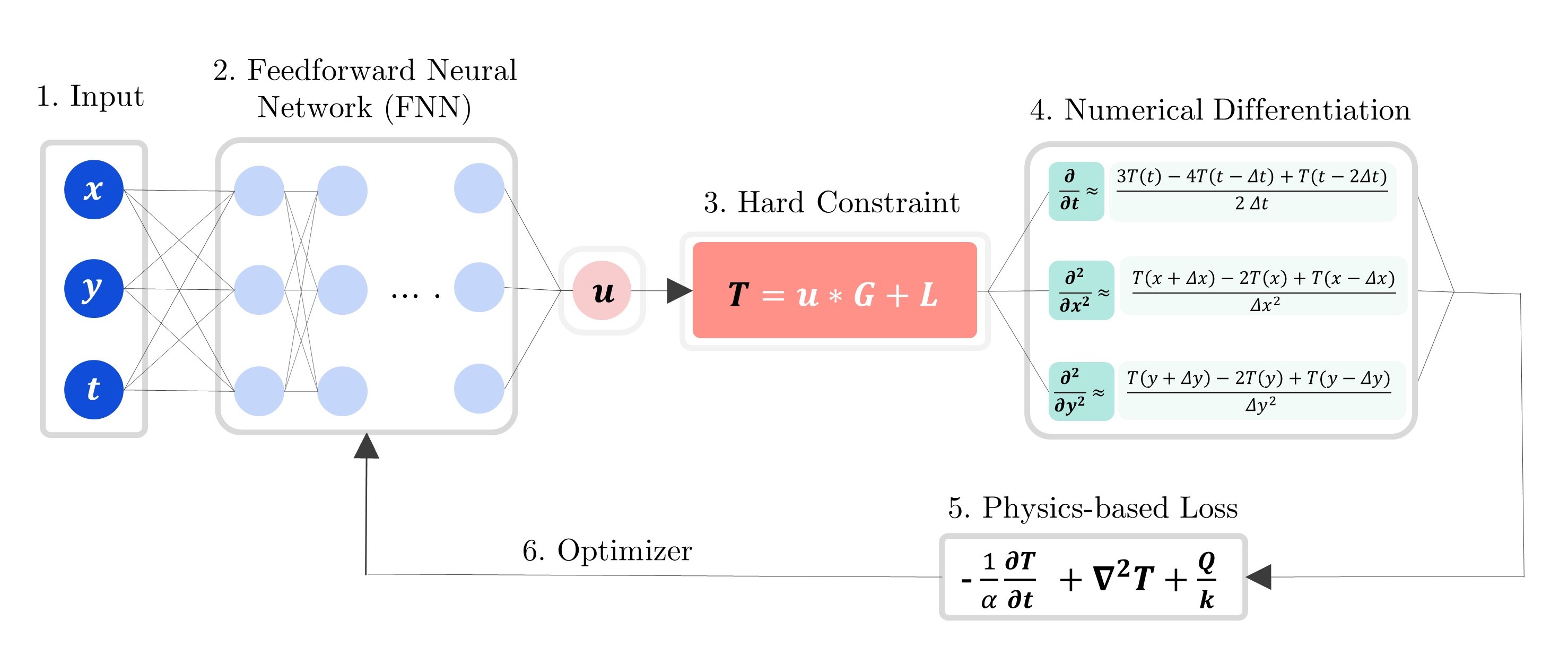}}
		\caption{Pipeline of the proposed PINN-based CFD approach }
		\label{fig:PINN model}
	\end{figure}

	A detailed mathematical formulation and numerical methodology of the proposed PINN formulation is presented below, \revised{for a 2D heat diffusion problem}, based on the different parts of the PINN pipeline shown in Figure \ref{fig:PINN model}.

	\newpage
	\subsection{Input Space and Sampling} 
	
	Two dimensional unsteady heat diffusion PDE, solved in the current study, is given as\\
	
	\begin{equation} 	\label{1}
		-\frac{1}{\alpha} \frac{\partial T}{\partial t} + \frac{\partial^2 T}{\partial x^2} + \frac{\partial^2 T}{\partial y^2} + \frac{Q(t)}{k}= 0, \quad t \in [0, t_{max}], \quad x, y \in \Omega,	
	\end{equation}\\
	
	\noindent subject to the initial and boundary conditions\\
	
	\begin{equation}
		T(x, y, 0) = f_{IC}(x, y), \quad x, y \in \Omega
	\end{equation}\\

	\begin{equation} \label{3}
		\mathcal{B}[T(x, y, t)]) = L(x, y, t), \quad t \in [0, t_{max}] \quad x, y \in \partial \Omega,
	\end{equation}\\
	
	\noindent where $\mathcal{B}[\cdot]$ is the boundary operator corresponding to the Dirichlet or periodic conditions ($L(x, y, t)$) on the 2D spatial domain $\Omega$. \textcolor{black}{Further, $k$ is the thermal conductivity and $\alpha$ is the thermal diffusivity of the medium.}
	Also, $T(x, y, t)$ describes the unknown transient solution that is governed by the PDE system of Eqn.\ref{1}.\\
	
	The unknown solution for temperature $T(x, y, t)$ is 
	approximated by a deep neural network given as  $T_{NN}(x, y, t; \textbf{w})$, following the pioneering work of Raissi et al. \cite{Raissi2019}. Here, the subscript \emph{`NN'} refers to the neural network. 
	The variable $\textbf{w}$ denotes the tunable parameters --- the weights and biases of the neural network model intended to be trained through an optimization process that aims to minimize a multi-objective loss function; given as\\
	
	\begin{equation} \label{4}
		\mathcal{L}(\textbf{w}) = \lambda_{IC} \mathcal{L}_{IC}(\textbf{w}) + \lambda_{BC} \mathcal{L}_{BC}(\textbf{w}) + \lambda_{PDE} \mathcal{L}_{PDE}(\textbf{w}) ,
	\end{equation}\\
	
	\noindent	where \\
	
	\begin{equation}
		\mathcal{L}_{IC}(\textbf{w}) = \frac{1}{N_{IC}} \Sigma_{i=1}^{N_{IC}} (T_{NN}(x_i, y_i, t_i;\textbf{w}) - T_{IC}(x_i, y_i, t_i))^2,
	\end{equation}\\

	\begin{equation}
		\mathcal{L}_{BC}(\textbf{w}) = \frac{1}{N_{BC}} \Sigma_{j=1}^{N_{BC}} (T_{NN}(x_j,y_j, t_j;\textbf{w}) - T_{BC}(x_j, y_j, t_j))^2,
	\end{equation}\\

	\begin{equation} \label{7}
		\mathcal{L}_{PDE}(\textbf{w}) = \frac{1}{N_{r}} \Sigma_{k=1}^{N_{r}} \left(\left[-\frac{1}{\alpha}\frac{\partial}{\partial t} + \frac{\partial^2}{\partial x^2} + \frac{\partial^2}{\partial y^2}   \right]T_{NN}(x_k, y_k, t_k;\textbf{w}) + \frac{Q(t_k)}{k}\right)^2 ,
	\end{equation}\\
	
	\noindent	where $\{ \lambda_{IC}, \lambda_{BC}, \lambda_{PDE}  \}$ are hyperparameters or coefficients used to balance the interplay between the different loss terms during model training process. The values of these  hyperparameters $\lambda$ are predefined and fixed throughout the context of this study. \revised{The subscripts \emph{`IC'} and \emph{`BC'} refer to initial condition and to boundary condition, respectively.}
	Here, the randomly distributed points $ \{ x_i, y_i, t_i \}_{i=1}^{N_{IC}} $ denote an initial condition  at time $t_i = 0$ in the full 2D spatial domain. The points $ \{ x_j, y_j, t_j \}_{j=1}^{N_{BC}} $ correspond to randomly distributed points, sampled at all or some of the boundaries $\partial \Omega$ of the domain. Finally, the points  $ \{ x_k, y_k, t_k \}_{k=1}^{N_{r}} $ denote the collocation points $N_r$ --- sampled in the spatio-temporal domain in a specific manner; discussed below.\\
	
\begin{figure}
		\hbox{\hspace{-0.5em}
			\includegraphics[width=175mm,scale=5]{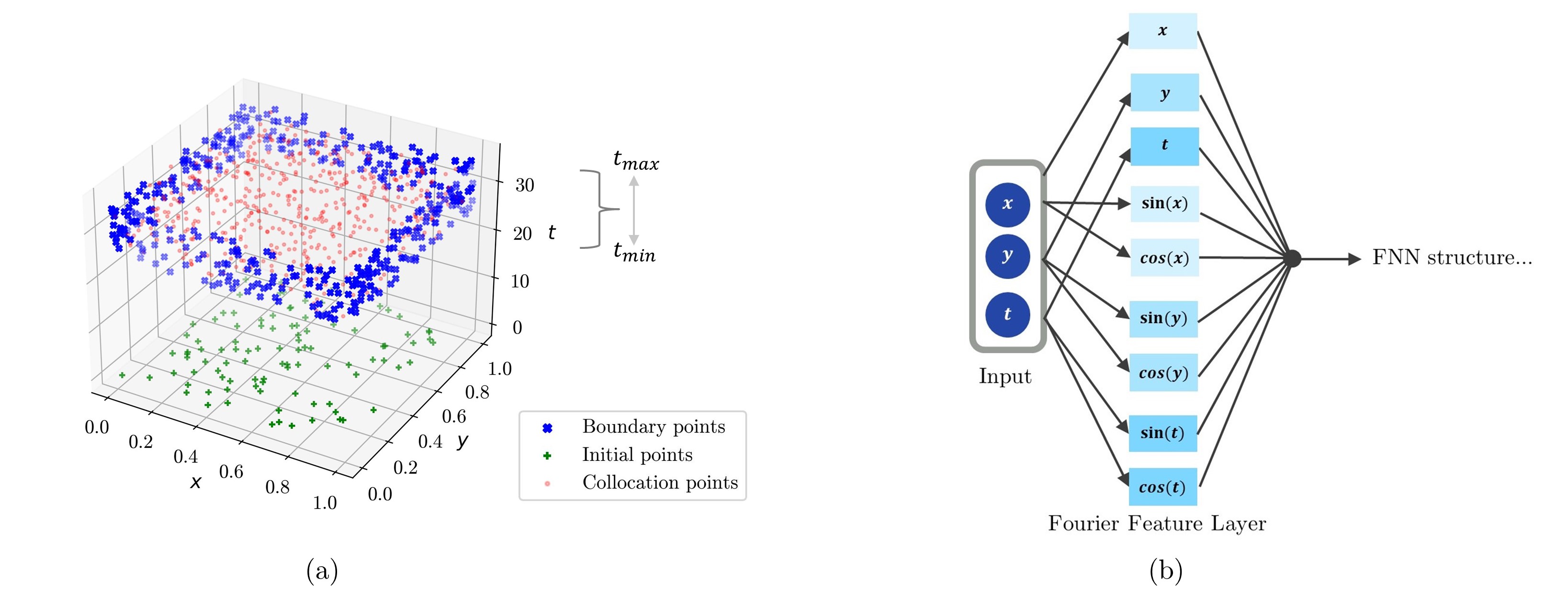}}
		\caption{(a) Illustration of various types of randomly distributed points, where the present sampling strategy is restricted to a narrow time span of one time-period $t_P = t_{max} - t_{min}$, for the periodic simulation (b) Fourier Feature Embedding (FFE) of the input space, which expands the 3 dimensional physical input to an intermediate 9 dimensional layer before being passed to the feedforward neural network (FNN)}
		\label{fig:Input Space_2}
	\end{figure}

	For a \emph{randomly} distributed set of points $(x, y, t)$ in the 3D domain, the novel sampling strategy involves starting the temporal dimension at a reasonably large time, $t = t_{min}$; instead of starting with the initial time $t = 0$ for the transient-to-periodic solution.
	Thus, the present work considers a narrow time span $t \in [t_{min} , t_{max}] $, that is exactly equal to one time-period of the periodic solution, as shown in Figure \ref{fig:Input Space_2}(a).
	For the random collocation points in the spatial dimensions $x$ and $y$, Latin Hypercube Sampling (LHS) is performed over the entire 2D spatial domain. The LHS method is reported to produce effective results \cite{Wu2023}, and is useful in reducing the number of simulations needed to quantify response uncertainty \cite{Dutta2020}. 
	Here, \emph{time} is regarded as an extra coordinate in which the spatial collocation point cloud (analogous to a grid in $x$ and $y$) is sampled only over one time period $t_P$, as the present interest is in the periodic solutions only. The solutions at any future time instant could be easily computed using the periodicity conditions of the system. \\

	\subsection{Feedforward Neural Network Structure}
	
	A simple feedforward neural network architecture is used in this work, ranging from 2 layers up to 8 layers, with each layer containing neurons that vary between 3 and 100. The impact of using both shallow and deep neural networks is examined and presented later in Section \ref{Performance Study}. An activation function, used in the present PINN model, is a sinusoidal function $sin(z)$. Such a choice has been seen more recently in the work of Chiu et al. \cite{Chiu2022}, and shown to have a more accurate solution by Wang et al and Wong et al. \cite{Wang2021, Wong2024}. A “$linear$”  activation function is used in the final (output) layer. In the present work, the weights and biases of the first layer are initialized using \emph{Glorot normal} and all the
	subsequent layers are initialized using \emph{He normal} as seen in the work of Chiu et al. \cite{Chiu2022}.\\

	The input layer, before being passed to the FNN structure (Figure \ref{fig:PINN model}), is expanded to higher-dimensional layer --- using the Fourier basis functions of the physical inputs $(x, y, t)$, as shown in Figure \ref{fig:Input Space_2}(b). For this work, only the first harmonic (principal frequency) is utilized. \textcolor{black}{This choice is motivated by the fact that for the periodic diffusion problems studied here, the fundamental mode dominates the response while higher-order harmonics decay rapidly due to the dissipative nature of diffusion. Hence, including additional harmonics was observed to provide negligible improvement relative to the added model complexity. Retaining only the fundamental mode results in a more compact and trainable network while still capturing the essential periodic behavior of the solution. We acknowledge that in problems with richer harmonic content, inclusion of higher harmonics may improve fidelity, and needs to be considered in future study.}
	This expansion transforms the original 3-dimensional input into a 9-dimensional input, which enhances the network's ability to capture the nonlinear behavior of the PDE solution  as described by Chen et al. \cite{Chen2021}. This approach, called Fourier Feature Embedding (FFE), improves the network's representation of complex dynamics inherent in the problem.\\

	\subsection{Hard Constraints} \label{Hard Constraints}

	During the training process, a common issue that PINNs encounter is an imbalance among various loss terms --- boundary condition loss, initial condition loss, and PDE residual loss (Eqn.\ref{4} - \ref{7}). While several loss balancing strategies have
	been explored in the literature \cite{McClenny2023,Li2022}, these approaches often suffer from errors and high computational time.	An alternative approach is to partially or completely eliminate the loss terms associated with boundary and initial conditions, by modifying the network architecture to strictly impose these conditions. This method not only ensures exact satisfaction of the boundary conditions, reducing computational cost, but also
	simplifies implementation as compared to that for the loss function-based approaches.	Inspiration for such a hard constraint-based formulation is taken from Lu et al. \cite{Lu2021},
	where the method has been used for steady state cases. The modified formulation for the present unsteady case is presented below.\\
	
	Consider continuous and smooth real-values functions $G(x,y,t)$ and $L(x,y,t)$ such that \\
	
	\begin{equation} \label{G_8}
		G(x_i,y_i,0) = 0 \quad \forall \quad  (x_i, y_i) \in \Omega ,  \quad {i \in [1, N_{IC}]} , \\
	\end{equation}

	\begin{equation} \label{G_9}
		G(x_j,y_j,t_j) = 0 \quad \forall \quad  (x_j, y_j, t_j) \in \partial \Omega , \quad {j \in [1, N_{BC}]}. \\
	\end{equation}\\
	
	Next, consider the function $L(x, y, t)$ from Eqn.\ref{3}  such that \\
	
	\begin{equation} \label{L_10}
		L(x_i,y_i,0) =  f_{IC}(x_i,y_i) \quad \forall \quad  (x_i,y_i) \in \Omega ,  \quad {i \in [1, N_{IC}]} , \\
	\end{equation}
	
	\begin{equation} \label{L_11}
		L(x_j,y_j,t_j) =  \mathcal{B}[T(x_j, y_j, t_j)] \quad \forall \quad  (x_j,y_j,t_j) \in \partial \Omega ,  \quad {j \in [1, N_{BC}]} , \\
	\end{equation}\\

	\noindent	where the arguments $(x_i, y_i, t_i)$ correspond to the number $N_{IC}$ of initial condition points (sampled at constant $t_i = 0$), and  $(x_j, y_j, t_j)$ correspond to the  number $N_{BC}$ of boundary condition points (sampled at the domain boundary $\partial \Omega $). \revised{The unsteady temperature field is now constructed using the FNN output ($u_{NN}(x, y, t)$ from Figure \ref{fig:PINN model}) as} \\

	\begin{equation} \label{12}
		\begin{aligned}
			T_{NN}(x,y,t) = u_{NN}(x,y,t) \cdot G(x,y,t) + L(x,y,t), \\
			\forall \quad t \in [t_{min}, t_{max}]  \quad (x,y,t) \in \Omega  \\
		\end{aligned}
	\end{equation}\\
	
	The elegance of the above equation lies in its inherent embedding of the boundary conditions as well as the initial condition into the formulation itself. Upon closer inspection, it's evident that at $t = 0$, the function $G(x, y, t)$ becomes zero (Eqn.\ref{G_8}), allowing function $L(x, y, t)$ to ensure the approximate temperature field $T_{NN}(x,y,t)$ exactly obeys the initial condition $f_{IC}(x_i,y_i)$ (Eqn.\ref{L_10}). Similarly, at the domain's boundary points ($\partial \Omega$), function $G(x, y, t)$ vanishes (Eqn.\ref{G_9}) and the function $L(x, y, t)$ guarantees that the temperature field strictly adheres to the boundary conditions (Eqn.\ref{L_11}) outlined in Eqn.\ref{3}. This rigorous constraint-based implementation enables the exact enforcement of boundary and initial conditions, leading to enhanced accuracy and reduced computational time. The specific functional forms of $G(x, y, t)$ and $L(x, y, t)$ will be presented in Section \ref{Performance Study}, for certain specific time-periodic unsteady heat diffusion problems.\\

	\subsection{Numerical Differentiation}
	
	The next component of the present PINN model, as shown in Figure \ref{fig:PINN model}, is \emph{numerical differentiation} that is required for computing the PDE operators of Eqn.\ref{7}.
	The PDE operators for the physics-based loss term are conventionally computed using Automatic Differentiation (AD), outlined originally in the work of Raissi et al. \cite{Raissi2019}. While the AD has the advantage of computing exact gradients at any point, PINNs often require very large numbers of collocation points and computational time to achieve high accuracy. Without sufficient collocation points, AD-based PINNs may converge towards unphysical solutions even when the training losses have been optimized to a very small value \cite{Chiu2022}. \\

\begin{figure}
		\hbox{\hspace{2em}
			\includegraphics[width=160mm,scale=5]{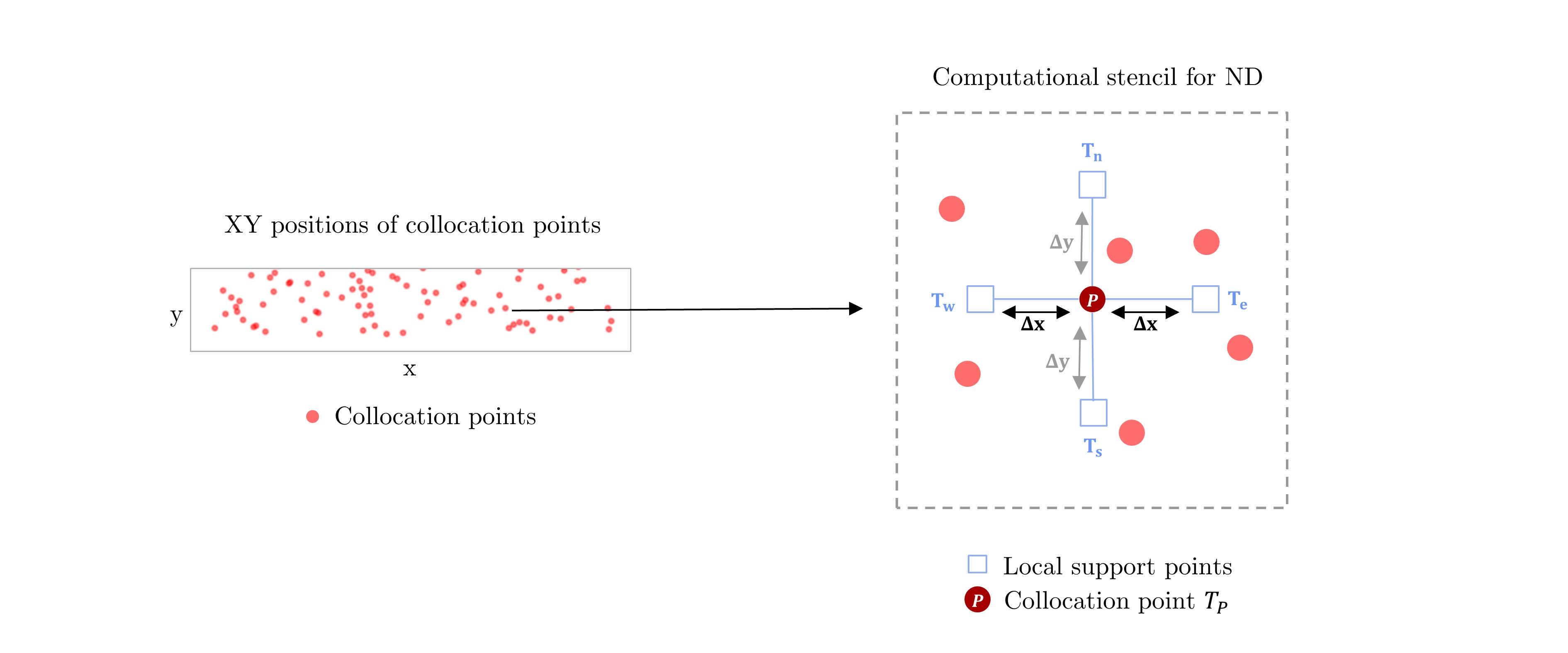}}
		\caption{Computational stencil for numerical differentiation at collocation point \emph{P}}
		\label{fig:ND}
	\end{figure}

	\textcolor{black}{In the current work, Numerical Differentiation (ND) is employed to construct the PDE residuals of the loss function, drawing inspiration from the original work of Chiu et al. \cite{Chiu2022}. By linking neighboring collocation points through ND stencils, the method enables efficient training even with relatively sparse samples.}	
	The AD is then used exclusively for the optimization process, where the weights and biases (the tunable parameters) of the neural network are updated via gradient descent, with gradients computed using the AD. This hybrid strategy balances the strengths of both AD and ND methods, improving training speed and accuracy. For the ND, Figure \ref{fig:ND} presents a 2D computational stencil that corresponds to a central difference scheme for the diffusion term.\\

	The ND is employed, instead of the AD, for the computation of the differential operators required in the physics-based training loss (Figure \ref{fig:PINN model}). As a very reliable and robust method, numerical differentiation is widely used in scientific computing and the computational physics community. The fundamental idea of numerical differentiation is to approximate the derivative terms by means of local support points (Figure \ref{fig:ND}). \\

	The second order numerical derivatives for the spatial terms and the first order numerical derivative for the temporal term, used in the loss function, for 2D unsteady heat diffusion (Eqn.\ref{1}) at point $P$ at $(x, y, t)$  are given as,\\

	\begin{equation} \label{13}
		\left(\frac{\partial^2 T}{\partial x^2}\right)_{p} \approx \frac{T_e - 2T_p + T_w}{\Delta x^2}  = \frac{ T_{NN}(x + \Delta x, y, t) - 2 T_{NN}(x, y, t) +  T_{NN}(x - \Delta x, y, t)}{\Delta x^2} ,
	\end{equation}\\

	\begin{equation} \label{14}
		\left(\frac{\partial^2 T}{\partial y^2}\right)_{p} \approx \frac{T_n - 2T_p + T_s}{\Delta y^2}  = \frac{ T_{NN}(x, y + \Delta y, t) - 2 T_{NN}(x, y, t) +  T_{NN}(x, y - \Delta y, t)}{\Delta y^2} ,
	\end{equation} \\

	\begin{equation} \label{15}
		\left(\frac{\partial T}{\partial t}\right)_{p} \approx  \frac{3  T_{NN}(x, y, t) - 4 T_{NN}(x, y, t - \Delta t) +  T_{NN}(x, y, t - 2\Delta t)}{2\Delta t} ,
	\end{equation}\\

	\noindent where $T_e$, $T_w$, $T_n$ and $T_s$ are the temperatures evaluated by the neural network at the neighboring support points (Figure \ref{fig:ND}). Further, the~\rom{2}-order central difference scheme is adopted for the spatial derivatives (Eqn.\ref{13} and Eqn.\ref{14}), whereas the~\rom{2}-order backward difference scheme is adopted for the temporal  derivatives (Eqn.\ref{15}). \\
	
	Some important features of the present ND formulation are as follows :
	
	\begin{itemize}
		\item The grid or point spacings $\Delta x$, $\Delta y$ and $\Delta t$ are hyperparameters now --- that means they require a careful \emph{a priori} selection for the training process. Appropriate
		choice of these free parameters is imperative in generating accurate and physical results. The performance of the present PINN model with respect to these point spacings is presented in greater detail in subsection \ref{point_spacing}.
		
		\item \textcolor{black}{To ensure stability and correctness, we have carefully controlled the point sampling process by restricting it to a subset of the valid physical domain. Specifically, for spatial and temporal domains defined by $[x_{\min}, x_{\max}]$, $[y_{\min}, y_{\max}]$, and $[t_{\min}, t_{\max}]$, the sampled collocation points are selected within the subdomains $[x_{\min} + \Delta x, x_{\max} - \Delta x]$, $[y_{\min} + \Delta y, y_{\max} - \Delta y]$, and $[t_{\min} + 2\Delta t, t_{\max}]$, respectively. By offsetting the sampling boundaries inward by small distances $\Delta x$, $\Delta y$, and $\Delta t$, we avoid placing any sampled points outside the physically valid region while performing numerical differentiation. This careful choice of sampling domain guarantees that all collocation points remain well within the simulation boundaries, thereby ensuring both numerical stability and the physical correctness of the computed solution.}
		
		\item The presented ND-based PINN approach shares the same advantages as AD-based PINNs, particularly in being mesh-free and producing a continuous function $T_{NN}(x, y, t)$ as the solution.  This continuity arises from the neural network's inherent structure, where $T_{NN}$ is represented as a differentiable function of the input variables $x, y$ and $t$ owing to the smooth activation functions used in the network \cite{Raissi2019}. The model minimizes the loss function at a finite number of collocation $N_r$, boundary $N_{BC}$, and initial $N_{IC}$ points (Eqn.\ref{4}-\ref{7}); however, the resulting function $T_{NN}(x, y, t)$ can predict the temperature field at any point within the solution domain.
		Unlike traditional FVM solvers, which provide discrete temperature values only at specific grid points, the neural network outputs a \emph{continuous} approximation that can be evaluated at any arbitrary point in space and time, allowing for predictions with effectively infinite resolution.

		\item The present ND-based formulation relies on local or neighboring support points. However, unlike the mesh in FVM/FDM, the collocation points here don't need to align precisely. This flexibility makes the ND-based PINN approach robust in selecting collocation points, enabling it to learn accurate physical solutions even with sparsely sampled data, as demonstrated later in Section \ref{Performance Study}. \\
	\end{itemize}

	\subsection{Physics-based Loss}
	
	The optimization problem along with the overall loss of (Eqn.\ref{4}) is converted from a multi-objective to a single-objective optimization problem by using a fully hard constraint-based formulation (Eqn.\ref{12}), owing to the elimination of the separate initial and boundary condition loss terms. The resulting single objective optimization problem is given as\\
	
	\begin{equation}
		\textbf{min}_w \{  Loss(\textbf{w}) \},
	\end{equation}\\
	
	\noindent	where 
	
	\begin{equation} \label{17}
		Loss(\textbf{w}) =  \frac{1}{N_{r}} \Sigma_{k=1}^{N_{r}} \left(\left[-\frac{1}{\alpha}\frac{\partial}{\partial t} + \frac{\partial^2}{\partial x^2} + \frac{\partial^2}{\partial y^2}   \right]T_{NN}(x_k, y_k, t_k;\textbf{w}) + \frac{Q(t_k)}{k}\right)^2 ,
	\end{equation}\\

	for the tunable weights and biases tensor $\textbf{w}$.\\

	\subsection{Optimizer}

	The PINN model requires an optimizer to iteratively update the weights and biases ($\textbf{w}$) of the neural network (trainable parameters) based on the gradients of the loss function (Eqn.\ref{17}), which reflects the degree to which the predicted solution violates the governing equation. This process allows the neural network to improve its approximation until it converges to an accurate solution, or until the maximum number of  iterations are completed.\\
	
	The optimization algorithm used in this study is \emph{Adam} \cite{Kingma2014}, which is a stochastic gradient descent based optimizer. For the Adam-based optimization during the implementation of the present PINN-based solver, our choices are as follows : \\

	\begin{itemize}
		\item {Iterations:} The model undergoes 10,000 iterations during training for all the tackled problems regardless of their geometry, FNN architecture or number of collocation points. These iterations allow the network to adjust its
		parameters and reduce the overall loss
		
		\item {Learning Rate Schedule:} An exponentially decaying learning rate schedule
		is employed during the training. The learning rate decreases over time, allowing for
		finer adjustments as the optimization process proceeds that helps in achieving
		convergence smoothly
		
		\item {Learning Rate:} The initial learning rate is set anywhere between $[0.02, 0.08]$ depending on the problem studied here
		
		\item {Decay Rate : }The learning rate is decayed by 80\% at regular intervals during training. This decay ensures that the optimization process becomes more focused and precise over time
		
		\item {Decay Steps :} The learning rate is decayed at intervals of 400 iterations
	\end{itemize}

	All the deep learning-based implementation in this study, including the use of Adam optimizer, has been done in \emph{Python} using the open source machine learning library \emph{TensorFlow} \cite{Adam, tf}.

	\subsection{Solution Algorithm for PINN-based Periodic Solver}

	Using the above discussed mathematical formulation and numerical methodology, solution algorithm for PINN-based periodic heat-diffusion solver is as follows\\
	
	\begin{enumerate}
		
		\item Generate $N_r$ collocation points  $ \{ x_k, y_k, t_k \}_{k=1}^{N_{r}} $ sampled using Latin Hypercube Sampling (LHS) from the spatio-temporal domain $[t_{min} , t_{max}] \times \Omega$, $N_{IC}$ initial condition points $ \{ x_i, y_i, t_i \}_{i=1}^{N_{IC}} $ sampled from $\{t_i=0\}$$\times \Omega$, and $N_{BC}$ boundary points $ \{ x_j, y_j, t_j \}_{j=1}^{N_{BC}} $ from $[t_{min} , t_{max}] \times \partial\Omega$; as per the solution domain of Eqn.\ref{1}.\\
		
		\item Select the feedforward neural network (FNN) model's architecture (number of layers $n_L$, number of neurons per layer $n_N$ and non-linear activation function). Additionally, incorporate a Fourier Feature Embedding intermediate layer to transform the input space to a higher dimensional space.\\
		
		\item Initialize the FNN model's trainable parameters $\textbf{w$^0$}$, and use \emph{Glorot normal} initialization for the first layer and \emph{He normal} initialization for all the subsequent layers.\\
		
		\item Define appropriate functions  $G(x,y,t)$ and $L(x,y,t)$ for embedding the initial and boundary constraints either fully or partially, within the network architecture, as per Eqns.\ref{G_8}-\ref{12}. Also specify hyperparameters $\{ \lambda_{IC}, \lambda_{BC}, \lambda_{PDE} \}$ in case of a soft constraint-based implementation as per Eqn.\ref{4}.\\
		
		\item Set point spacing hyperparameters $\Delta x$, $\Delta y$ and $\Delta t$ for the numerical differentiation; required for computing the PDE operators in Eqn.\ref{7}. \\
		
		\item Construct the loss function $\mathcal{L}(\textbf{w})$ (Eqn.\ref{4}), based on a general multi-objective optimization approach. \\
		
		\item Minimize the loss function $\mathcal{L}(\textbf{w})$ via the Adam optimizer and update the weights $\textbf{w$^i$}$ after every iteration \textbf{$i$} ($i \in [1, Iter_{max}]$). Adam-specific hyperparameters, including the initial learning rate, decay rate, and decay steps, should be appropriately chosen to ensure stable and efficient convergence. \\
		
	\end{enumerate}

	\section{Computational Setup and Implementation Details for Test Problems} \label{Computational Setup}
	
	\revised{	For the present PINN-based periodic solver, comprehensive results and discussion are presented below for three different 2D heat diffusion problems and one 2D incompressible fluid flow problem; subject to a periodic boundary condition. \emph{Problem-specific} computational setup and implementation details for the present PINN-based periodic solver are presented in this section.\\}

\begin{figure}
		\hbox{\hspace{-0.2em}
			\includegraphics[width=170mm,scale=5]{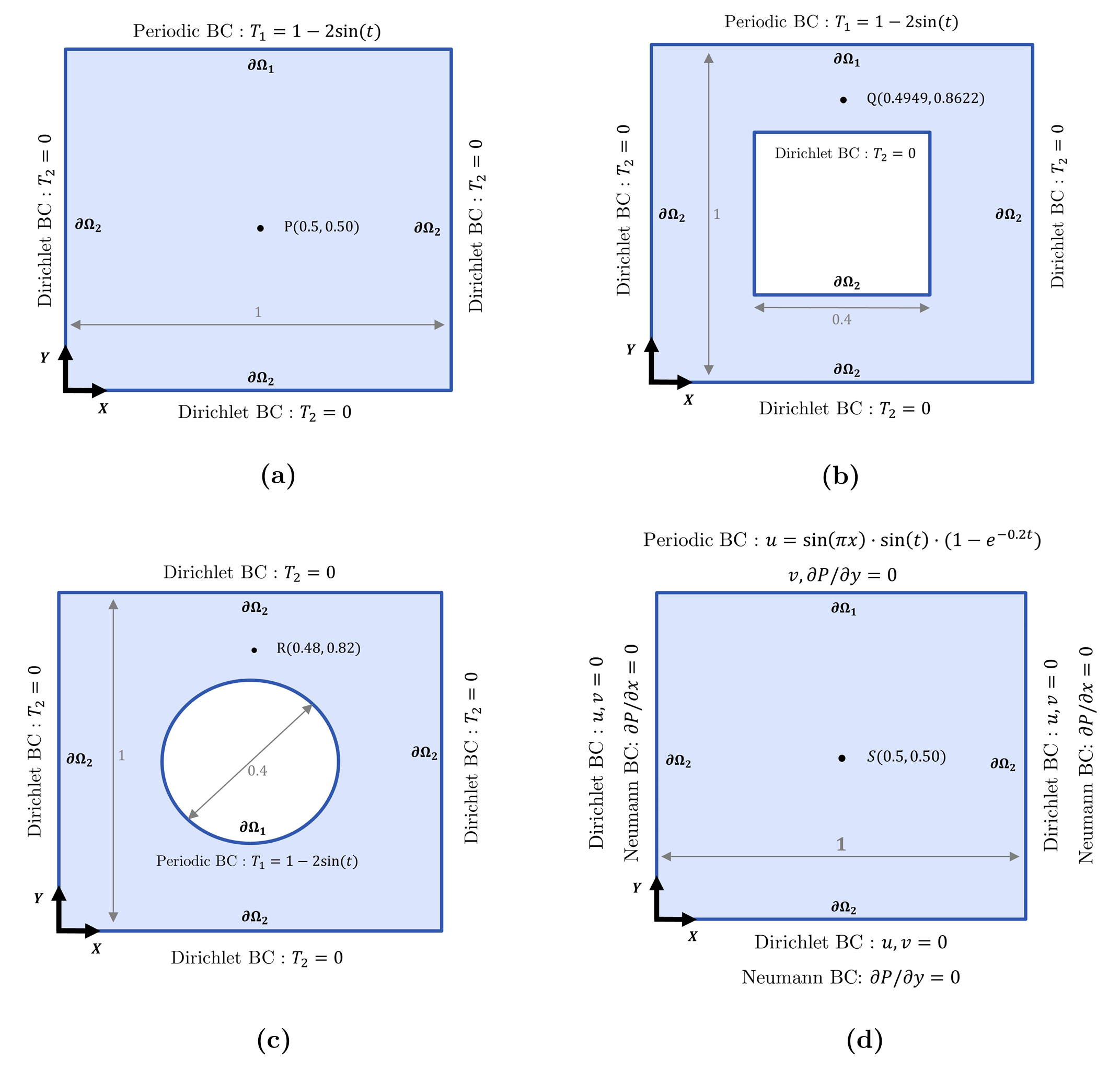}}
		\caption{Computational setup for a transient-to-periodic 2D heat diffusion in a square plate, with {(b}) square and {(c)} circular holes\revised{, and (d) 2D fluid flow in an oscillating lid driven cavity}}
		\label{fig:Domain 2D All 2}
	\end{figure}

	For the	three test problems on the periodic heat diffusion, Figure \ref{fig:Domain 2D All 2} shows geometric details and boundary conditions for heat diffusion in a plate --- with and without a hole. Note from the Figure \ref{fig:Domain 2D All 2} that all the problems involve a time-periodic Dirichlet boundary condition --- at the top wall in Figure \ref{fig:Domain 2D All 2}(a) and (b), and inner circular-boundary in Figure \ref{fig:Domain 2D All 2}(c). Also, each of the three heat diffusion problems considers an exponentially decaying volumetric heat generation rate  $Q(t) = Q_o\ e^{-t /\ \tau}$. Here, $Q_o = 25 \text{W}/m^3$ for the first problem, and $Q_o = 500 \text{W}/m^3$ for the other two problems, along with  $\tau = 5s$ for all the problems. 
	Also, for all the problems, the temporal domain $[t_{min}, t_{max}]$ is chosen such that $ t_{max} -  t_{min} = 2\pi$ s (time period of the \revised{time-periodic Dirichlet boundary conditions} in Figure \ref{fig:Domain 2D All 2}), with $t_{min} = 28.72s$ and $t_{max} = 35s$ for the first problem while $t_{min} = 43.72s$ and $t_{max} = 50s$ for the other problems. \\
	
	\revised{For the 2D incompressible fluid flow problem, Figure \ref{fig:Domain 2D All 2}(d) shows the geometric details and boundary conditions for an oscillating lid-driven cavity (LDC) flow problem. The figure shows a periodic boundary condition for the $u$-velocity at the top wall, with all other walls subjected to a no-slip boundary condition of $u=v=0$.  The temporal domain $[t_{min}, t_{max}]$ is chosen such that $ t_{max} -  t_{min} = 2\pi$ s, with $t_{min} = 26.72s$ and $t_{max} = 33s$. The non-dimensional parameter governing the flow physics is Reynolds number ($Re = u_c L / \nu$, where characteristic velocity $u_c$ is space-averaged velocity of the lid), with $Re = 10, 50$ and $100$ for the present study.\\}

	After presenting the computational setup here, implementation details on the types of constraints (for the present PINN-based approach) pertaining to the three heat diffusion and one fluid flow problems are presented in separate subsections below. 
	
	\subsection{Implementation Details for Heat Diffusion in a Plate}

	For the problems in Figures \ref{fig:Domain 2D All 2}(a) and (b), a \emph{hybrid} constraint-based PINN approach is utilized where the hard constraint (Section \ref{Hard Constraints}) is used for the initial condition and the soft constraint is applied for the boundary conditions. Further, for a hybrid constraint, the functional forms of $G(x, y, t)$ and $L(x, y, t)$ are chosen here as\\
	
	\begin{equation} \label{18_G_t}
		G(x, y, t) = t, \quad L(x, y, t) = 0
	\end{equation}\\
	
	\noindent with the final temperature field $T_{NN}(x, y, t)$ given as\\

	\begin{equation} \label{25}
		\begin{aligned}
			T_{NN}(x,y,t) = u_{NN}(x,y,t) \cdot t, \quad
			\forall \quad t \in [t_{min}, t_{max}],  \quad (x,y,t) \in \Omega 
		\end{aligned}
	\end{equation}\\

	It is important to highlight our choice of the soft constraint for the boundary conditions, which led us to use the hybrid constraint approach, for the couple of problems in Figures \ref{fig:Domain 2D All 2}(a) and (b). The choice is due to the discontinuous nature of the periodic boundary conditions at the north-west and north-east corners of the 2D square domain in these problems. At these corners, the temperature abruptly jumps from a zero value at the left and right boundaries to a non-zero value at the top wall. As a result of this singularity in boundary conditions, continuous and smooth distance functions for $G(x, y, t)$ and $L(x, y, t)$, that would exactly conform to the singular boundary conditions, cannot be ensured by the hard constraint.
	\textcolor{black}{For the hybrid-constraint formulation, the PDE residual loss was assigned a weight of $\lambda_{PDE} = 1$, while the boundary condition loss was given $\lambda_{BC} = 15$ for the problems in Figures \ref{fig:Domain 2D All 2}(a) and (b). The initial condition was enforced as a hard constraint, and hence no additional weight was required. The value of $\lambda_{BC}$ was chosen empirically: weights in the range of $10$-$20 \times \lambda_{PDE}$ ensured strong satisfaction of boundary conditions while maintaining stable training. In contrast, smaller weights led to boundary mismatch, whereas excessively large weights hindered convergence. The reported values were found to provide a good balance in the present problems.
	}
	\\

	For the problem in Figure \ref{fig:Domain 2D All 2}(c), since the periodic boundary condition is applied on the inner continuous circular boundary and there is no singularity in the boundary conditions, a fully hard constraint-based approach is utilized here, and the functional forms of $G(x, y, t)$ and $L(x, y, t)$  are chosen as\\
	
	\begin{equation} \label{20_G_f_t}
		G(x, y, t) = t \cdot (1-f(x,y)) \cdot f(x,y), \quad L(x, y, t) = f(x,y) \cdot (1 - 2sin(t)),
	\end{equation}\\
	
	\noindent where the function $f(x, y)$ is defined as\\
	
	\begin{equation}
		f(x, y) = \frac{d_2(x,y)}{d_1(x,y) + d_2(x,y)},
	\end{equation}\\
	
	\noindent and auxiliary distance functions $d_1$ and $d_2$ are constructed such that they are exactly zero at the boundaries $\partial \Omega_1$ and $\partial \Omega_2$ respectively, as shown in Figure \ref{fig:Domain 2D All 2}(c). These functions are defined as
	
	\begin{equation}
		d_1(x, y) = \frac{\sqrt{(x-0.5)^2 + (y-0.5)^2} - 0.2}{0.30},
	\end{equation}\\

	\begin{equation}
		d_2(x, y) = \frac{x (1-x) y (1-y)}{0.0525}
	\end{equation}\\

	\noindent that results in the final temperature field $T_{NN}(x, y, t)$ as\\

	\begin{equation} \label{25}
		\begin{aligned}
			T_{NN}(x,y,t) = u_{NN}(x,y,t) \cdot G(x,y,t) + L(x,y,t), \quad
			\forall \quad t \in [t_{min}, t_{max}],  \quad (x,y,t) \in \Omega 
		\end{aligned}
	\end{equation}\\

	\textcolor{black}{In this work, the auxiliary functions $G(x,y,t)$ and $L(x,y,t)$ were chosen in simple polynomial and Eulerian distance-like forms to satisfy vanishing conditions while remaining smooth and bounded across the non-dimensional spatio-temporal domain. For instance, in Eqn.\ref{18_G_t}, $G(x, y, t)=t$ vanishes at the initial condition and stays within $[0,1]$, while in Eqn.\ref{20_G_f_t}, $G(x, y, t) = t \cdot f \cdot (1-f)$ with $f \in [0,1]$ ensures boundedness and smoothness. Keeping $G$ and $L$ bounded avoids extreme scaling that would otherwise burden the feedforward neural network by over- or under-compensating in the ansatz $T = u_{NN} \cdot G + L$.  Although these choices are problem-specific, they act only as smooth multiplicative masks, guiding constraint satisfaction without hindering convergence. This is consistent with prior work where geometry-aware multipliers enforce exact boundaries. For example, Sukumar and Srivastava \cite{Sukumar2022} proposed a systematic distance-function framework for constructing such multipliers in general geometries. Such distance-function approach provides a natural extension of the current framework, which can be considered in future work on more general geometries.}

	\revised{
	
	\subsection{Implementation Details for Oscillating Lid Driven Cavity Flow}
	
	While the preceding sections focus on time-periodic unsteady heat diffusion problems, the proposed PINN-based method is extended here for two-dimensional incompressible Navier-Stokes (NS) equations. The FVM-based flow solver, used for obtaining reference solutions in the 2D periodic studies, is an in-house implementation that is independently verified against steady state results of Ghia et al. \cite{Ghia1982}, for a LDC flow problem; presented in Appendix \hyperref[Appendix_B]{B.}\\
	
	For 2D unsteady flow, the governing Navier-Stokes equations are given as\\

	\begin{equation}\label{continuity}
		\frac{\partial u}{\partial x} + \frac{\partial v}{\partial y} = 0,
		\end{equation}  
		
		\begin{equation} \label{X_mom}
	\frac{\partial u}{\partial t}
	+ u \frac{\partial u}{\partial x}
	+ v \frac{\partial u}{\partial y}
	= -\frac{\partial P}{\partial x}
	+ \frac{1}{Re} \left(
	\frac{\partial^2 u}{\partial x^2}
	+ \frac{\partial^2 u}{\partial y^2}
	\right),
	\end{equation}  
	
	\begin{equation} \label{Y_mom}
	\frac{\partial v}{\partial t}
	+ u \frac{\partial v}{\partial x}
	+ v \frac{\partial v}{\partial y}
	= -\frac{\partial P}{\partial y}
	+ \frac{1}{Re} \left(
	\frac{\partial^2 v}{\partial x^2}
	+ \frac{\partial^2 v}{\partial y^2}
	\right),
	\end{equation} \\

	\noindent where $Re = \frac{u_c \cdot L_c}{\nu}$ is the Reynolds number.\\

	In this subsection, a periodic lid-driven cavity (LDC) flow (Figure \ref{fig:Domain 2D All 2}(d)) is considered as a representative benchmark problem for the verification and performance study. 	
	The PINN-based incompressible flow solver employs the same overall pipeline as the heat diffusion-based solver shown in Figure \ref{fig:PINN model}. Separate FNNs are used to approximate each of the velocity components and the pressure field, namely $u(x,y,t)$, $v(x,y,t)$, and $P(x,y,t)$. The corresponding hard-constrained ansatz for the solution fields is defined as\\
	
	\begin{equation}
	u(x, y, t) = u_{NN}(x, y, t) \cdot x(1-x) \cdot y(1-y) \cdot t  + sin(\pi x) sin(t) (1-e^{-0.2t}) \cdot y ,
	\end{equation}
	
	\begin{equation}
	v(x, y, t) = v_{NN}(x, y, t) \cdot x(1-x) \cdot y(1-y) \cdot t,
	\end{equation}
	
	\begin{equation}
	P(x, y, t) = P_{NN}(x, y, t) \cdot t,
	\end{equation}\\
	
	\noindent where $u_{NN}$, $v_{NN}$, and $P_{NN}$ denote the outputs of the respective FNNs. This formulation enforces the initial conditions $u(x,y,0)=v(x,y,0)=P(x,y,0)=0$ exactly through the multiplicative time factor. In addition, the spatial factors impose hard boundary constraints consistent with the LDC geometry of Figure \ref{fig:Domain 2D All 2}(d). Specifically, the pressure field is enforced with a hard initial condition only, while its Neumann boundary conditions are imposed as soft constraint. 
	The oscillating lid (dictated by $u-$velocity BC) induces time-periodically varying flow properties and helps in building a challenging test case for the present PINN-based periodic solution methodology.\\

	The PINN-based NS solver presented in this section incorporates two additional penalty terms to explicitly enforce temporal periodicity over the solution interval $[t_{min}, t_{max}]$. These terms penalize discrepancies between the predicted flow field and its temporal derivative at the beginning and end of one time period as,

	\begin{equation} \label{loss_periodic}
	\mathcal{L}^{per}_{\textbf{u}}(\textbf{w}) = \frac{1}{N_{r}} \Sigma_{k=1}^{N_{r}} \left(\textbf{u}_{PINN}(x_k, y_k, t_{min};\textbf{w}) - \textbf{u}_{PINN}(x_k, y_k, t_{max};\textbf{w})\right)^2 ,
	\end{equation}\\
	
	\begin{equation} \label{loss_periodic_del_t}
	\mathcal{L}^{per}_{\mathbf{\dot u}}(\textbf{w}) = \frac{1}{N_{r}} \Sigma_{k=1}^{N_{r}} \left(\mathbf{\dot u}_{PINN}(x_k, y_k, t_{min};\textbf{w}) - \mathbf{\dot u}_{PINN}(x_k, y_k, t_{max};\textbf{w})\right)^2 ,
	\end{equation}\\

	\noindent where  $\mathbf{u}_{PINN}$ is the velocity vector $\mathbf{u} = [u, v]$ predicted by the entire PINN model, and $\mathbf{\dot u}_{PINN}$ is the corresponding temporal derivative obtained using numerical differentiation. These penalty terms enforce equality of both the velocity field and its temporal derivative at the start and end of the time interval, thereby ensuring that the learned solution satisfies the time-periodic condition associated with the physical system. Consequently, the optimizer is guided toward solutions that are consistent with the underlying periodic dynamics of the flow. Since the PINN-based flow solver requires soft constraints for selected boundary conditions (for $P$) as well as the governing equations, the overall loss function is given as\\

	\begin{equation} \label{loss_flow}
	\mathcal{L}(\textbf{w}) = 
	\lambda_{BC} \mathcal{L}_{BC}(\textbf{w}) + \lambda_{R} ( \mathcal{L}_{R1}(\textbf{w}) + \mathcal{L}_{R2}(\textbf{w}) + \mathcal{L}_{R3}(\textbf{w})) + \lambda_{period} (\mathcal{L}^{per}_{\mathbf{u}}(\textbf{w})  + \mathcal{L}^{per}_{\mathbf{\dot u}}(\textbf{w}))  ,
	\end{equation}\\
	
	\noindent where $\mathcal{L}_{BC}$ represents the pressure BC loss, and $\mathcal{L}_{R1}$, $\mathcal{L}_{R2}$, and $\mathcal{L}_{R3}$ correspond to the PDE residual losses associated with the continuity equation (Eqn.\ref{continuity}), the $x$-momentum equation (Eqn.\ref{X_mom}), and the $y$-momentum equation (Eqn.\ref{Y_mom}), respectively. For the numerical differentiation, \rom{2}-order central difference scheme is adopted for all the spatial derivatives while \rom{2}-order backward difference scheme is adopted for the temporal derivatives. \\
	
}

	\section{Results and Discussion} \label{Performance Study}

	\revised{In this section, using our in-house PINN-based solvers, results are presented in separate subsections below on verification and performance studies for the PINN-based periodic diffusion and flow solvers as compared to FVM-based transient-to-periodic solvers.  Note that a FVM-based solver is also developed in \emph{Python}, for the present work,  to demonstrate relative accuracy and computational time of our novel PINN-based solvers as compared to the traditional FVM-based solvers. 	} Also note that the relative verification and performance is demonstrated here for a periodic solution --- obtained by the FVM-based \emph{transient-to-periodic} simulations and PINN-based \emph{direct periodic} simulations. All the simulations performed in this study, using the PINN-based as well the FVM-based solvers, are on an Intel i7-2700K system running with 8 logical cores. \revised{The reported computational times correspond to the CPU execution time consumed during the training phase of the PINN-based solver, measured using per-core user and system CPU times and reported as the effective parallel runtime. This choice was made to minimize the influence of operating system overhead and background processes inherent to shared environments (e.g., Jupyter notebooks under Windows OS). For full transparency, wall-clock time and detailed CPU profiling (including per-core and total CPU usage) are provided in the accompanying open-source repository.} Testing/inference is nearly instantaneous and therefore not included.\\

	For the verification and performance study, computational time and $L_2$ norm error are presented as the mean value of five independent runs for the PINN solver, along with the corresponding standard deviation (\emph{SD}) to indicate variability. The mean-$L_2$ norm error is the main figure of merit to compare the error between the PINN generated
	solution ($T_{NN}$) and a reference solution ($T_{ref}$). The reference solution here is a grid-independent fully explicit FVM-based solution unless otherwise stated. Using a PINN-based solution $T_{NN}$ and reference solution $T_{ref}$, the $L_2$ norm error is given as,\\
	
	\begin{equation}
		{L_2} \text{ error } (\%) = \sqrt{\frac{\Sigma_i^{N_u}|T_{NN}(x_i, y_i, t_i, \Delta x, \Delta y, \Delta t) - T_{ref}(x_i, y_i, t_i)|^2}{\Sigma_i^{N_u}|T_{ref}(x_i, y_i, t_i)|^2}} \times 100,
	\end{equation}\\
	
	\noindent and absolute $L_1$ error is given as,
	
		\begin{equation}
		{L_1(x_i, y_i, t_i)}  = {|T_{NN}(x_i, y_i, t_i, \Delta x, \Delta y, \Delta t) - T_{ref}(x_i, y_i, t_i)|},
	\end{equation}\\

	\noindent where $T_{NN}(x_i, y_i, t_i)$ is obtained at grid points $(x_i, y_i)$ and time instants $t_i$---corresponding to the high fidelity grid-independent FVM-based solver. Further, $N_u$ represents the total number of spatio-temporal points in the FVM-based solver. 
	\\

	\subsection{Verification, Hyperparameters, and Performance Studies: PINN-based Periodic Heat Diffusion Solver}	
	
	For the current PINN-based time-periodic unsteady heat diffusion solver, verification study is presented in separate subsections below, for both 1D and 2D heat diffusion problems. Further, for the verification study, the reference solution is analytical solution for the 1D diffusion while a grid-independent \emph{periodic} numerical solution (obtained from a transient-to-periodic simulation by our FVM-based solver) is used for the 2D diffusion. \revised{The FVM-based diffusion solver used as a reference in the 2D periodic studies is an in-house code, which is independently verified against analytical solution for 2D heat diffusion, presented in Appendix \hyperref[Appendix_A]{A.}}

	\subsubsection{Verification Study: 1D Steady Heat Diffusion}

	For 1D unsteady heat diffusion, the governing equation and spatio-temporal temperature field are given as\\
	
	\begin{equation} \label{18-anal}
		-\frac{1}{\alpha} \frac{\partial T}{\partial t} + \frac{\partial^2 T}{\partial x^2} = 0, \quad t \in [0, 0.2], \quad x \in [0, 1],	
	\end{equation}\\

	\noindent subject to the initial and boundary conditions\\
	
	\begin{equation} \label{19-anal}
		T(x, 0) = 2sin(\pi x), \quad x \in [0, 1]
	\end{equation}
	
	\begin{equation} \label{20-anal}
		T(x = 0, t) = T(x = 1, t) = 0, \quad t \in [0, 0.2].
	\end{equation}\\
	
	For the above set of equations, the resulting analytical solution is given as \cite{Ramabathiran2021}\\
	
	\begin{equation}\label{21}
		T(x,t) = 2sin(\pi x) \cdot exp(-\pi^2 t)
	\end{equation}\\
	
\begin{figure}[h]
		\hbox{\hspace{-0.0em}
			\includegraphics[width=175mm,scale=5]{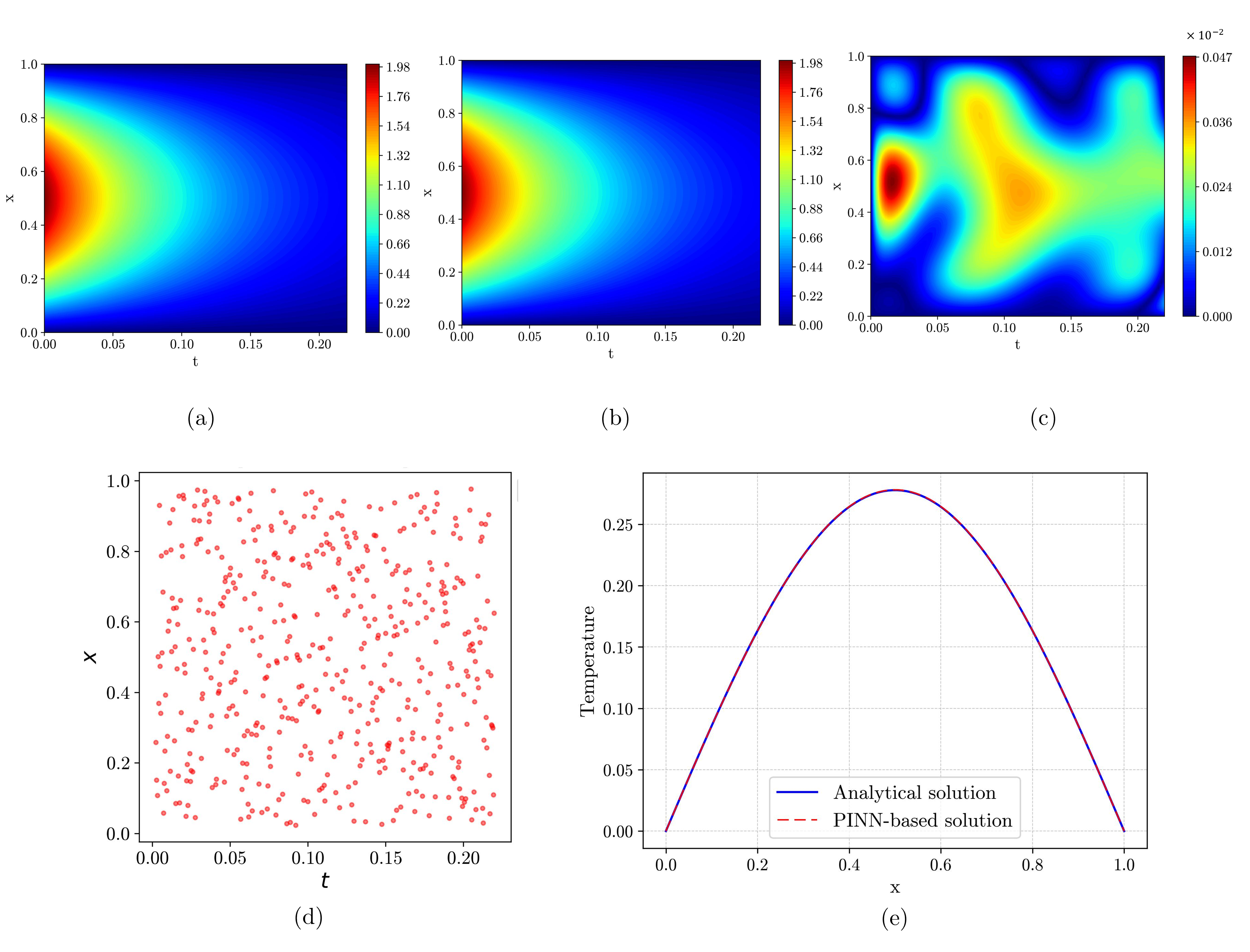}}
		\caption{	Spatio-temporal temperature contour obtained from {(a)} analytical solution (Eqn.\ref{21}), {(b)} our in-house PINN solver ($n_{L} = 3$, $n_{N} = 10$, $N_r = 500$, $\Delta x =  0.023$, and $\Delta t = 0.005$), and (c) absolute error ($L_1$). Further, {(d)} $N_r = 500$ collocation points sampled in the spatio-temporal domain using Latin Hypercube Sampling (LHS), and {(e)} comparison of the instantaneous temperature (at $t = 0.2s$) obtained from the analytical and numerical solution
		}
		\label{fig:Combined Plot - Validation}
	\end{figure}

	For the verification of the present results (from our PINN-based solver) with the above exact-solution, number of layers $n_{L}=3$ and number of neurons per layer $n_{N}=10$ are considered  in our FNN architecture. Further, the number of collocation points sampled are $N_r = 500$---with a point spacing $\Delta x = 0.023$ and $\Delta t = 0.005$ for the numerical differentiation, \revised{as depicted in the point cloud of Figure \ref{fig:Combined Plot - Validation}(d)}. Also, a fully hard constraint-based formulation is used with $G(x, t) = x(1-x) \cdot t$ and $L(x, t) = 2sin(\pi x) $, such that\\
	
	\begin{equation}
		T_{NN}(x, t) = u_{NN}(x, t) \cdot x(1-x) \cdot t + 2sin(\pi x),
	\end{equation}\\
	
	\noindent where  $u_{NN}(x, t)$ is the output of the neural network model as shown in Figure \ref{fig:PINN model}. The function $G(x, t)$ is chosen such that it vanishes at the spatial boundaries, $x = 0$ and $x = 1$, as well as at the initial time, $t = 0$. This ensures that the PINN model exactly adheres to the prescribed boundary and initial conditions.
	\\
	
	For the above transient-to-steady heat diffusion problem, Figures \ref{fig:Combined Plot - Validation}(a) and \ref{fig:Combined Plot - Validation}(b) show the spatio-temporal temperature contour, representing the 1D unsteady temperature field. \revised{Further, for the transient-to-steady results, absolute error ($L_1$) plot of Figure \ref{fig:Combined Plot - Validation}(c) shows an excellent agreement between the analytical solution and our PINN-based solution---with the mean $L_2$ norm error $\bar{L_2} = 0.038\%$ and standard deviation $SD = 0.002\%$.} Further excellent agreement is seen in Figure \ref{fig:Combined Plot - Validation}(e) for an instantaneous temperature profile at $t = 0.2s$---with $\bar{L_2} = 0.096\%$ and $SD = 0.002\%$. \\

	\textcolor{ black}{To ensure a fair comparison with the \emph{SPINN solver} reported by Ramabathiran and Ramachandran \cite{Ramabathiran2021}, we note that their approach used 100 internal nodes, resulting in 400 trainable parameters (100 kernel positions for $(x,t)$, 100 kernel widths, and 100 kernel weights), with 400 sampled points in total. These computational settings closely match those of our current PINN-based solver, which also uses a comparable number of trainable parameters and collocation points. Finally, Table \ref{Table2} shows that the $L_{\infty}$ norm error from our PINN-based solver is two orders of magnitude smaller than that reported by \cite{Ramabathiran2021}, highlighting the enhanced accuracy of our approach under equivalent computational settings and resource consumption.}

	\begin{table}[h!] \small
		\caption{ For the 1D unsteady heat diffusion (Eqn.\ref{18-anal}-\ref{20-anal}), error norms for the present and published \cite{Ramabathiran2021} PINN-based solutions as compared to the analytical solution (Eqn.\ref{21}). Here, SD represents standard deviation.
		}
		\centering
		
		\begin{tabular}{||c c c||} 
			
			\hline
			
			\textbf{Author} & \textbf{$L_\infty$ error} &  \textbf{$\bar{L_2}$ error}\\ [0.5ex] 
			\hline\hline
			\makecell{Ramabathiran and Ramachandran \cite{Ramabathiran2021}} & $O(10^{-2})$& $-$
			\\ \hline 
			Present Work & \makecell{$3.2e-04$ \\ ($SD=1.0e-4$)}& \makecell{$9.6e-04$ \\ ($SD=2.0e-4$)}
			\\  [1ex] 
			\hline
		\end{tabular}
		\label{Table2}
	\end{table}

	\subsubsection{Verification Study: 1D Periodic Non-Linear Heat Diffusion}
	
	For 1D unsteady non-linear heat diffusion, the governing equation and spatio-temporal temperature field are given as\\
	
	\begin{equation} \label{31-eqn}
		- \frac{\partial T}{\partial t} + \frac{\partial }{\partial x}\left(K(T) \frac{\partial T}{\partial x}\right) = 0, \quad t \in [0, 20], \quad x \in [0, L],	
	\end{equation}\\

	\noindent subject to the boundary conditions\\
	
	\begin{equation} \label{32-eqn}
		T(x = 0, t) = T_m + T_a cos(\omega t), \quad T(x = L, t) = T_m,  \quad t \in [0, 20].
	\end{equation}\\
	
	For the above set of equations, analytical solution is given \cite{Shepherd1994} as\\
	
	\begin{equation}\label{33 - nonlinear analytical}
		\begin{split}
			T(x,t) = T_m + T_a e^{(-\alpha_m x)} cos(\omega t - \alpha_m x) 		
			+ \frac{1}{2} \frac{nT_a^2}{T_m} \{e^{(-\sqrt{2}\alpha_m x)} cos(2\omega t - \sqrt{2}\alpha_m x) - e^{(-2\alpha_m x)} cos(2\omega t - 2\alpha_m x)\}	\\ 		
			+ \frac{1}{4} \frac{nT_a^2}{T_m} \{1 - e^{(-2\alpha_m x)}\} ,
		\end{split}
	\end{equation}\\
	
	\noindent where $\alpha_m$ and $K(T)$ take the form, \\
	
	\begin{equation} \label{K(T)}
		\alpha_m = \sqrt{\frac{\omega}{2K(T_m)}}, \quad K(T) = c T^n.
	\end{equation}\\

	For the above non-linear heat diffusion, $T_m = 2$, $T_a=0.1$, $\omega = 5 s^{-1}$, $L = 8$, $c = 1$ and  $n = 3$ are considered in the present simulations. \\

	The general form of the above non-linear diffusion equation makes it applicable to a wide range of physical phenomena, including heat conduction by electrons in plasma, viscous gravity currents, and flow in thin saturated regions of porous media, as discussed by Diez et al. \cite{Diez1992}. Shepherd et al. \cite{Shepherd1994} further demonstrated that, for this system, any initial condition becomes entrained by the surface oscillations at the boundary and does not influence the long-term behavior of the solution. Accordingly, no initial condition is prescribed for the PINN-based solver in this verification case. The analytical solution is considered over one complete period $T = 2\pi/\omega = 1.26s$. The temporal window is selected as 
	$[t_{min}, t_{max}] = [18.74, 20]$, corresponding to a sufficiently large post-transient regime where the solution exhibits periodicity.\\

\begin{figure}
			\hbox{\hspace{2em}
				\includegraphics[width=147mm,scale=5]{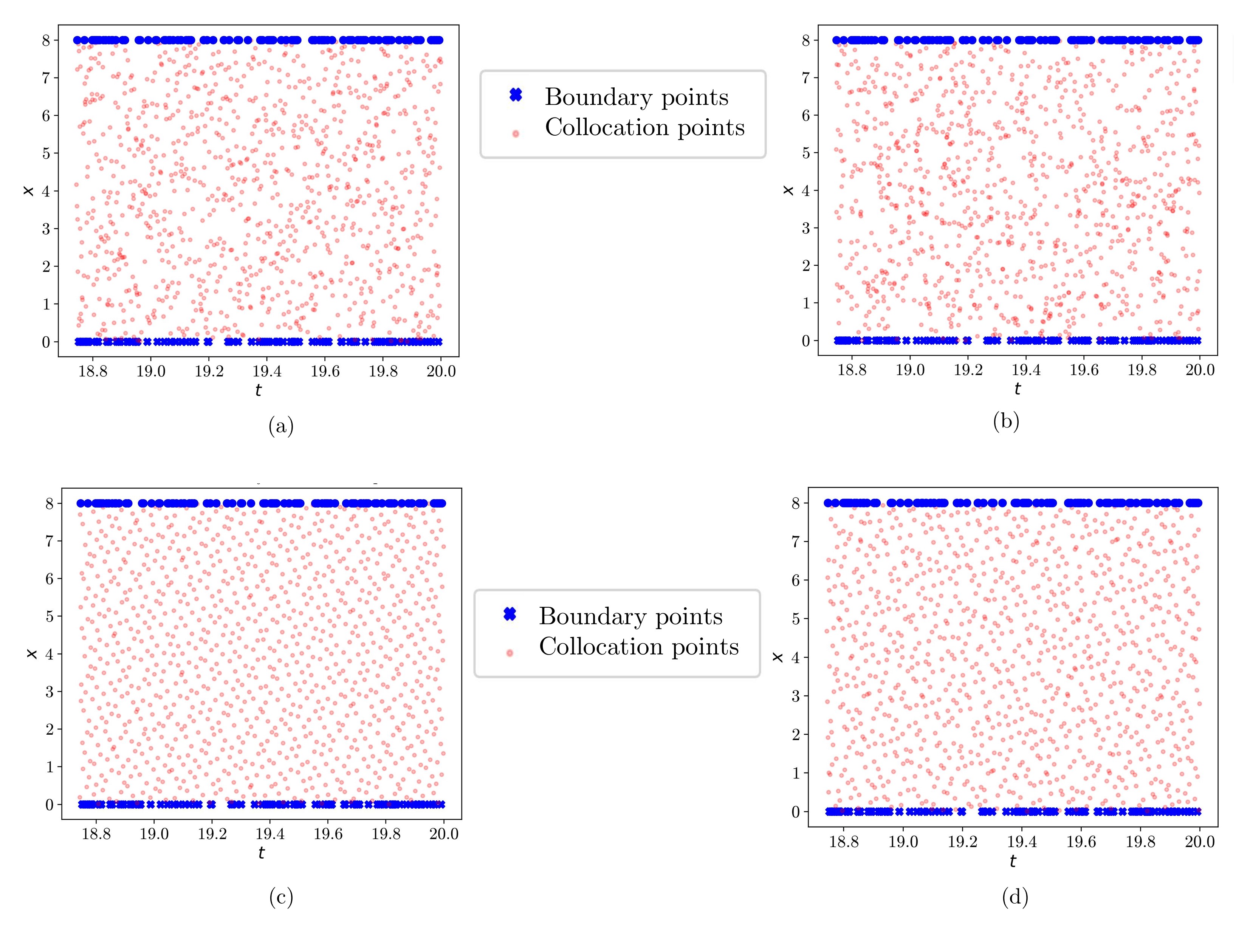}}
			\caption{Collocation and boundary points ($N_r = 1024$ and $N_{BC} = 200$) sampled in the spatio-temporal domain using the (a) LHS, (b) uniform random, (c) Sobol, and (d) Halton sampling techniques for solving Eqn.\ref{31-eqn} using the standard PINN and current PINN-based solvers  }
			\label{fig:Combined Plot - Point Clouds - 1D Nonlinear}
		\end{figure}

		The simulation results for the above system are first obtained using a baseline PINN framework, as introduced by Raissi et al. \cite{Raissi2019}. This standard approach employs automatic differentiation (AD) for computing the PDE residuals, utilizes soft constraints for both residual and boundary condition losses, and does not include any Fourier Feature Embedding (FFE) transformation of the input variables $(x, t)$. For the standard PINN-based solver \cite{Raissi2019}, the number of collocation and boundary points chosen are $N_r = 1024$ and $N_{BC} = 200$, respectively. The FNN architecture is chosen, as before, with $n_N=10$ and $n_L=3$. To evaluate the solver's robustness across different point selection strategies, simulations are performed using four sampling techniques: Latin Hypercube Sampling (LHS), uniform random, Sobol, and Halton sequences, as illustrated in Figure \ref{fig:Combined Plot - Point Clouds - 1D Nonlinear}. \\

		The 1D non-linear heat diffusion system is also simulated using the PINN-based solver proposed here, with the same FNN structure and point cloud density as the previous standard PINN-based solver : $n_N=10$, $n_L=3$ and $N_r = 1024$. The point spacing $\Delta x = 0.05$ and $\Delta t = 0.005$ was used for the numerical differentiation.
		Also, a fully hard constraint-based formulation is used with $G(x, t) = x(L-x) \cdot sin(\omega t)$ and $L(x, t) = T_m + T_a cos(\omega t) \cdot e^{-10x} $, such that\\
		
		\begin{equation}
			T_{NN}(x, t) = u_{NN}(x, t) \cdot x(L-x) \cdot sin(\omega t)  + T_m + T_a cos(\omega t) \cdot e^{-10x},
		\end{equation}\\
		
		\noindent where  $u_{NN}(x, t)$ is the output of the neural network model as shown in Figure \ref{fig:PINN model}. The function $G(x, t)$ is chosen such that it vanishes at the spatial boundaries, $x = 0$ and $x = L$. This ensures that the PINN model exactly adheres to the prescribed boundary conditions of Eqn.\ref{32-eqn}.\\

		Figure \ref{fig:Combined Plot - Nonlinear analytical, std, inhouse} shows the spatio-temporal temperature contours, representing the 1D periodic temperature field. The analytical solutions over the full temporal domain and over one time-period $[18.74, 20]$ are shown in Figure \ref{fig:Combined Plot - Nonlinear analytical, std, inhouse}(a) and (b), respectively. Using the LHS technique, the standard PINN-based solver ($N_r = 1024$, $N_{BC} = 200$, $n_N=10$, $n_L=3$) attains poor performance with mean $L_2$ error $\bar{L_2} = 4.7\%$ while $SD = 1.3\%$ and maximum error $=13.3\%$, as depicted in Figure \ref{fig:Combined Plot - Nonlinear analytical, std, inhouse}(c). For the current PINN-based solution, excellent agreement is seen with the analytical solution in Figure \ref{fig:Combined Plot - Nonlinear analytical, std, inhouse}(d), resulting in mean $L_2$ error $\bar{L_2} = 0.12\%$, $SD = 0.03\%$ and maximum error $=0.5\%$. To achieve acceptable accuracy with the standard PINN, a deeper and wider neural network ($n_N=20$ and $n_L=8$) along with a denser point cloud ($N_r = 7000$ and $N_{BC}=1400$) was required. The resulting prediction, shown in Figure \ref{fig:Combined Plot - Nonlinear analytical, std, inhouse}(e), significantly improves the solution quality, with $\bar{L_2} = 0.25\%$, $SD = 0.05\%$ and maximum error $=2.2\%$. \\

		\begin{figure}[htbp]
			\hbox{\hspace{-1em}
				\includegraphics[width=175mm,scale=5]{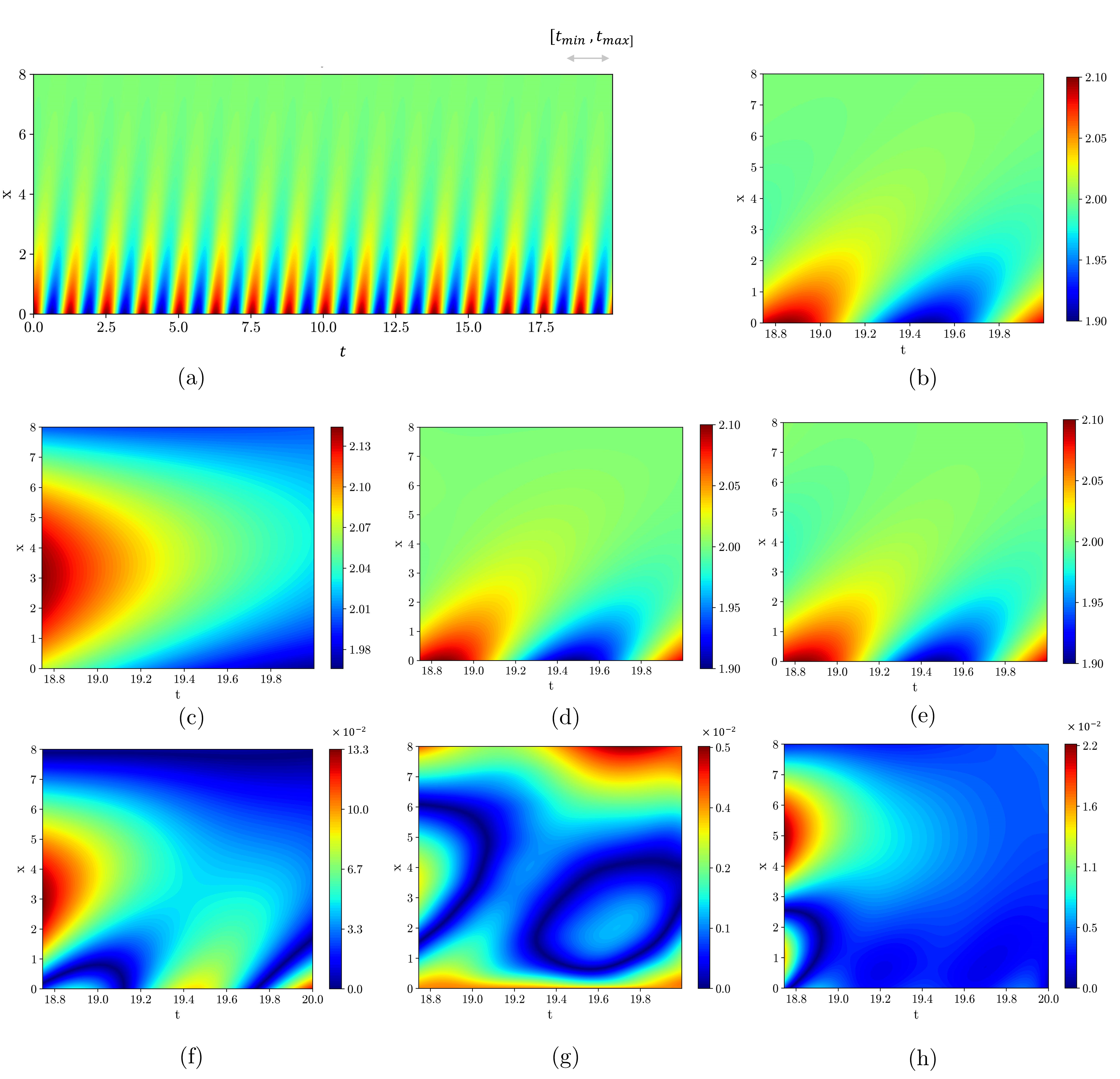}}
			\caption{ (a) Transient-to-periodic analytical solution (Eqn. \ref{33 - nonlinear analytical}) and periodic (b) analytical and (c)-(e) PINN-based numerical  solutions over one time-period [18.74, 20]. The numerical solutions correspond to (c) standard PINN \cite{Raissi2019} ($N_r = 1024$, $N_{BC} = 200$, $n_N=10$, $n_L=3$), (d) present PINN ($N_r = 1024$, $n_N=10$, $n_L=3$, $\Delta x = 0.05$, $\Delta t = 0.005$), and (e) standard PINN ($N_r = 7000$, $N_{BC} = 1400$, $n_N=20$, $n_L=8$), using the LHS technique. The corresponding absolute error ($L_1$) plots (f, g, h) are obtained by subtracting the PINN-based solution from the analytical solution }
			\label{fig:Combined Plot - Nonlinear analytical, std, inhouse}
		\end{figure}

\begin{figure}
			\hbox{\hspace{0em}
				\includegraphics[width=170mm,scale=5]{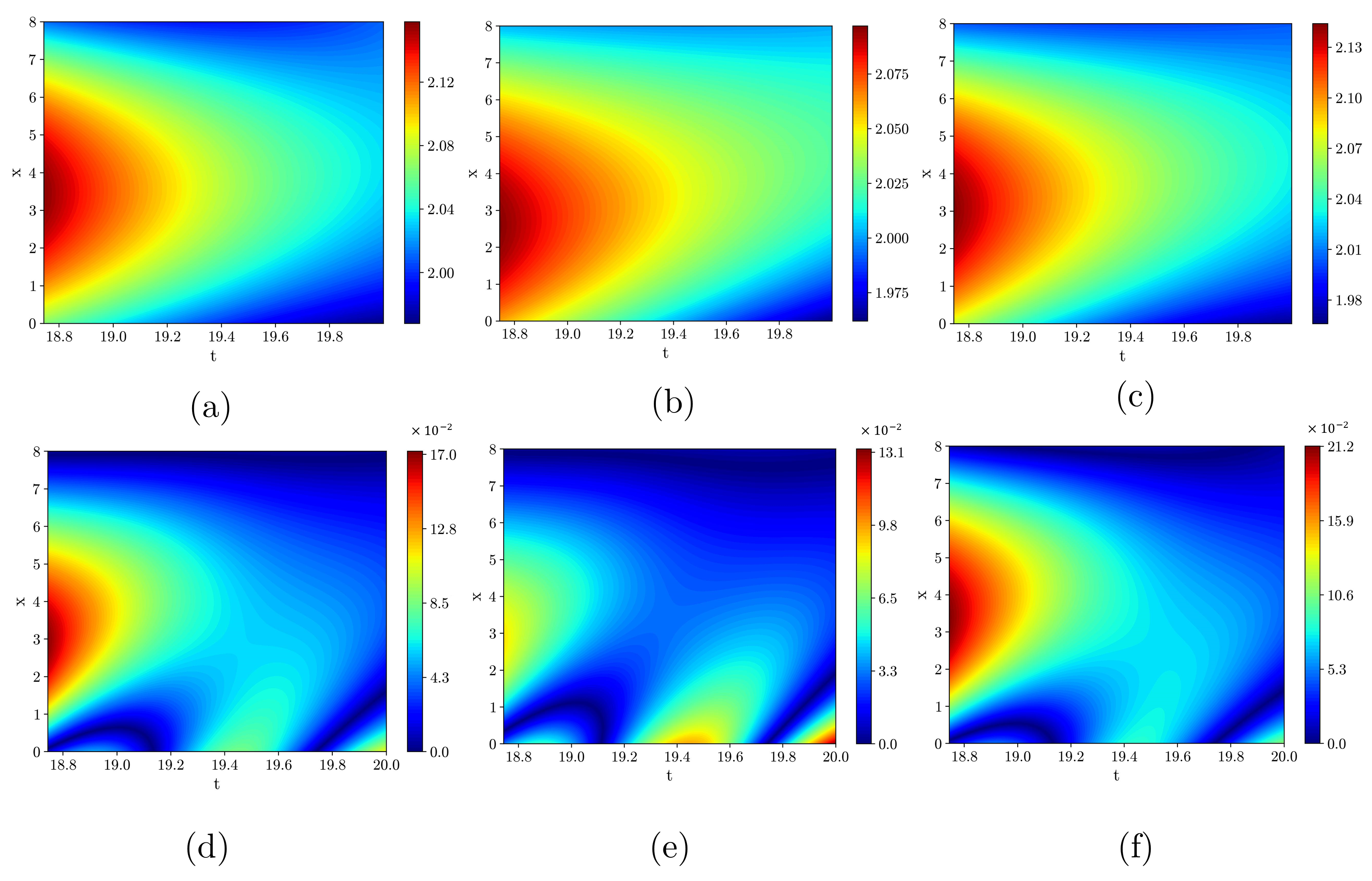}}
			\caption{ Spatio-temporal temperature contours obtained using the standard PINN \cite{Raissi2019} ($N_r = 1024$, $N_{BC} = 200$, $n_N=10$, $n_L=3$) and the corresponding (d, e, f) absolute error ($L_1$) plots using (a) uniform random, (b) Sobol, and (c) Halton sampling techniques }
			\label{fig:Combined Plot - Contour - Vanilla nonlinear - 2}
		\end{figure}

\begin{figure}
			\hbox{\hspace{0em}
				\includegraphics[width=170mm,scale=5]{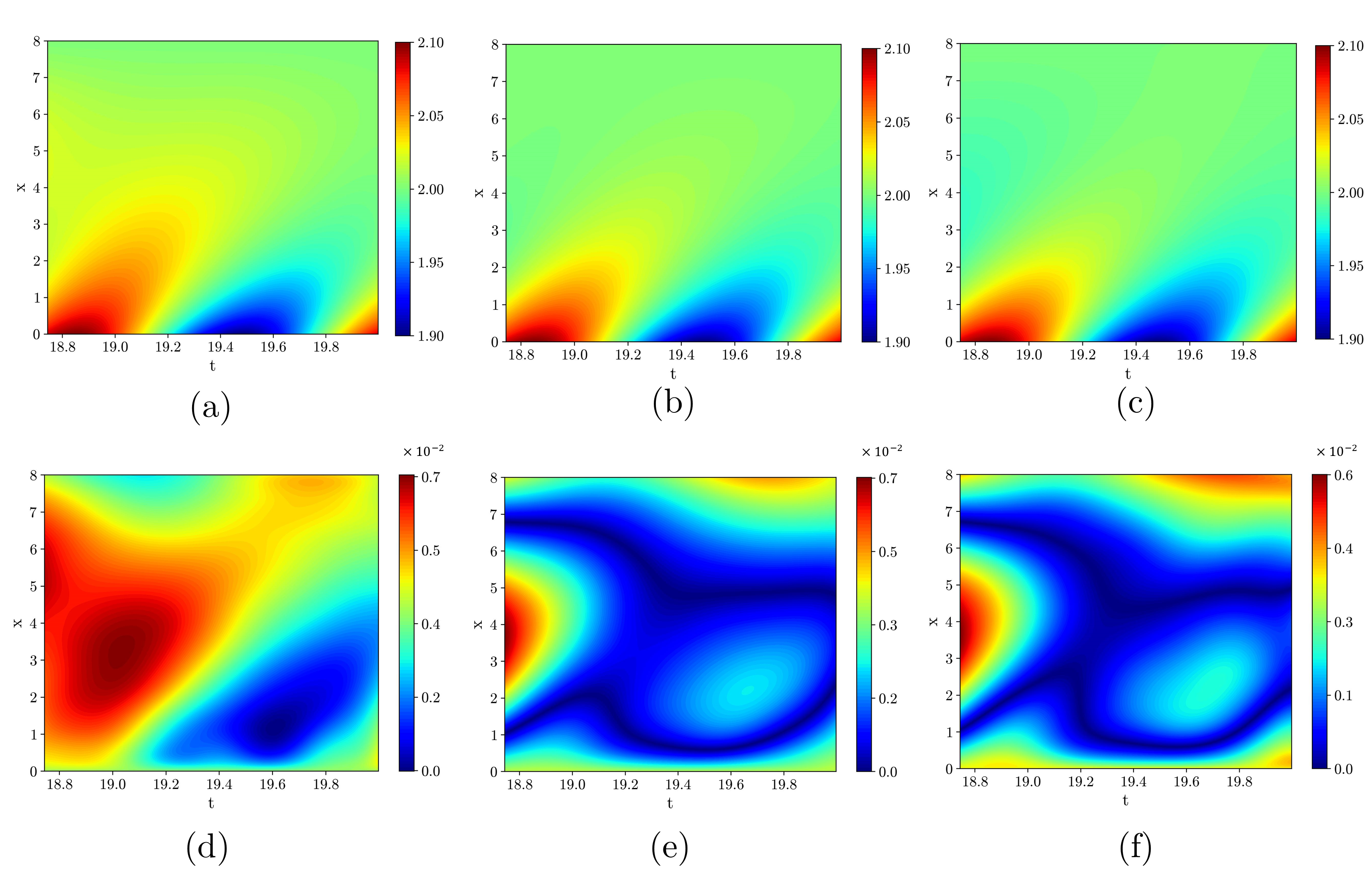}}
			\caption{ Spatio-temporal temperature contours obtained using the present PINN-based solver ($N_r = 1024$, $n_N=10$, $n_L=3$, $\Delta x = 0.05$, $\Delta t = 0.005$) and the corresponding (d, e, f) absolute error ($L_1$) plots using (a) uniform random, (b) Sobol, and (c) Halton sampling techniques}
			\label{fig:Combined Plot - Contour ND nonlinear - 2}
		\end{figure}

		Spatio-temporal temperature contours for the standard PINN-based solver ($N_r = 1024$, $N_{BC} = 200$, $n_N=10$, $n_L=3$) using uniform random, Sobol and Halton sampling techniques are shown in Figure \ref{fig:Combined Plot - Contour - Vanilla nonlinear - 2}. Across all three sampling strategies, the solver demonstrates poor agreement with the analytical solution. The mean $L_2$ norm error $\bar{L_2} = 4.5\%, 5.6\%$ and $5.3\%$, while the standard deviation $SD = 2.2\%, 1.0\%$ and $2\%$; for the three different sampling techniques in Figure \ref{fig:Combined Plot - Contour - Vanilla nonlinear - 2}(a), (b) and (c), respectively. For the present PINN-based solver, Figure \ref{fig:Combined Plot - Contour ND nonlinear - 2} shows the spatio-temporal temperature contours, with agreement between our present PINN-based solution and the analytical solution being excellent regardless of the sampling strategy used. The mean $L_2$ norm error $\bar{L_2} = 0.52\%, 0.13\%$ and $0.23\%$, while $SD = 0.08\%, 0.04\%$ and $0.03\%$; for the three different sampling techniques in Figure \ref{fig:Combined Plot - Contour ND nonlinear - 2}(a), (b) and (c) respectively. These results show that the present PINN-based solver delivers robust performance across various sampling strategies, effectively capturing the time-periodic behavior of the nonlinear diffusion system. In contrast, the standard PINN-based solver required a significantly denser point cloud and a deeper network to achieve comparable accuracy, resulting in higher computational cost and memory usage. \\

		\subsubsection{Verification Study: 2D Periodic Heat Diffusion}  \label{fixed}

		For verification study on 2D heat diffusion, accurate transient-to-periodic temperature field is generated using our FVM-based solver implemented in \emph{Python}, as discussed above. Further, a grid independence study was performed for all the 2D unsteady heat diffusion problems in Figure \ref{fig:Domain 2D All 2}. Grid-independent FVM-based solution was achieved on a Cartesian grid size of \emph{100$\times$100} for the problems in Figures \ref{fig:Domain 2D All 2}(a) and (b), and on a curvilinear grid size of \emph{96$\times$96} for the problem in Figure \ref{fig:Domain 2D All 2}(c). As discussed above, the FVM-based grid-independent periodic solution is used for the verification of accuracy of the present PINN-based periodic diffusion solver.\\

\begin{figure}
			\hbox{\hspace{-0.0em}
				\includegraphics[width=175mm,scale=5]{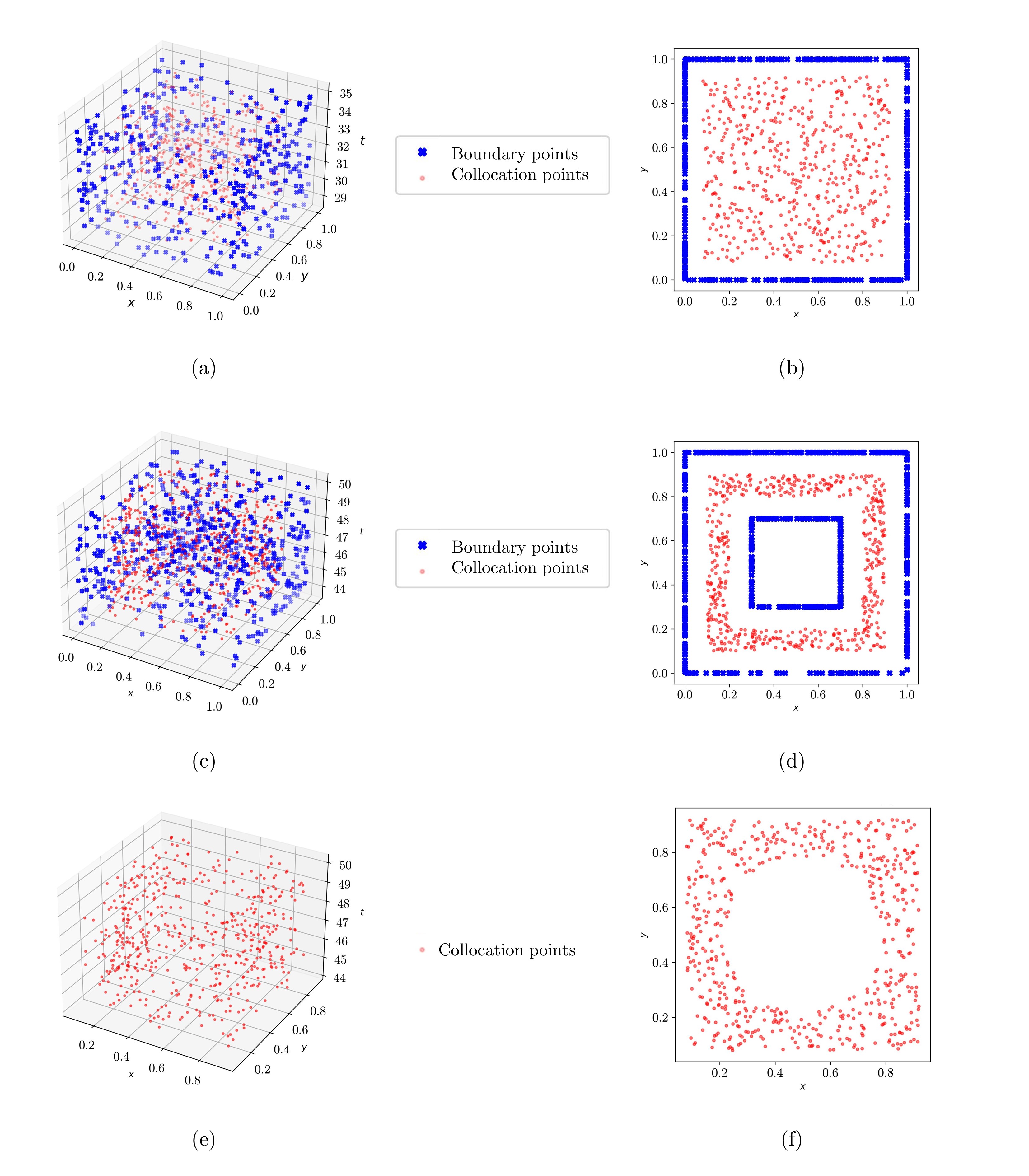}}
			\caption{ Collocation and boundary points sampled in the spatio-temporal domain using the LHS and the corresponding 2D projections of the point clouds, for the problems in {(a,b)} Figure \ref{fig:Domain 2D All 2}(a), {(c,d)} Figure \ref{fig:Domain 2D All 2}(b) and {(e,f)} Figure \ref{fig:Domain 2D All 2}(c)
			}
			\label{fig:Combined Plot - Point Clouds}
		\end{figure}

		For the 2D verification study, the present PINN-based periodic-diffusion solver considers a specific set of hyperparameters---number of layers $n_{L} = 3$, number of neurons per layer $n_{N} = 10$, number of collocation points $N_r = 550$, point spacing for numerical differentiation $\Delta x = \Delta y = 0.08$, and $\Delta t = 0.005$. Further, number of boundary points $N_{BC} = 400$ and $600$ are considered for the problems in Figure \ref{fig:Domain 2D All 2}(a) and (b), respectively. Whereas, due to a hard constraint-based formulation of the boundary conditions for the problem \ref{fig:Domain 2D All 2}(c), note that no boundary points $N_{BC}$ are required. The sampled collocation and boundary points, forming the spatio-temporal point cloud, are depicted in Figure \ref{fig:Combined Plot - Point Clouds}(a)-(c). The above values of hyperparameters are chosen due to its efficient balance, which results in a very small computational time (of a few seconds) while maintaining $\bar{L_2}$ $<$ 4\% across all the test problems. For the FVM-based solution on the Cartesian and Curvilinear grid systems, the present numerical methodology can be found in  a recent textbook on CFD by Sharma \cite{Sharma2021}.\\

\begin{figure}
			\hbox{\hspace{-0.5em}
				\includegraphics[width=175mm,scale=5]{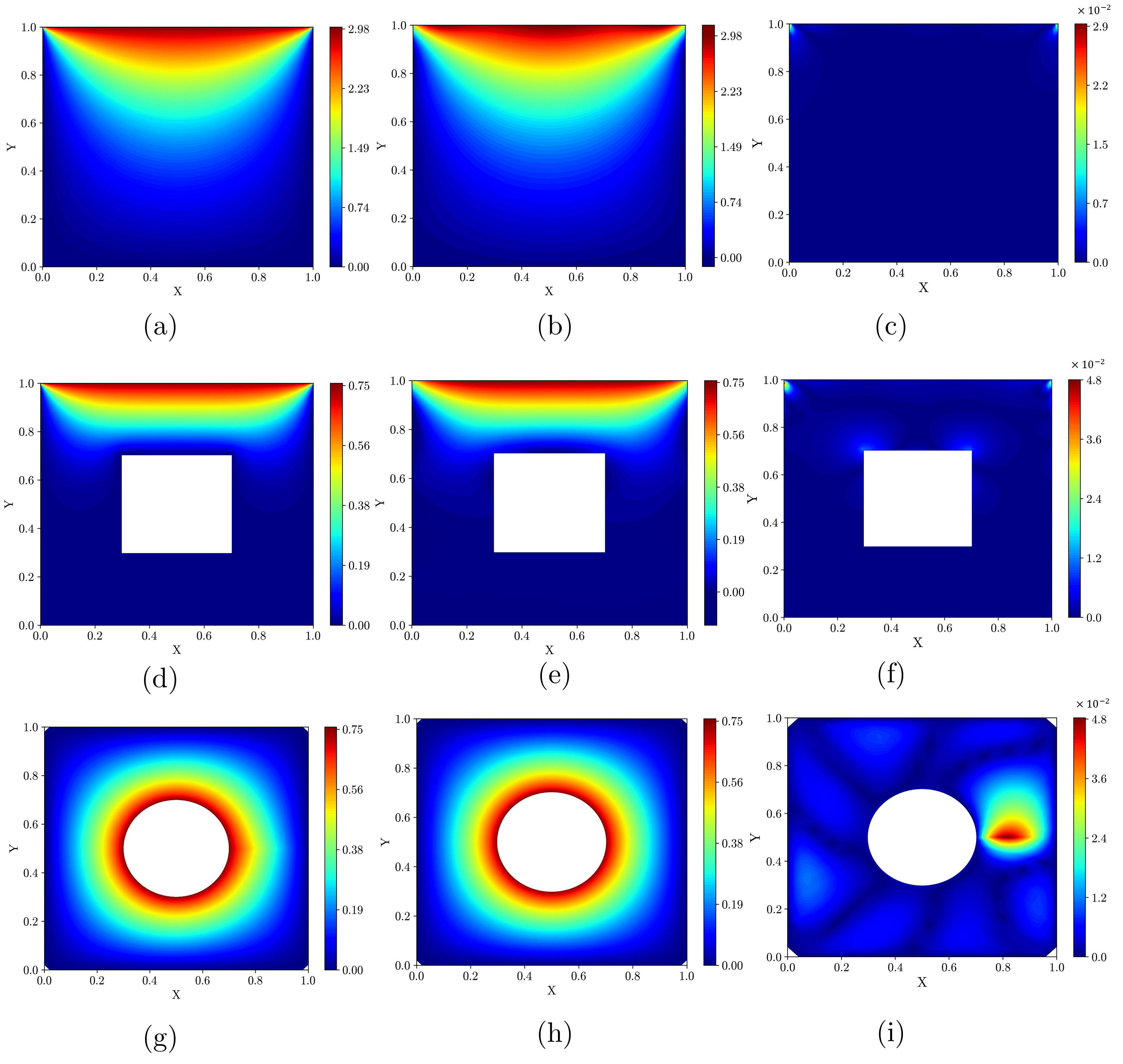}}
			\caption{ For the three 2D problems in Figure \ref{fig:Domain 2D All 2}, contour plots of instantaneous temperature obtained by {(a,d,g)} FVM-based transient-to-periodic solver, {(b,e,h)} PINN-based periodic solver and {(c, f, i)} absolute error ($L_1$) plots. The temperature contour corresponds to $t = 30s$ for the heat diffusion in a square plate (Figure \ref{fig:Domain 2D All 2}(a)), and $t = 47s$ for the plate with square and circular holes (Figure \ref{fig:Domain 2D All 2}(b) and (c)) 
			}
			\label{fig:Combined Plot - Contour}
		\end{figure}

\begin{figure}
			\hbox{\hspace{-0.0em}
				\includegraphics[width=175mm,scale=5]{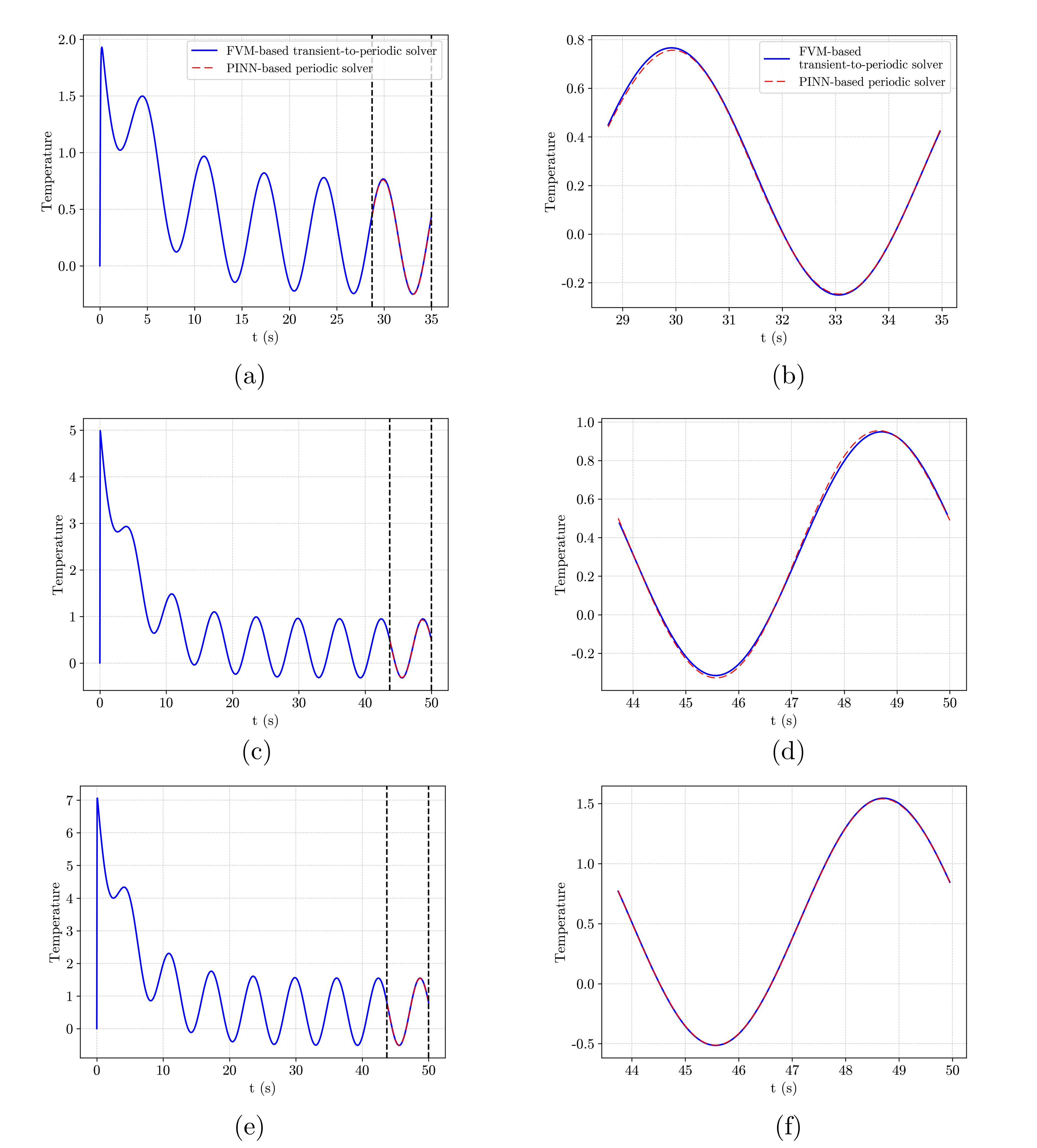}}
			\caption{  Comparison between grid independent FVM-based transient solution and PINN-based periodic solution for temporal evolution of temperature at points {(a, b)} P, {(c, d)} Q, and {(e, f)} R --- marked in Figure \ref{fig:Domain 2D All 2}(a), (b) and (c), respectively. }
			\label{fig:Combined Plot - Temporal Plots}
		\end{figure}

		For the PINN-based solution as compared to the grid-independent FVM-based solution, excellent agreement is seen in Figure \ref{fig:Combined Plot - Contour} (for instantaneous temperature contours) and Figure \ref{fig:Combined Plot - Temporal Plots} (for periodic temperature profiles) at points P, Q and R in Figure \ref{fig:Domain 2D All 2}. Further, for the temperature contour, the mean $L_2$ norm error $\bar{L_2} = 4.7\%, 7.7\%$ and $2.7\%$, while the standard deviation $SD = 0.3\%, 1.1\%$ and $0.2\%$; for the problems in Figure \ref{fig:Domain 2D All 2}(a), (b) and (c), respectively. 
		Whereas, the periodic temperature profiles for the respective problems resulted in $\bar{L_2} = 1.6\%, 3.2\%$ and $0.46\%$, while $SD = 0.3\%, 0.3\%$ and $0.01\%$.
		\textcolor{black}{Finally, the computational time required by our PINN-based periodic solver is substantially less as compared to that required by our FVM-based transient-to-periodic solver. For the problems in Figure \ref{fig:Domain 2D All 2}(a), (b) and (c), our PINN-based periodic solver takes a mean computational time of $18.5s$ $(SD = 0.3s)$, $19.8s$ $(SD = 0.8s)$ and $16.2s$ $(SD = 0.6s)$, respectively.}
		\\

		The above verification study demonstrated physically realistic numerical results from our PINN-based periodic-diffusion solver that led us to an investigation on effect of choice of specific values of hyperparameters in the solver; presented in the next section. However, for the above verification study, the specific values of hyperparameters are chosen such that it leads to both computational efficiency and reasonable accuracy --- making this choice as reference values for subsequent analysis.
		\\

		\subsubsection{Effect of Hyperparameters on the PINN-based Diffusion Solver}	
		
		After presenting the verification study, performance of the PINN-based periodic-diffusion solver is systematically explored with variation in input values for the various hyperparameters. This parametric study is conducted by varying the number of layers, $n_L$, from 2 to 8, and the number of neurons per layer, $n_N$, from 3 to 100; thereby, exploring both shallow and deep neural network structures. This range of $n_L$ and $n_N$ ensures coverage from simple networks to more complex FNN architectures, as briefly demonstrated in the work by Raissi et al. \cite{Raissi2019}. Thereafter, the number $N_r$ of collocation points is varied significantly, from 50 to 5000, to examine whether the proposed PINN model can maintain reasonable accuracy even with highly sparse point clouds. This approach contrasts with the use of very dense collocation points in works like those of Putra et al. \cite{Putra2022} and Cai et al. \cite{Cai2021}. Finally, the point spacing used for numerical differentiation, ($\Delta x, \Delta y$), is also varied from a minimal initial value of 0.01 (comparable to that seen in \cite{Chiu2022}) up to the largest practical values allowed by the domain size; roughly in the range of 0.1 to 0.3. This comprehensive exploration of the hyperparameter variation aims to assess our PINN-based periodic-diffusion solver's robustness across a range of inputs, shedding light on its adaptability to the various problem setups while balancing accuracy and computational efficiency.\\

\begin{figure}
			\hbox{\hspace{-0.5em}
				\includegraphics[width=175mm,scale=5]{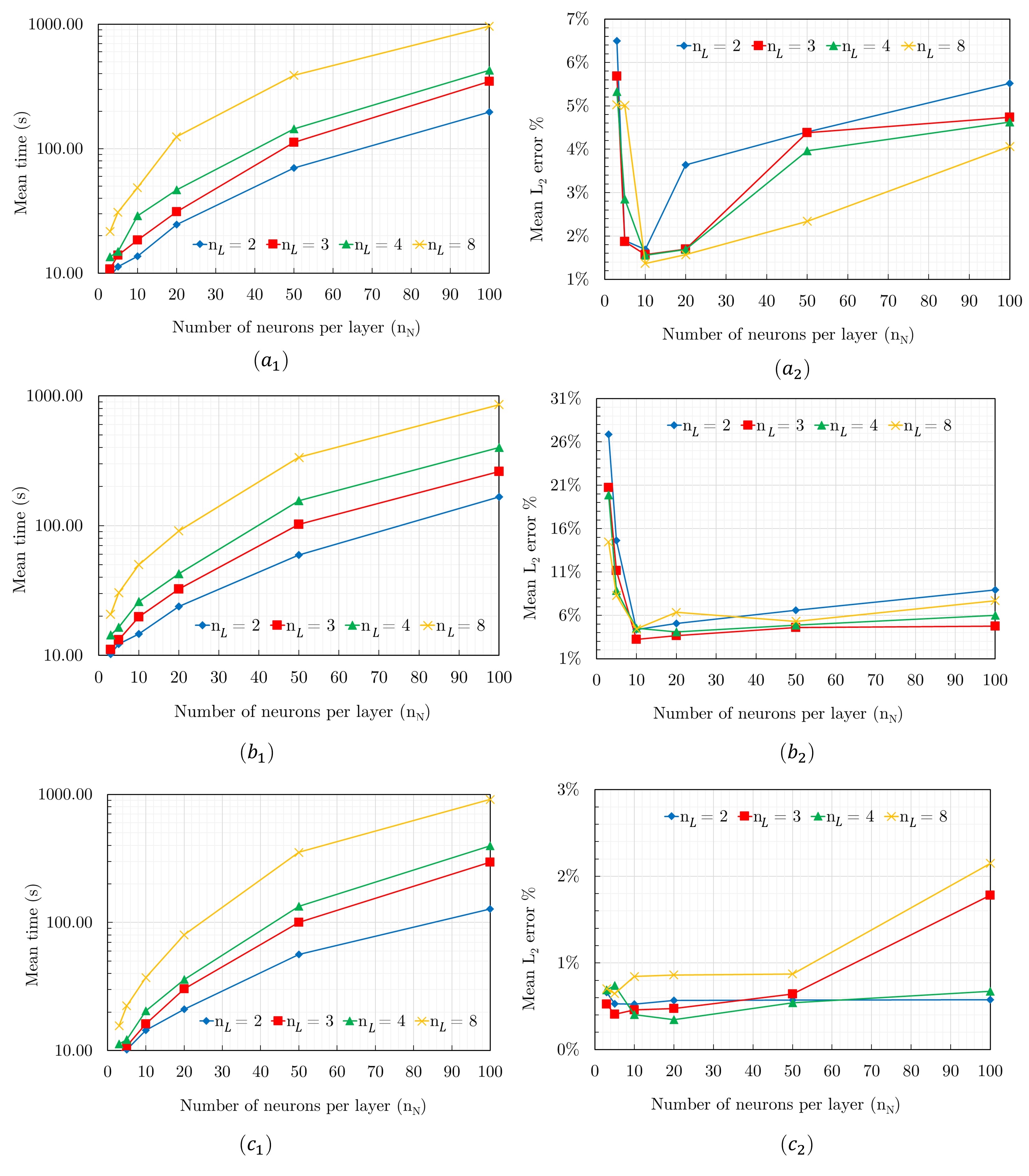}}
			\caption{ 	For various number of layers $n_L = 2-8$ of the feedforward neural network, variation of ($a_1$, $b_1$, $c_1$) mean computational time of the PINN-based solver and its ($a_2$, $b_2$, $c_2$) mean $\bar{L_2}$ error (with respect to the grid independent FVM solution), with increasing number of neurons per layer $n_N$. The above results in subfigures $(a_1, a_2)$, $(b_1, b_2)$ and $(c_1, c_2)$ correspond to the periodic results at points P, Q, R in Figure \ref{fig:Domain 2D All 2}(a), (b), and (c), respectively.}	
			\label{fig:Combined Plot - FNN}
		\end{figure}

		\emph{Effect of variation of feedforward neural network (FNN) architecture: }	 		
		For the variation of hyperparameters involved in the feedforward neural network (FNN) architecture, this section presents effect of variation of number of layers $n_L = 2-8$, and number of neurons per layer $n_N = 5-100$. This study is done at a constant value of $N_r = 550$, $\Delta x = \Delta y = 0.08$, and $\Delta t = 0.005$. The constant values of boundary points are $N_{BC} = 400$ and $600$ for the problems in Figure \ref{fig:Domain 2D All 2}(a) and \ref{fig:Domain 2D All 2}(b) respectively. The associated results are shown in Figure \ref{fig:Combined Plot - FNN}($a_1, a_2$), ($b_1, b_2$) and ($c_1, c_2$), for the problems in Figure \ref{fig:Domain 2D All 2}(a), \ref{fig:Domain 2D All 2}(b) and \ref{fig:Domain 2D All 2}(c), respectively. Figure \ref{fig:Combined Plot - FNN} shows the effect of $n_L$ and $n_N$ on computational performance of our PINN-based periodic diffusion solver in terms of the mean computational time and mean $L_2$ norm error.\\
		
		Figure \ref{fig:Combined Plot - FNN}($a_1$)-($c_1$), show that the mean computational time for the PINN-based solver increases with increasing $n_L$ and with increasing $n_N$. This is because increasing  $n_L$ and $n_N$ adds more trainable parameters, making the neural network deeper and thus computationally demanding. Further, for the problems with the periodic boundary condition on the top wall of the plate, with increasing $n_N$, Figure \ref{fig:Combined Plot - FNN}($a_2$) and ($b_2$) shows that the mean $L_2$ error decreases sharply---reaching a minimum at $n_N=10$ (for all values of $n_L$) before gradually rising again. Whereas, for the problem with the periodic boundary condition on the circular hole, Figure \ref{fig:Combined Plot - FNN}($c_2$) shows very small values as well as range of variation of $\bar{L_2}$ $(1\%-2.2\%)$ with increasing $n_N$ and $n_L$. Thus, using the hard constraint for this problem as compared to the hybrid constraint for the other problems, $n_N$ and $n_L$ seem to have a negligible effect on the mean $L_2$ error although there is a substantially large effect on the computational time (Figure \ref{fig:Combined Plot - FNN}($c_1$)). Note from Figure \ref{fig:Combined Plot - FNN}($a_1$)-($c_1$) that the variation of computational time, with increasing $n_N$, is similar for all the three problems.\\

\begin{figure}
			\hbox{\hspace{-0.5em}
				\includegraphics[width=175mm,scale=5]{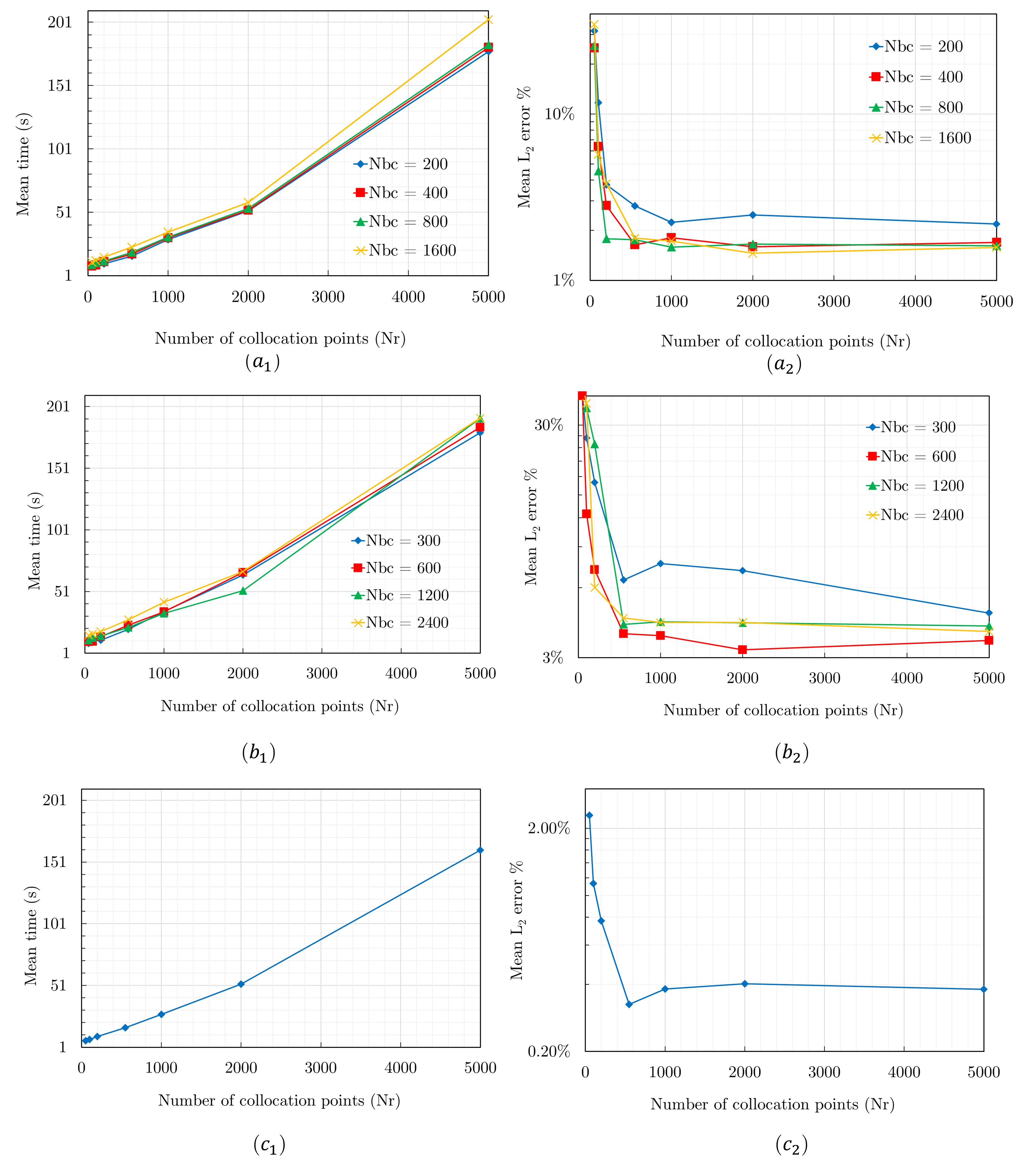}}
			\caption{ For various number of boundary points $N_{BC} = 200-2400$, variation of ($a_1$, $b_1$, $c_1$) mean computational time of the PINN-based solver and its ($a_2$, $b_2$, $c_2$) mean $\bar{L_2}$ error (with respect to the grid independent FVM solution), with increasing number of collocation points $N_r$.  The above results in subfigures $(a_1, a_2)$, $(b_1, b_2)$ and $(c_1, c_2)$ correspond to the periodic results at points P, Q, R in Figure \ref{fig:Domain 2D All 2}(a), (b), and (c), respectively.	}	
			\label{fig:Combined Plot - Nr}
		\end{figure}

		\emph{Effect of variation of collocation ($N_r$) and boundary points ($N_{BC}$): }	
		Figure \ref{fig:Combined Plot - Nr} shows the computational performance under the effect of variation of $N_r = 50-5000$ and $N_{BC} = 200-2400$, at a constant value of $n_L = 3$, $n_N = 10$, $\Delta x = \Delta y = 0.08$, and $\Delta t = 0.005$.  Note the $N_{BC}$ is varied here for the problem with square plate with no-hole and square-hole, but not for the problem with circular-hole. This is due to the usage of hard constraints for the problem with circular-hole as compared to hybrid constraints for the other problems.
		\\

		Figure \ref{fig:Combined Plot - Nr}($a_1$)-($c_1$) shows that the mean computational time increases sharply with increasing $N_r$ while it increases slightly with increasing $N_{BC}$. Further, with increasing $N_r$, Figure \ref{fig:Combined Plot - Nr}($a_2$) shows that the mean $L_2$ error decreases rapidly for all the values of $N_{BC}$, eventually stabilizing within a range of 1\% to 3\% when $N_r$ exceeds 550. Similarly, Figure \ref{fig:Combined Plot - Nr}($b_2$) shows a significant reduction in the ${L_2}$ error with increasing $N_r$, stabilizing within a range of 3\% to 4\% for all $N_{BC}$ values except when $N_{BC} = 300$. Lastly, Figure \ref{fig:Combined Plot - Nr}($c_2$) shows that the $\bar{L_2}$ error stabilizes between 0.20\% and 0.40\% for $N_r$ $>$ 550, showing consistent behavior across the tested problems.
		\\

\begin{figure}
			\hbox{\hspace{-0.5em}
				\includegraphics[width=175mm,scale=5]{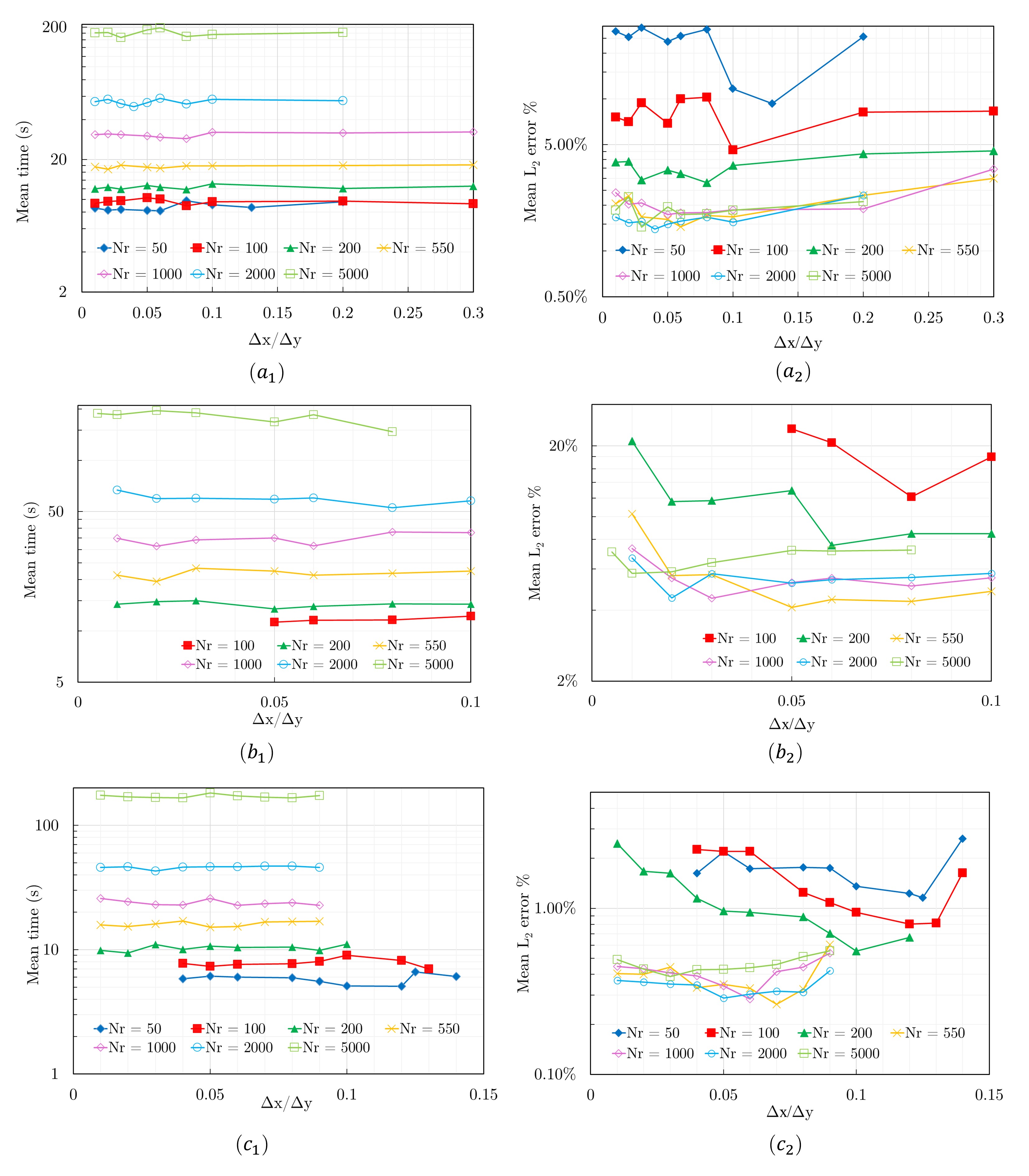}}
			\caption{For various number of collocation points $N_{r} = 50-5000$, variation of ($a_1$, $b_1$, $c_1$) mean computational time of the PINN-based solver and its ($a_2$, $b_2$, $c_2$) mean $\bar{L_2}$ error (with respect to the grid independent FVM solution), with increasing value of point spacing ($\Delta x, \Delta y$) for numerical differentiation.  The above results in subfigures $(a_1, a_2)$, $(b_1, b_2)$ and $(c_1, c_2)$ correspond to the periodic results at points P, Q, R in Figure \ref{fig:Domain 2D All 2}(a), (b), and (c), respectively.	}
			
			\label{fig:Combined Plot - del_x}
		\end{figure}

		\emph{Effect of variation of point spacing ($\Delta x , \Delta y$) for numerical differentiation and collocation points ($N_r$): } \label{point_spacing}		
		For the variation of hyperparameters involved in numerical differentiation for the PDE operators, Figure \ref{fig:Combined Plot - del_x} shows the computational performance under the effect of variation of the point spacing $\Delta x = \Delta y =  0.01-0.30$, and the number of collocation points $N_r = 50-5000$. This study is done at a constant value of $n_N = 10$, $n_L = 3$, and $\Delta t = 0.005$. Further, the constant values of boundary points are $N_{BC} = 400$ and $600$ for the problems in Figure \ref{fig:Domain 2D All 2}(a) and (b), respectively.
		\\

		Figure \ref{fig:Combined Plot - del_x}($a_1$)-($c_1$) show that the mean computational time remains nearly constant with variations in $\Delta x$ and $\Delta y$ at a constant value of $N_r$. However, as $N_r$ increases, the computational time increases, following the same pattern as seen above in Figure \ref{fig:Combined Plot - Nr}($a_1$), ($b_1$) and ($c_1$). For the variation of error with decreasing number $N_r$ of collocation points, note from the values on the y-axis  of Figure \ref{fig:Combined Plot - del_x}($a_2$)-($c_2$) that the range of variation of the error (with variations in $\Delta x$ and $\Delta y$) is very small except for the error in Figure \ref{fig:Combined Plot - del_x}($b_2$) at a coarse $N_r \le 200$. Anyway, for the problem in Figure \ref{fig:Domain 2D All 2}(b), the solutions on a reasonably coarser collocation point cloud ($N_r \le 200$) are not accurate enough (presented below in Figure \ref{fig:Combined Plot - FVM vs PINN}(b) with the error $\ge 7.2\%$) that results in the larger error variation for $N_r \le 200$. Thus, for the $N_r$ that results in a reasonably accurate solution, the $\Delta x$ and $\Delta y$ for numerical differentiation does not seem to have much effect on the accuracy of the solution. \\

		\textcolor{black}{We note that, unlike traditional discretization-based solvers where the grid spacing or step size directly controls the overall accuracy, the situation for PINN-based solvers is more nuanced. Following Goraya et al. \cite{Goraya2023}, the total error of a PINN-based solution can be decomposed into three components: approximation error (due to the finite representational capacity of the neural network architecture), training/estimation error (due to finite sampling set of collocation points), and optimization error (arising from the non-convexity of the loss function). Once a sufficient number of collocation points are employed, the discretization component ceases to be dominant, and the total error is governed mainly by the training residual and approximation capacity of the network. This behavior is observed in our experiments: the $\bar{L_2}$ error saturates for $N_r > 500$ across all problems (Figure \ref{fig:Combined Plot - Nr}), and varying the numerical differentiation point spacing produces only marginal differences in accuracy (Figure \ref{fig:Combined Plot - del_x}). Our conjecture is that, although finite-difference truncation errors are introduced by the numerical differentiation scheme, they are small relative to the network error and can be partially absorbed during training through weight adjustments by the optimizer. Consequently, the solver exhibits only weak dependence on the $\Delta x/\Delta y$ for differentiation as compared to conventional solvers, making it relatively robust to moderate variations in this parameter.}

		\subsubsection{Performance Study : PINN v/s FVM-based Heat Diffusion Solver}\label{PINN_vs_FVM_Heat}
		
		This section presents performance of the present PINN-based solver with increasing number of collocation points $N_r$ as compared to the FVM-based solver with increasing grid sizes; for the three problems presented in Figure \ref{fig:Domain 2D All 2}. \textcolor{black}{The number of collocation points chosen for the PINN solver are $N_r = 100$, $200$, and $550$. This selection is based on a specific reasoning --- the ratio of the number of collocation points along a single spatial dimension, approximated as $N_r^{\frac{1}{3}}$ (due to the use of LHS in the $x$, $y$, and $t$ dimensions), is comparable to the ratio of grid points along the same dimension for successive FVM solvers. This provides a logical basis for aligning the collocation points of the PINN solver with the grid sizes of the FVM solvers. Table \ref{Table3} summarizes this rationale. For instance, the ratio of grid points for successive FVM solvers, such as $50/40 = 1.25$, is approximately equal to the ratio of collocation points along a single dimension for the PINN solver, which is $5.8 / 4.6 \approx 1.26$. This relationship holds for the other choices of collocation points as well, such as $N_r = 200$ and $N_r = 550$, ensuring a consistent comparison between the PINN and FVM solvers across different grid sizes. All the other hyperparameters of the PINN solvers, except $N_r$, are fixed and the same as used in subsection \ref{fixed}.}
		\textcolor{black}{For the PINN-based and FVM-based solvers (on the Cartesian grid size $40\times40$, $50\times50$ and $80\times80$), the $\bar{L_2}$ error is computed with respect to the grid-independent FVM-based solution. The grid-independent solution is obtained on a $100\times100$ grid for the problems in Figures \ref{fig:Domain 2D All 2}(a) and \ref{fig:Domain 2D All 2}(b), and on a grid size of $96\times96$ for the problem in Figure \ref{fig:Domain 2D All 2}(c).}
			\\

		\begin{table} \small
			\caption{Summary of the number of grid points used in the coarse mesh FVM solvers and the corresponding number of collocation points chosen for the PINN solver. The collocation points $N_r = 100$, $200$, and $550$ for the PINN solver are selected based on the approximate equivalence of the ratios of grid points in the FVM solvers and collocation points in the PINN solver, along a single spatial dimension.}
			\centering
			\begin{tabular}{||c c c||} 
				\hline
				\textbf{Solver} & \textbf{\makecell{Number of points along a single \\ linear spatial dimension}} & \textbf{Ratio of points} \\ [0.5ex] 
				\hline\hline
				\emph{40X40} FVM & \makecell{$40$}  & \makecell{$40 /\ 40 = 1$}  \\ \hline
				
				\emph{50X50} FVM & \makecell{$50$} & \makecell{$50 /\ 40 = 1.25$}  \\ \hline 
					
				\emph{80X80} FVM & \makecell{$80$} & \makecell{$80 /\ 50 = 1.6$, $80 /\ 56 \approx 1.42$} \\ \hline 
				
				$N_r = 100$ PINN  & \makecell{$N_r^{\frac{1}{3}}  \approx 4.6$ } & \makecell{$4.6 /\ 4.6 = 1$} \\ \hline 
				
				$N_r = 200$ PINN  & \makecell{$N_r^{\frac{1}{3}}  \approx 5.8$  } & \makecell{$5.8 /\ 4.6 \approx 1.26$} \\ \hline 
				$N_r = 550$ PINN & \makecell{$N_r^{\frac{1}{3}}  \approx 8.2$ } & \makecell{$8.2 /\ 5.8 \approx 1.42$} \\ [1ex]  		\hline
			\end{tabular}
			\label{Table3}
		\end{table}

\begin{figure}
			\hbox{\hspace{-0.5em}
				\includegraphics[width=175mm,scale=5]{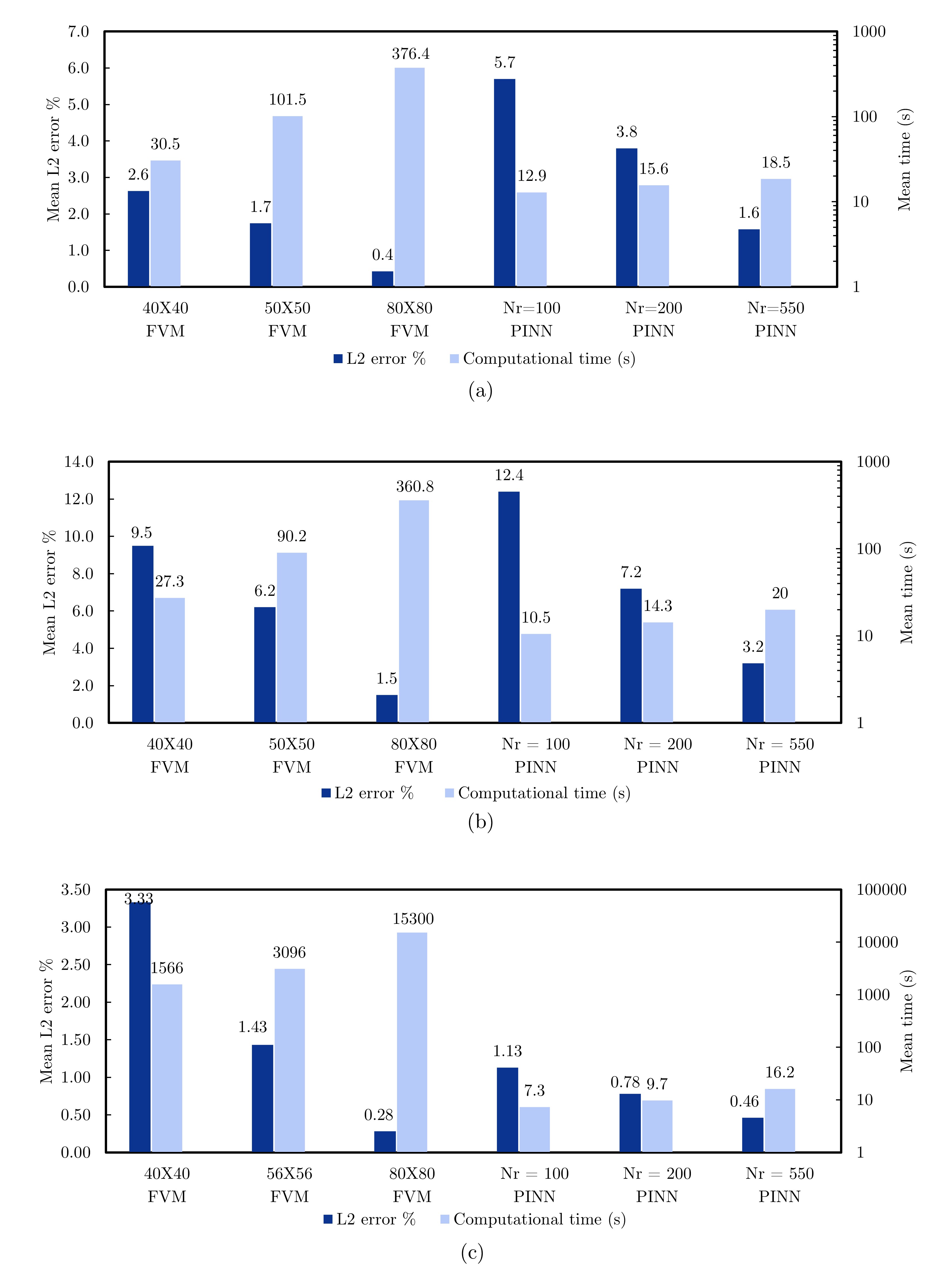}}
			\caption{ 	Comparison of computational time and $\bar{L_2}$ error for the present PINN-based periodic and FVM-based transient-to-periodic diffusion solvers, on various collocation points $N_r$ and grid sizes, respectively. The computational performance in subfigures (a), (b), and (c) corresponds to problems in Figure \ref{fig:Domain 2D All 2}(a), (b), and (c), respectively. Further, for both the solvers, the error is computed with respect to a grid-independent FVM-based solution on a grid size of $100\times100$. For the PINN-based periodic solver, the above results are obtained with $n_{L} = 3$, $n_{N} = 10$, $\Delta x = \Delta y = 0.08$, $\Delta t = 0.005$, $N_{BC} = 400$ for (a), and $N_{BC} = 600$ for (b).
			}
			\label{fig:Combined Plot - FVM vs PINN}
		\end{figure}

		For the FVM-based unsteady-to-periodic solutions of the various 2D problems (Figure \ref{fig:Domain 2D All 2}), Figure \ref{fig:Combined Plot - FVM vs PINN} shows a much larger increase in computational time as compared to the decrease in the $\bar{L_2}$ error with grid refinement. In contrast, with increasing $N_r$ for the PINN-based direct periodic solution, the figure shows a much smaller increase in computational time as compared to the decrease in the $\bar{L_2}$ error. For example, with increasing grid size from $40\times40$ to $80\times80$ for the heat diffusion in a square plate, Figure \ref{fig:Combined Plot - FVM vs PINN}(a) shows a $12.34$ times increase in the computational time and $84\%$ decrease in the $\bar{L_2}$ error. Whereas, with $N_r$ increasing from $100$ to $550$, the figure shows $1.43$ times increase in the computational time and $72\%$ decrease in the $\bar{L_2}$ error.
		Further, comparing FVM-based solution on $50\times50$ grid size and PINN-based solution at $N_r = 550$, Figure \ref{fig:Combined Plot - FVM vs PINN}(a) shows that the present PINN-based periodic solver results in almost same accuracy in a substantially reduced (atmost $82\%$) computational time. \\

		Similarly, with increasing grid size from $40\times40$ to $80\times80$ for the heat diffusion in a square plate with a square-hole, Figure \ref{fig:Combined Plot - FVM vs PINN}(b) shows a $13.22$ times increase in the computational time and $81\%$ decrease in the $\bar{L_2}$ error for the FVM-based solvers. Further, with increasing $N_r = 100$ to $550$, the figure shows $1.90$ times increase in the computational time and $74\%$ decrease in the $\bar{L_2}$ error. Further, comparing FVM-based solution on $50\times50$ grid size and PINN-based solution at $N_r = 550$, Figure \ref{fig:Combined Plot - FVM vs PINN}(b) shows that the present PINN-based periodic solver results in the error norm that is almost half (more accurate) in a substantially reduced (atmost $78\%$) computational time. \\

		Lastly, for the problem on square plate with a circular-hole, Figure \ref{fig:Combined Plot - FVM vs PINN}(c) shows a $9.77$ times increase in the computational time and $92\%$ decrease in the $\bar{L_2}$ error, with increase in grid size from $40\times40$ to $80\times80$. Whereas, with $N_r$ increasing from $100$ to $550$, the figure shows $2.21$ times increase in the computational time and $60\%$ decrease in the $\bar{L_2}$ error. Further, comparing FVM-based solution on $50\times50$ grid size and PINN-based solution at $N_r = 550$, Figure \ref{fig:Combined Plot - FVM vs PINN}(c) shows that the present PINN-based periodic solver results in the error norm that is almost one-third (much more accurate) in a substantially reduced (atmost $99\%$) computational time. \\
		
		\textcolor{black}{For the FVM-based solver (for all grid sizes – $40\times40$, $50\times50$, $80\times80$, and the grid-independent $100\times100$) computational time is reported here for transient-to-periodic simulations starting at $t=0$, and ending at a terminal time $t_{max} = t_{\min}+t_P$. Here,  $t_{min}$ is the transient time required to reach periodic state (less than 1\% variation in peak amplitude of temporal temperature profile beyond  $t_{min}$) and $t_P$ is the time period. Whereas, for the PINN-based solver, the periodic simulations (over the time period $t_P$) start with the transient time $t_{min}$ to ensure a fair comparison of the computational time. For the periodic PINN-based solver, although $t_{min}$ is taken here equal to the transient time (obtained by the FVM-based solver), a smaller $t_{min}$ is found to capture the transient result accurately unless $t_{min}$ is too small. A systematic study on the choice of $t_{min}$  needs to be explored in future work, particularly in systems with multi-frequency periodicity.
		}

		\subsubsection{Discussion} \label{Discussion_Heat}

		The present study investigates how different hyperparameters of the proposed PINN model influence computational time and $L_2$ error. As evident in Figure \ref{fig:Combined Plot - FNN}($a_2$)-($c_2$), the $L_2$ error reaches a minimum when varying $n_N$ for a given $n_L$. This minimum occurs at approximately $n_N = 5$ to $20$ across all values of $n_L$ for the three different 2D periodic heat-diffusion problems examined in the present work. When $n_N$ is small, the neural network has few trainable parameters, reducing its expressiveness and limiting its ability to capture the complex nonlinear PDE solutions. Conversely, when $n_N$ is too large (greater than 20), the number of trainable parameters becomes excessively large (greater than $O(10^3)$), and the fixed number of Adam optimizer iterations may not be sufficient to fine-tune the network's weights and biases effectively. Therefore, using the proposed PINN-based periodic-diffusion solver, a shallow neural network architecture with 10-20 neurons per layer, yielding around $O(10^2)$ trainable parameters, is optimal for achieving the best accuracy in a periodic heat-diffusion problem. The appearance of such a minima was also previously reported in the study by Xing et al. \cite{Xing2023}, in their work on anisotropic heat diffusion problems.
		\\
		
		The behaviour of the proposed PINN-based periodic solver resembles that of traditional FVM-based transient-to-periodic solvers when examining the $L_2$ error as a function of the number of collocation points $N_r$. As shown in Figure \ref{fig:Combined Plot - Nr}($a_2$)-($c_2$), the $L_2$ error in the PINN-based solver decreases and stabilizes within a tolerance range of <3\% as $N_r$ increases. This saturation behavior is typical of numerical methods like the FVM, where increasing the resolution or grid size generally leads to diminishing returns in error reduction. However, between the solvers based on PINN and FVM, a vital distinction lies in their computational efficiency. For FVM-based solvers, when the grid size is doubled, the number of cells increases fourfold, which directly impacts the computational load. Moreover, to maintain numerical stability, particularly with explicit time-stepping schemes, the time step needs to be reduced as the grid is refined. This combination of increased cell count and reduced time step leads to a substantial increase in computational time as the grid is made finer. In contrast, the PINN-based solver exhibits a nearly linear increase in computational time as $N_r$ increases. Unlike FVM-based solvers, the PINN-based approach does not require time-marching or explicit time-stepping schemes, which are the primary reasons for the substantial increase in the computational time in the FVM-based methods. Thus, the PINN-based solver maintains a relatively low computational cost even at high point densities, making it a more efficient choice for problems that would otherwise demand enormous computational resources using traditional FVM-based solvers.\\
		
		The current study also offers a unique perspective by drawing a comparison between the computational efficiency of the PINN-based solver with increasing $N_r$ and FVM-based solvers with increasing grid size; for the three periodic heat-diffusion problems. Remarkably, the proposed PINN-based solver, with a sparse point cloud of just $N_r = 550$, consistently outperforms the \emph{$40\times40$} and \emph{$50\times50$ }coarse mesh FVM-based solvers in both simulation time and accuracy (Figure \ref{fig:Combined Plot - FVM vs PINN}). Although the \emph{$80\times80$} FVM-based solver achieves a lower $L_2$ error, it is crucial to note that this solver operates on a time-marching scheme with a significantly larger spatial grid size (\emph{$80\times80$}, totalling 6400 points in the X-Y plane alone). In contrast, the PINN-based solver runs over 18-20 times faster across all test cases, using only 550 spatio-temporal points while still maintaining small enough $L_2$ errors. This makes the proposed PINN-based approach an excellent alternative to traditional FVM-based approach, especially in scenarios where computational resources and time are critical constraints for attaining time-periodic states of heat-diffusion systems.\\

		\textcolor{black}{It is worth noting that the present formulation bypasses the transient dynamics and directly approximates the periodic state within $[t_{\min}, t_{\max}]$. This is distinct from time-marching PINN-based frameworks such as the Runge-Kutta PINN proposed by Raissi et al.{\cite{Raissi2019}}, which are designed to advance solutions across large time steps in transient problems. Such approaches provide a valuable complementary direction, especially for problems where the transient evolution itself is of interest, and could be incorporated into future extensions of this work.
		}

		\revised{

		\subsection{Verification and Performance Studies:  PINN-based Periodic Flow Solver}	\label{2D_flow_subsection}
		
		\subsubsection{Verification Study}		
		
	Verification of the present PINN-based periodic flow solver is presented here for the 2D periodic LDC problem of Figure \ref{fig:Domain 2D All 2}(d). For the verification study, the reference solution is a grid-independent periodic result that is obtained from a FVM-based transient-to-periodic flow solver. Further, verification study for the FVM-based flow solver is presented in Appendix \hyperref[Appendix_B]{B.}
	\\

		 The specific set of hyperparameters considered are number of layers $n_{L} = 3$ and number of neurons per layer $n_{N} = 20$ for each of the three FNNs. Further, number of collocation points $N_r = 5500$, number of boundary points $N_{BC} = 800$, point spacing for numerical differentiation $\Delta x = \Delta y = 0.005$, and $\Delta t = 0.002$. The loss coefficients from Eqn.\ref{loss_flow} are $\lambda_{BC} = 1$, $\lambda_{R} = 20$ and $\lambda_{period} = 1$.  In contrast to the PINN-based periodic diffusion solver, the present PINN-based flow solver has a larger number of collocation points and trainable parameters as 2D incompressible NS is a more complicated system of PDEs to solve numerically.
		 Reference solution used here is grid-independent \emph{transient-to-periodic} FVM-based solution, on a grid size of $150 \times 150$, for $Re = 10, 50$ and $100$.\\

	For the PINN-based solution as compared to the grid-independent FVM-based solution, excellent agreement is seen in Figure \ref{fig:Fig_18} for instantaneous velocity magnitude contours, and in Figure \ref{fig:Fig_19} for periodic $u$-velocity profiles at point $S(0.5, 0.5)$ in Figure \ref{fig:Domain 2D All 2}(d). Further, for the total velocity magnitude contour, the error $\bar{L_2} = 1.6\%, 1.2\%$ and $1.6\%$, and the standard deviation $SD = 0.3\%, 0.1\%$ and $0.7\%$; at $Re =10, 50$ and $100$, respectively. Whereas, the periodic $u$-velocity profiles for the respective $Re$ resulted in $\bar{L_2} = 1.7\%, 1.1\%$ and $1.8\%$, while $SD = 0.2\%, 0.3\%$ and $0.5\%$.
	Finally, the computational time required by our PINN-based periodic solver is substantially less as compared to that required by our FVM-based transient-to-periodic solver. For the three $Re$ numbers, our PINN-based periodic solver takes a mean computational time of $137s$ $(SD = 4s)$, $138s$ $(SD = 3s)$ and $139s$ $(SD = 5s)$, respectively. On the other hand, the grid-independent FVM-based solver takes a mean computational time of $3204s$ $(SD = 7s)$, $3510s$ $(SD = 4s)$ and $3807s$ $(SD = 8s)$, at $Re =10, 50$ and $100$, respectively. 	\\

		\begin{figure}
			\hbox{\hspace{-0.5em}
				\includegraphics[width=175mm,scale=5]{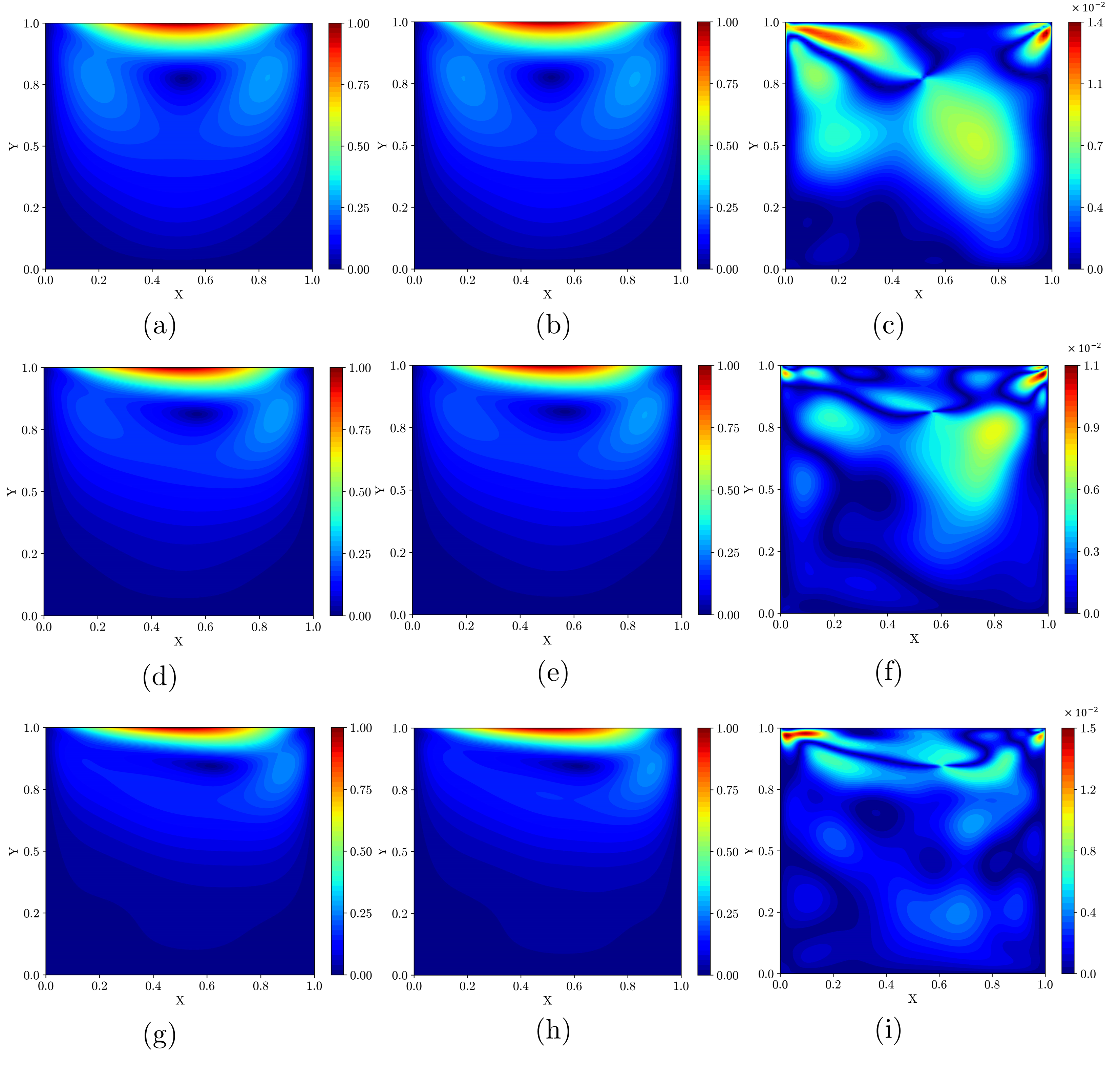}}
			\caption{ \revised{For the periodically oscillating LDC flow problem in Figure \ref{fig:Domain 2D All 2}(d), contour plots of instantaneous resultant velocity magnitude obtained by {(a,d,g)} FVM-based transient-to-periodic solver, {(b,e,h)} PINN-based periodic solver and {(c, f, i)} absolute error ($L_1$) plots, respectively. The velocity contour corresponds to $t = 33s$, for Reynolds number of (a-c) $10$, (d-f) $50$, and (g-i) $100$.}
			}
			\label{fig:Fig_18}
		\end{figure}

		\begin{figure}
			\hbox{\hspace{-0.0em}
				\includegraphics[width=175mm,scale=5]{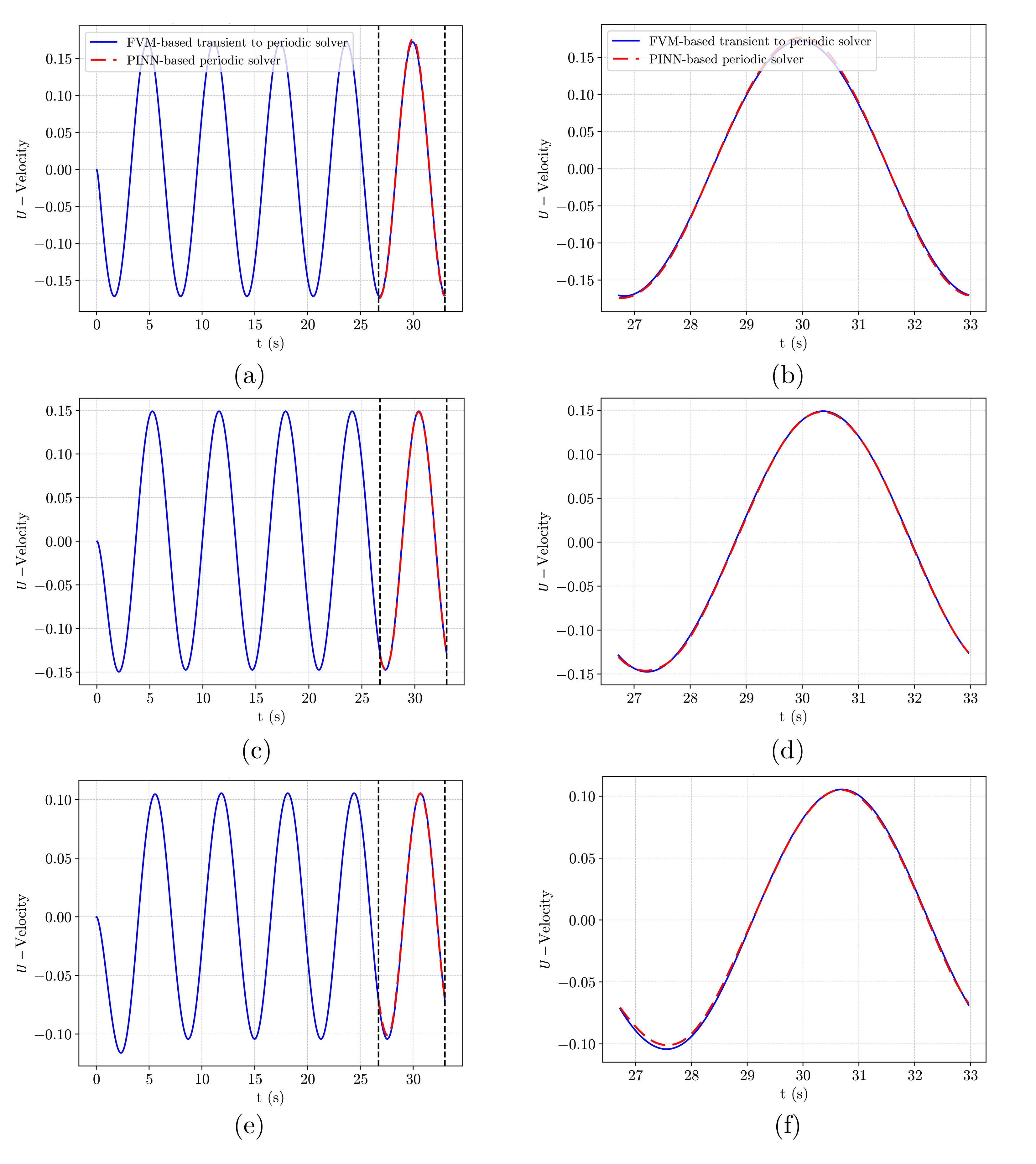}}
			\caption{\revised{Comparison between grid-independent FVM-based transient solution and PINN-based periodic solution for temporal evolution of u-velocity at point $S$ (marked in Figure \ref{fig:Domain 2D All 2}(d)) at a Reynolds number of (a,b) $10$, (c,d) $50$, and (e,f) $100$.}
			}
			\label{fig:Fig_19}
		\end{figure}

\subsubsection{Performance Study : PINN v/s FVM-based Navier-Stokes Solver}\label{PINN_vs_FVM_Flow}

		This section presents performance of the present PINN-based flow solver with increasing number of collocation points $N_r$, as compared to the FVM-based solver with increasing grid sizes for the periodic LDC problem in Figure \ref{fig:Domain 2D All 2}(d). The number of collocation points chosen for the PINN solver are $N_r = 1000$, $2000$, and $5500$. This selection is based on the same reasoning as the one outlined in subsection \ref{PINN_vs_FVM_Heat}. For the PINN-based and FVM-based solvers, the $\bar{L_2}$ error is computed with respect to the grid-independent FVM-based solution on a grid size of $150\times150$.\\

		\begin{figure}
			\hbox{\hspace{-0.5em}
				\includegraphics[width=175mm,scale=5]{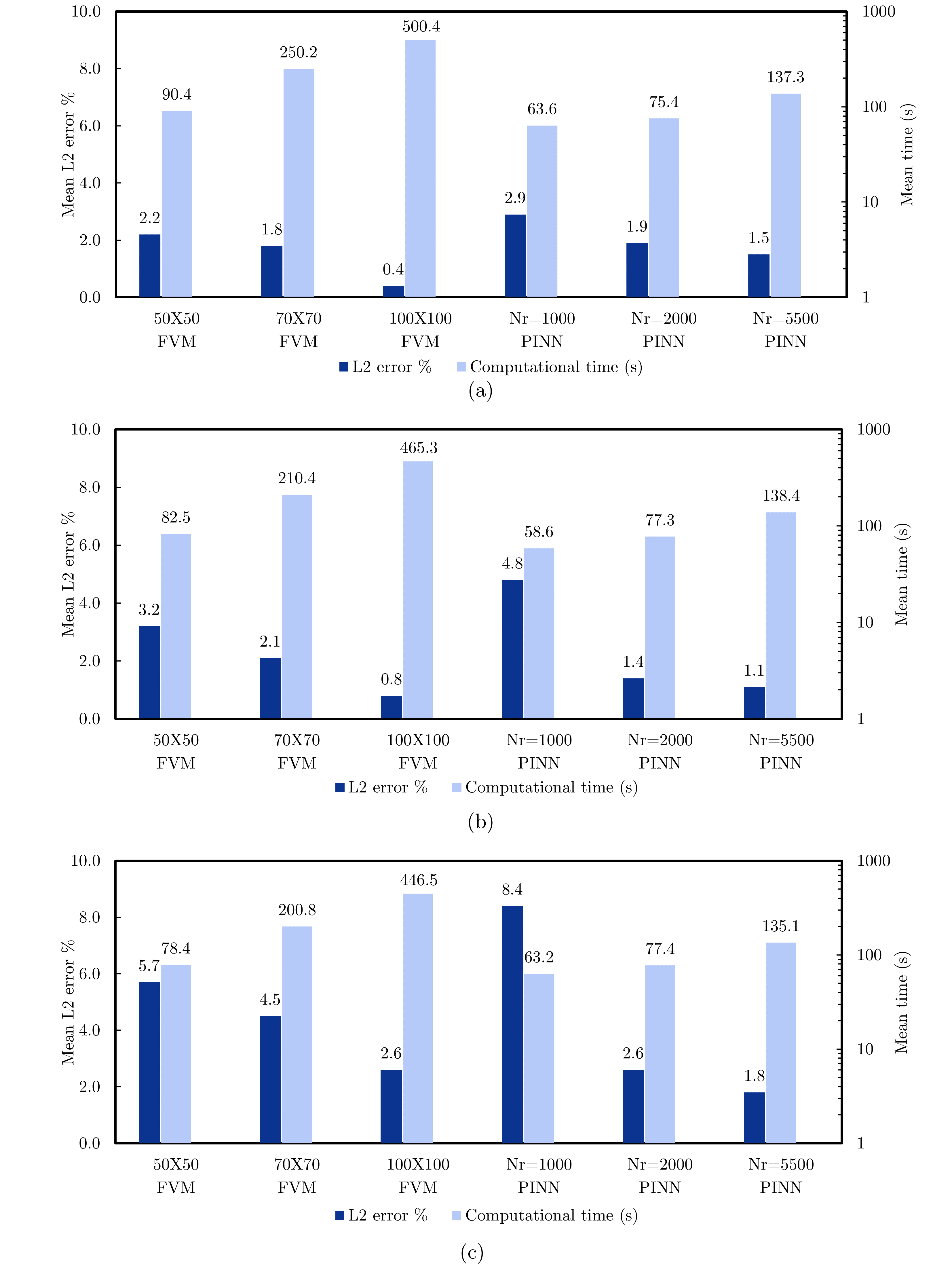}}
			\caption{ 	\revised{Comparison of computational time and $\bar{L_2}$ error for the present PINN-based periodic and FVM-based transient-to-periodic flow solvers on various collocation points $N_r$ and grid sizes, respectively. The computational performance in subfigures (a), (b), and (c) correspond to $Re = 10, 50$ and $100$, respectively. Further, for both the flow solvers, the error is computed with respect to a grid-independent FVM-based solution on a grid size of $150\times150$. For the PINN-based periodic solver, the above results are obtained with $n_{L} = 3$, $n_{N} = 20$, $\Delta x = \Delta y = 0.005$, $\Delta t = 0.002$, $N_{BC} = 800$.}
			}
			\label{fig:Fig_20}
		\end{figure}

		For the FVM-based transient-to-periodic solutions of the 2D LDC problem (Figure \ref{fig:Domain 2D All 2}(d)), Figure \ref{fig:Fig_20} shows a much larger increase in computational time as compared to the decrease in the $\bar{L_2}$ error with grid refinement. In contrast, with increasing $N_r$ for the PINN-based periodic solution, the figure shows a much smaller increase in computational time as compared to the decrease in the $\bar{L_2}$ error. For example, with increasing grid size from $50\times50$ to $100\times100$ at $Re = 10$, Figure \ref{fig:Fig_20}(a) shows a $5.55$ times increase in the computational time and $82\%$ decrease in the $\bar{L_2}$ error. Whereas, with $N_r$ increasing from $1000$ to $5500$, the figure shows $2.15$ times increase in the computational time and $71\%$ decrease in the $\bar{L_2}$ error.
		Further, comparing FVM-based solution on $70\times70$ grid size and PINN-based solution at $N_r = 2000$, Figure \ref{fig:Fig_20}(a) shows that the present PINN-based periodic flow solver results in almost same accuracy in a substantially reduced (atmost $70\%$) computational time. \\

		Similarly, with increasing grid size from $50\times50$ to $100\times100$ at $Re = 50$, Figure \ref{fig:Fig_20}(b) shows a $5.64$ times increase in the computational time and $75\%$ decrease in the $\bar{L_2}$ error for the FVM-based solvers. Further, with increasing $N_r = 1000$ to $5500$, the figure shows $2.36$ times increase in the computational time and $77\%$ decrease in the $\bar{L_2}$ error. Further, comparing FVM-based solution on $70\times70$ grid size and PINN-based solution at $N_r = 5500$, Figure \ref{fig:Fig_20}(b) shows that the present PINN-based periodic solver results in the error norm that is almost half (more accurate) in a substantially reduced (atmost $40\%$) computational time. \\

		Lastly, for $Re = 100$, Figure \ref{fig:Fig_20}(c) shows a $5.69$ times increase in the computational time and $54\%$ decrease in the $\bar{L_2}$ error, with increase in grid size from $50\times50$ to $100\times100$. Whereas, with $N_r$ increasing from $1000$ to $5500$, the figure shows $2.13$ times increase in the computational time and $79\%$ decrease in the $\bar{L_2}$ error. Further, comparing FVM-based solution on $70\times70$ grid size and PINN-based solution at $N_r = 2000$, Figure \ref{fig:Fig_20}(c) shows that the present PINN-based periodic solver results in a relative $\bar{L_2}$ error that is almost half (much more accurate) in a reduced (atmost $62\%$) computational time. \\

	\subsubsection{Discussion}
	
	The present study demonstrates the efficacy of the current PINN-based periodic approach to solve 2D incompressible fluid flow problem using a lid driven cavity (LDC) flow problem, across three Reynolds numbers. The comparison between the computational efficiency of the PINN-based solver with increasing $N_r$ and FVM-based solver with increasing grid size is also presented here. The proposed PINN-based solver, with  a point cloud of just $N_r = 2000$, consistently outperforms the $50 \times 50$ and $70 \times 70$ coarse-mesh FVM-based solvers in both simulation time and accuracy (Figure \ref{fig:Fig_20}). Although the $100 \times 100$ FVM-based solver achieves a lower $L_2$ error, it should be noted that it operated on a time-marching scheme with an extremely larger spatial grid size ($10000$). As the PINN-based solver runs over $5-10$ times faster across all test cases, using only $1000-2000$ spatio-temporal points, it offers a viable alternative to traditional FVM-based approaches for attaining time periodic states of nonlinear fluid flow systems.

}
	
	
	\revised{	
		
		\section{Conclusions} \label{Conclusion}

		A Physics-Informed Neural Network (PINN)-based framework is proposed for accelerating the attainment of time-periodic snapshots of heat and fluid flow-fields and the periodic state-based engineering parameters, and has provided insights for future research on PINN-based CFD methodology and its applications in systems exhibiting periodic behavior. From the presented mathematical formulation, verification study, and performance study, conclusions drawn are as follows:
		
		\begin{enumerate}
			
			\item A novel methodology, for the PINN-based periodic solver, is discussed that combines hard/hybrid constraint enforcement along with numerical differentiation to accelerate the convergence to periodic states in simulations of unsteady heat and fluid flow problems, both in simple and complex geometries, achieving a $\bar{L_2}$ error norm of $O(10^{-2})$ across all cases.
			
			\item The present study systematically evaluates the PINN-based diffusion solver's performance by analyzing variations in key hyperparameters and their impact on computational time and accuracy. This highlights the solver’s robustness and adaptability across different problem setups.
			
			\item  A fresh perspective on the practical applicability and limitations of the PINN-based periodic solver as compared to traditional FVM-based transient-to-periodic solver is presented, highlighting the substantial difference in computational time and trade-offs in accuracy inherent in these methods.
		\end{enumerate}

		Collectively, these findings pave the way for further advancements in data-free machine learning and optimization techniques, with the potential to substantially enhance the computational efficiency while achieving the desirable accuracy in CFD problems.
		
	}

		\section*{Acknowledgement}

		We gratefully acknowledge that this work is supported by a grant (No. CRG/2022/008177) from Anusandhan National Research Foundation (ANRF), Department of Science and Technology (DST), New Delhi, India.

		\section*{ Appendix A. Verification of the FVM-based Diffusion Solver} \label{Appendix_A}
		
\begin{figure}
			\hbox{\hspace{0em}
				\includegraphics[width=170mm,scale=5]{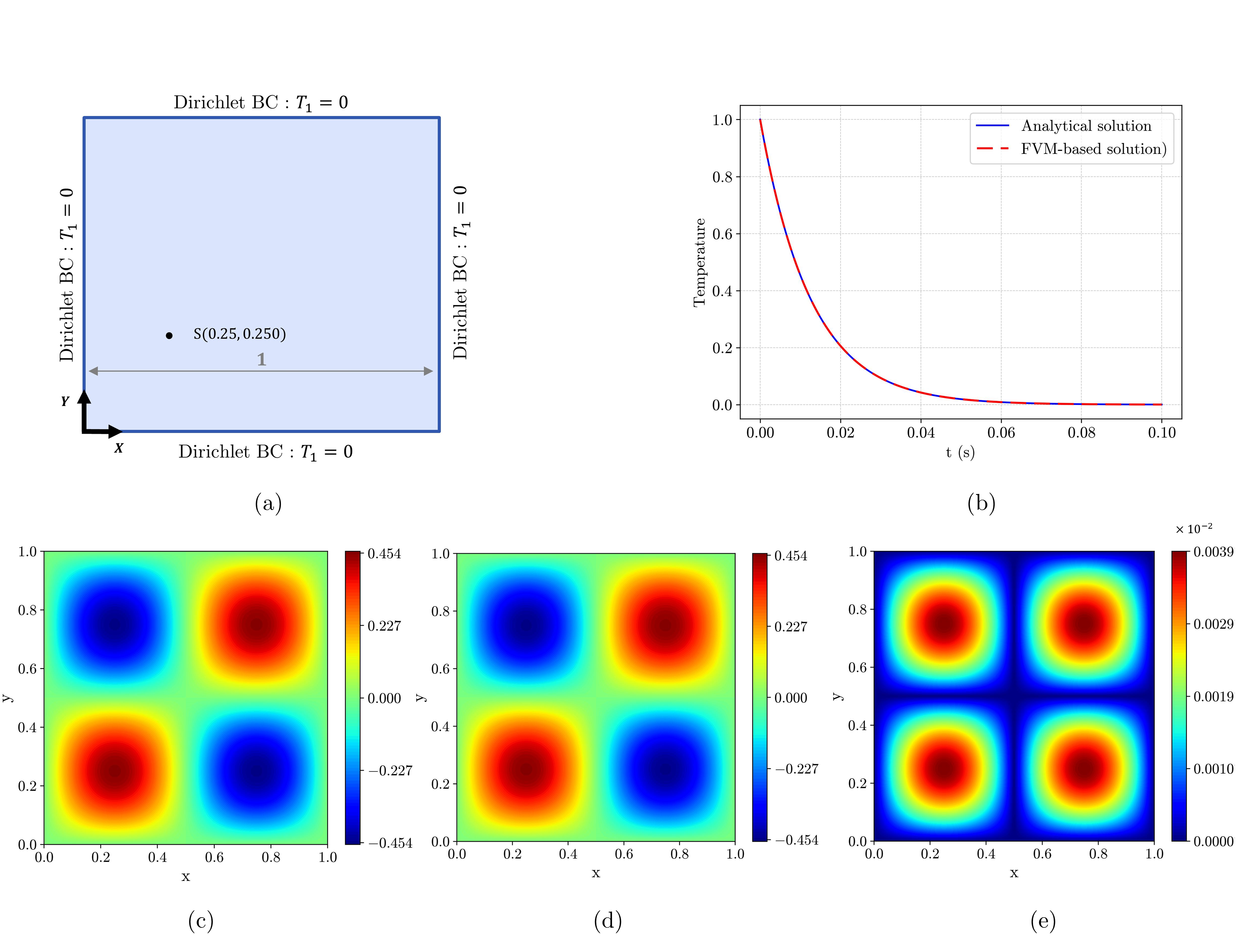}}
			\caption{ (a) Computational setup for 2D transient diffusion in a square plate, (b) comparison of $160 \times 160$ FVM-based transient solution and analytical solution at point $S(0.25, 0.25)$. Further, instantaneous temperature contours obtained from (c) analytical solution, (d) $160 \times 160$ FVM-based solution, and (e) absolute error, at time $t = 0.01s$}
			\label{fig:Combined Plot - FVM Verification Appendix A}
		\end{figure}

		Verification of the current FVM-based diffusion solver is presented in this section. The problem considered is 2D heat diffusion in a unit square plate, having homogeneous Dirichlet boundary conditions at the four walls, and a sinusoidal initial condition $T_{IC}$ as depicted in Figure \ref{fig:Combined Plot - FVM Verification Appendix A}(a). The PDE considered here is the same heat diffusion PDE of Eqn.\ref{1}, with $Q(t) = 0$, $\alpha = 1$ and $k = 1$. The system's temporal domain is limited to $t \in [0, t_{max}]$, where $t_{max} = 0.1s$.
		The analytical solution of such a system as given by Carslaw and Jaeger \cite{Carslaw1986} is\\
		
		\begin{equation}
			T(x, y, t) = sin\left(\frac{p \pi x}{L}\right) sin\left(\frac{q \pi y}{L}\right) exp\left(-\alpha \frac{\pi^2 (p^2 + q^2)}{L^2}t\right),
		\end{equation}\\
		
		\noindent for an initial condition\\
		
		\begin{equation}
			T_{IC}(x, y, t) = sin\left(\frac{p \pi x}{L}\right) sin\left(\frac{q \pi y}{L}\right),
		\end{equation}\\
		
		\noindent with integers $(p, q) \ge 1$. Values of p and q are both equal to $2$ for the current system.\\

		The FVM-based solver is developed using an explicit unsteady formulation on a uniform Cartesian mesh. Diffusive heat fluxes are discretized using second-order central differencing while the temporal derivative is advanced using explicit forward Euler scheme. This formulation yields a conditionally stable scheme, with the time-step restricted by the following stability criterion \\ 
		
		\begin{equation}
			\Delta t_{stable} \le \frac{1}{2 \alpha} \left[ \frac{1}{\Delta x^2} + \frac{1}{\Delta y^2}\right]^{-1}
		\end{equation}\\

		Detailed derivation of the finite volume formulation, implementation steps and solution algorithm can be found in Sharma \cite{Sharma2021}.\\
		
		Verification results are presented in Figure \ref{fig:Combined Plot - FVM Verification Appendix A}. Figure \ref{fig:Combined Plot - FVM Verification Appendix A}(b) shows excellent agreement between the analytical solution and the FVM-based solution for the temporal temperature profile at the point $S(0.25, 0.25)$, with $\bar{L_2} = 0.0074\%$. Further, the FVM-based solution is also compared with the analytical solution at an intermediate timestep of $t = 0.01s$. The analytical solution, FVM-based solution on a grid size of $160 \times 160$, and the absolute error plots are shown in Figure \ref{fig:Combined Plot - FVM Verification Appendix A}(c), \ref{fig:Combined Plot - FVM Verification Appendix A}(d) and \ref{fig:Combined Plot - FVM Verification Appendix A}(e), respectively. These results confirm that the FVM-based solver achieves excellent verification with the 2D heat diffusion benchmark problem.\\

		\revised{
		
		\section*{ Appendix B. Verification of the FVM-based Navier-Stokes Solver}  \label{Appendix_B}
		
		Verification of the current FVM-based incompressible Navier--Stokes solver, used as the reference solution in Section~\ref{2D_flow_subsection}, is presented in this section. The problem considered is also a lid driven cavity flow with uniform lid velocity in the x-direction, as depicted schematically in Figure \ref{fig:LDC_Ghia_FVM}(a). The PDE considered is the same as in Eqn.\ref{continuity}-\ref{Y_mom}, with $Re=100$. The system's temporal domain is $t \in [0, 10]s$, allowing it to reach a steady state. All the flow properties start from a $u=v=0$ initial condition for the present verification case. \\

		\begin{figure}
			\hbox{\hspace{-3em}
				\includegraphics[width=190mm,scale=5]{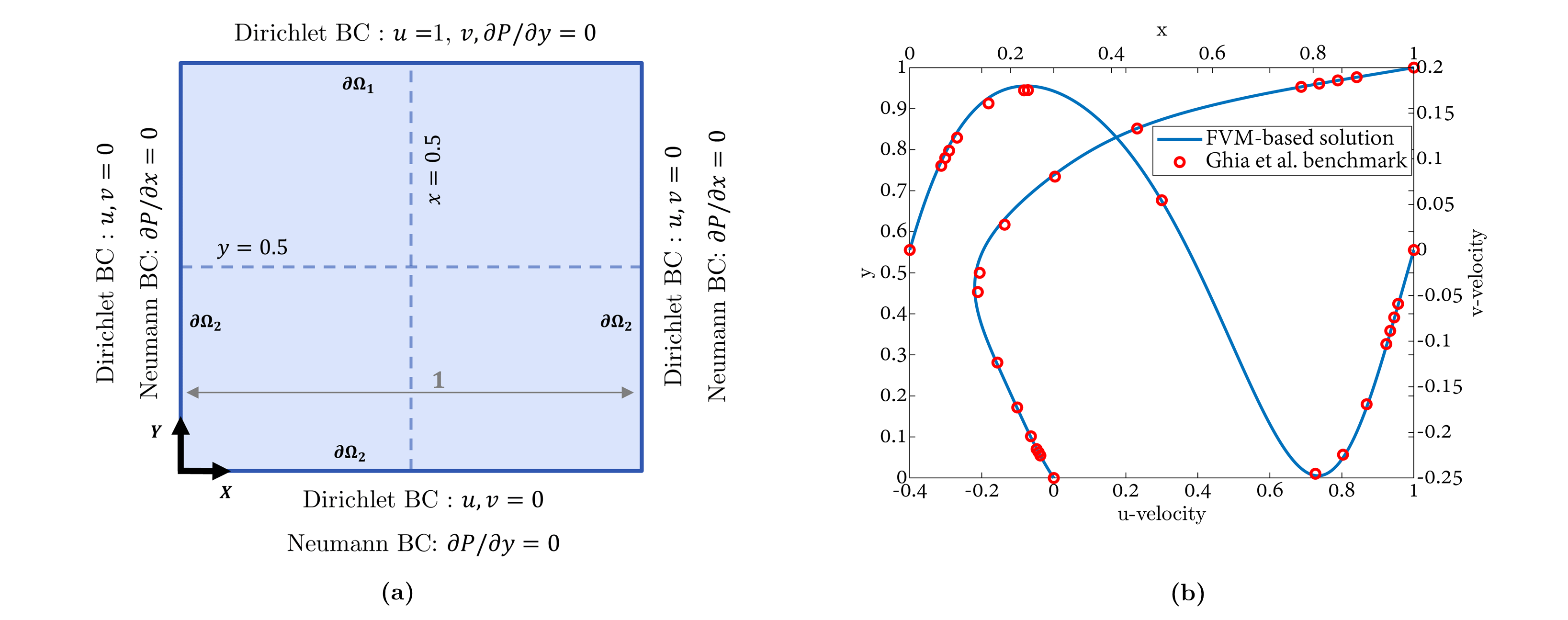}}
			\caption{\revised{(a) Computational setup for 2D LDC flow and (b) comparison of present and published results on variation of u-velocity along vertical centerline, and v-velocity along horizontal centerline. The present results are obtained from the in-house FVM-based flow solver on a grid size of $201\times201$.}
			}
			\label{fig:LDC_Ghia_FVM}
		\end{figure}

	    The solver is implemented on a staggered Cartesian grid and a semi-explicit predictor-corrector strategy is employed for time integration. 	A detailed derivation of the finite-volume discretization, implementation details, and solution algorithm can be found in Sharma~\cite{Sharma2021}. Figure \ref{fig:LDC_Ghia_FVM}(b) shows an excellent agreement between the present FVM-based velocity profiles as compared to the benchmark solution by Ghia et al. \cite{Ghia1982}.

	}

		
		\newpage
		\bibliographystyle{apacite}
		\bibliography{PINN_References_New}

\begin{thebibliography}{10}
\expandafter\ifx\csname url\endcsname\relax
  \def\url#1{\texttt{#1}}\fi
\expandafter\ifx\csname urlprefix\endcsname\relax\def\urlprefix{URL }\fi
\expandafter\ifx\csname href\endcsname\relax
  \def\href#1#2{#2} \def\path#1{#1}\fi

\bibitem{Colonius2011}
T.~Colonius, D.~R. Williams,
  \href{https://pubmed.ncbi.nlm.nih.gov/21382829/}{Control of vortex shedding
  on two- and three-dimensional aerofoils}, Philosophical transactions. Series
  A, Mathematical, physical, and engineering sciences 369 (2011) 1525--1539.
\newblock \href {https://doi.org/10.1098/RSTA.2010.0355}
  {\path{doi:10.1098/RSTA.2010.0355}}.
\newline\urlprefix\url{https://pubmed.ncbi.nlm.nih.gov/21382829/}

\bibitem{Morab2024}
S.~R. Morab, J.~S. Murallidharan, A.~Sharma,
  \href{/aip/pof/article/36/4/041906/3281081/Computational-hemodynamics-and-hemoacoustic-study}{Computational
  hemodynamics and hemoacoustic study on carotid bifurcation: Effect of
  stenosis and branch angle}, Physics of Fluids 36 (4 2024).
\newblock \href {https://doi.org/10.1063/5.0203193/3281081}
  {\path{doi:10.1063/5.0203193/3281081}}.
\newline\urlprefix\url{/aip/pof/article/36/4/041906/3281081/Computational-hemodynamics-and-hemoacoustic-study}

\bibitem{Mehrabi2012}
M.~Mehrabi, S.~Setayeshi,
  \href{https://onlinelibrary.wiley.com/doi/full/10.1155/2012/804765
  https://onlinelibrary.wiley.com/doi/abs/10.1155/2012/804765
  https://onlinelibrary.wiley.com/doi/10.1155/2012/804765}{Computational fluid
  dynamics analysis of pulsatile blood flow behavior in modelled stenosed
  vessels with different severities}, Mathematical Problems in Engineering 2012
  (2012) 804765.
\newblock \href {https://doi.org/10.1155/2012/804765}
  {\path{doi:10.1155/2012/804765}}.
\newline\urlprefix\url{https://onlinelibrary.wiley.com/doi/full/10.1155/2012/804765
  https://onlinelibrary.wiley.com/doi/abs/10.1155/2012/804765
  https://onlinelibrary.wiley.com/doi/10.1155/2012/804765}

\bibitem{McMullen2001}
M.~McMullen, A.~Jameson, J.~J. Alonso,
  \href{https://arc.aiaa.org/doi/10.2514/6.2001-152}{Acceleration of
  convergence to a periodic steady state in turbomachinery flows}, 39th
  Aerospace Sciences Meeting and Exhibit (2001).
\newblock \href {https://doi.org/10.2514/6.2001-152}
  {\path{doi:10.2514/6.2001-152}}.
\newline\urlprefix\url{https://arc.aiaa.org/doi/10.2514/6.2001-152}

\bibitem{Cao2022}
Y.~Cao, L.~Zhou, C.~Ou, H.~Fang, D.~Liu, 3d cfd simulation and analysis of
  transient flow in a water pipeline, Aqua Water Infrastructure, Ecosystems and
  Society 71 (2022) 751--767.
\newblock \href {https://doi.org/10.2166/AQUA.2022.023}
  {\path{doi:10.2166/AQUA.2022.023}}.

\bibitem{Xue2021}
D.~Xue, Y.~Liu, L.~Li,
  \href{https://onlinelibrary.wiley.com/doi/full/10.1002/htj.22237
  https://onlinelibrary.wiley.com/doi/abs/10.1002/htj.22237
  https://onlinelibrary.wiley.com/doi/10.1002/htj.22237}{Heat transfer
  performance of wet porous solar collectors under periodic conditions}, Heat
  Transfer 50 (2021) 7440--7453.
\newblock \href {https://doi.org/10.1002/HTJ.22237}
  {\path{doi:10.1002/HTJ.22237}}.
\newline\urlprefix\url{https://onlinelibrary.wiley.com/doi/full/10.1002/htj.22237
  https://onlinelibrary.wiley.com/doi/abs/10.1002/htj.22237
  https://onlinelibrary.wiley.com/doi/10.1002/htj.22237}

\bibitem{Anderson1995}
J.~D. Anderson,
  \href{https://books.google.com/books/about/Computational_Fluid_Dynamics.html?id=dJceAQAAIAAJ}{Computational
  Fluid Dynamics: The Basic with Application}, Vol.~1, McGraw-Hill, 1995.
\newline\urlprefix\url{https://books.google.com/books/about/Computational_Fluid_Dynamics.html?id=dJceAQAAIAAJ}

\bibitem{Sharma2021}
A.~Sharma, Introduction to Computational Fluid Dynamics: Development,
  Application and Analysis, Springer International Publishing, 2021.
\newblock \href {https://doi.org/10.1007/978-3-030-72884-7/COVER}
  {\path{doi:10.1007/978-3-030-72884-7/COVER}}.

\bibitem{Pfaller2021}
M.~R. Pfaller, J.~Pham, N.~M. Wilson, D.~W. Parker, A.~L. Marsden,
  \href{https://pubmed.ncbi.nlm.nih.gov/34169398/}{On the periodicity of
  cardiovascular fluid dynamics simulations}, Annals of biomedical engineering
  49 (2021) 3574--3592.
\newblock \href {https://doi.org/10.1007/S10439-021-02796-X}
  {\path{doi:10.1007/S10439-021-02796-X}}.
\newline\urlprefix\url{https://pubmed.ncbi.nlm.nih.gov/34169398/}

\bibitem{Johnson1993}
A.~A. Johnson, T.~E. Tezduyar, J.~Liou,
  \href{https://link.springer.com/article/10.1007/BF00350094}{Numerical
  simulation of flows past periodic arrays of cylinders}, Computational
  Mechanics 11 (1993) 371--383.
\newblock \href {https://doi.org/10.1007/BF00350094/METRICS}
  {\path{doi:10.1007/BF00350094/METRICS}}.
\newline\urlprefix\url{https://link.springer.com/article/10.1007/BF00350094}

\bibitem{Carte1995}
G.~Carte, J.~Dušek, P.~Fraunié, A spectral time discretization for flows with
  dominant periodicity, Journal of Computational Physics 120 (1995) 171--183.
\newblock \href {https://doi.org/10.1006/JCPH.1995.1157}
  {\path{doi:10.1006/JCPH.1995.1157}}.

\bibitem{Hall2012}
K.~C. Hall, J.~P. Thomas, W.~S. Clark,
  \href{https://arc.aiaa.org/doi/10.2514/2.1754}{Computation of unsteady
  nonlinear flows in cascades using a harmonic balance technique},
  https://doi.org/10.2514/2.1754 40 (2012) 879--886.
\newblock \href {https://doi.org/10.2514/2.1754} {\path{doi:10.2514/2.1754}}.
\newline\urlprefix\url{https://arc.aiaa.org/doi/10.2514/2.1754}

\bibitem{Ekici2012}
K.~Ekici, K.~C. Hall, \href{https://arc.aiaa.org/doi/10.2514/1.22888}{Nonlinear
  analysis of unsteady flows in multistage turbomachines using harmonic
  balance}, https://doi.org/10.2514/1.22888 45 (2012) 1047--1057.
\newblock \href {https://doi.org/10.2514/1.22888} {\path{doi:10.2514/1.22888}}.
\newline\urlprefix\url{https://arc.aiaa.org/doi/10.2514/1.22888}

\bibitem{R2007}
R.~Lübke, A.~Seidel-Morgenstern, L.~Tobiska, Numerical method for accelerated
  calculation of cyclic steady state of modicon–smb-processes, Computers \&
  Chemical Engineering 31 (2007) 258--267.
\newblock \href {https://doi.org/10.1016/J.COMPCHEMENG.2006.06.013}
  {\path{doi:10.1016/J.COMPCHEMENG.2006.06.013}}.

\bibitem{Zhang2011}
W.~Zhang, G.~Xi, C.~Zhang, Z.~Huang,
  \href{https://www.tandfonline.com/doi/abs/10.1080/10618562.2011.575369}{A
  high-accuracy temporal–spatial pseudospectral method for time-periodic
  unsteady fluid flow and heat transfer problems}, International Journal of
  Computational Fluid Dynamics 25 (2011) 191--206.
\newblock \href {https://doi.org/10.1080/10618562.2011.575369}
  {\path{doi:10.1080/10618562.2011.575369}}.
\newline\urlprefix\url{https://www.tandfonline.com/doi/abs/10.1080/10618562.2011.575369}

\bibitem{Richter2021}
T.~Richter, An averaging scheme for the efficient approximation of
  time-periodic flow problems, Computers \& Fluids 214 (2021) 104769.
\newblock \href {https://doi.org/10.1016/J.COMPFLUID.2020.104769}
  {\path{doi:10.1016/J.COMPFLUID.2020.104769}}.

\bibitem{Raissi2019}
M.~Raissi, P.~Perdikaris, G.~E. Karniadakis, Physics-informed neural networks:
  A deep learning framework for solving forward and inverse problems involving
  nonlinear partial differential equations, Journal of Computational Physics
  378 (2019) 686--707.
\newblock \href {https://doi.org/10.1016/J.JCP.2018.10.045}
  {\path{doi:10.1016/J.JCP.2018.10.045}}.

\bibitem{Dissanayake1994}
M.~W. Dissanayake, N.~Phan‐Thien,
  \href{https://onlinelibrary.wiley.com/doi/full/10.1002/cnm.1640100303
  https://onlinelibrary.wiley.com/doi/abs/10.1002/cnm.1640100303
  https://onlinelibrary.wiley.com/doi/10.1002/cnm.1640100303}{Neural-network-based
  approximations for solving partial differential equations}, Communications in
  Numerical Methods in Engineering 10 (1994) 195--201.
\newblock \href {https://doi.org/10.1002/CNM.1640100303}
  {\path{doi:10.1002/CNM.1640100303}}.
\newline\urlprefix\url{https://onlinelibrary.wiley.com/doi/full/10.1002/cnm.1640100303
  https://onlinelibrary.wiley.com/doi/abs/10.1002/cnm.1640100303
  https://onlinelibrary.wiley.com/doi/10.1002/cnm.1640100303}

\bibitem{Shah2024}
S.~Shah, N.~K. Anand,
  \href{/aip/pof/article/36/7/073620/3303820/Physics-informed-neural-networks-for-periodic}{Physics-informed
  neural networks for periodic flows}, Physics of Fluids 36 (7 2024).
\newblock \href {https://doi.org/10.1063/5.0216266/3303820}
  {\path{doi:10.1063/5.0216266/3303820}}.
\newline\urlprefix\url{/aip/pof/article/36/7/073620/3303820/Physics-informed-neural-networks-for-periodic}

\bibitem{McClenny2023}
L.~D. McClenny, U.~M. Braga-Neto, Self-adaptive physics-informed neural
  networks, Journal of Computational Physics 474 (2023) 111722.
\newblock \href {https://doi.org/10.1016/J.JCP.2022.111722}
  {\path{doi:10.1016/J.JCP.2022.111722}}.

\bibitem{Wang2022}
S.~Wang, X.~Yu, P.~Perdikaris, When and why pinns fail to train: A neural
  tangent kernel perspective, Journal of Computational Physics 449 (2022)
  110768.
\newblock \href {https://doi.org/10.1016/J.JCP.2021.110768}
  {\path{doi:10.1016/J.JCP.2021.110768}}.

\bibitem{Wang2021}
S.~Wang, H.~Wang, P.~Perdikaris, On the eigenvector bias of fourier feature
  networks: From regression to solving multi-scale pdes with physics-informed
  neural networks, Computer Methods in Applied Mechanics and Engineering 384
  (2021) 113938.
\newblock \href {https://doi.org/10.1016/J.CMA.2021.113938}
  {\path{doi:10.1016/J.CMA.2021.113938}}.

\bibitem{Nabian2021}
M.~A. Nabian, R.~J. Gladstone, H.~Meidani, Efficient training of
  physics‐informed neural networks via importance sampling, Computer‐Aided
  Civil and Infrastructure Engineering 36 (2021) 962--977.
\newblock \href {https://doi.org/10.1111/MICE.12685}
  {\path{doi:10.1111/MICE.12685}}.

\bibitem{Jagtap2022}
A.~D. Jagtap, Y.~Shin, K.~Kawaguchi, G.~E. Karniadakis, Deep kronecker neural
  networks: A general framework for neural networks with adaptive activation
  functions, Neurocomputing 468 (2022) 165--180.
\newblock \href {https://doi.org/10.1016/J.NEUCOM.2021.10.036}
  {\path{doi:10.1016/J.NEUCOM.2021.10.036}}.

\bibitem{Jagtap2020}
A.~D. Jagtap, K.~Kawaguchi, G.~E. Karniadakis, Adaptive activation functions
  accelerate convergence in deep and physics-informed neural networks, Journal
  of Computational Physics 404 (2020) 109136.
\newblock \href {https://doi.org/10.1016/J.JCP.2019.109136}
  {\path{doi:10.1016/J.JCP.2019.109136}}.

\bibitem{Mattey2022}
R.~Mattey, S.~Ghosh, A novel sequential method to train physics informed neural
  networks for allen cahn and cahn hilliard equations, Computer Methods in
  Applied Mechanics and Engineering 390 (2022) 114474.
\newblock \href {https://doi.org/10.1016/J.CMA.2021.114474}
  {\path{doi:10.1016/J.CMA.2021.114474}}.

\bibitem{Wu2023}
C.~Wu, M.~Zhu, Q.~Tan, Y.~Kartha, L.~Lu, A comprehensive study of non-adaptive
  and residual-based adaptive sampling for physics-informed neural networks,
  Computer Methods in Applied Mechanics and Engineering 403 (2023) 115671.
\newblock \href {https://doi.org/10.1016/J.CMA.2022.115671}
  {\path{doi:10.1016/J.CMA.2022.115671}}.

\bibitem{Dutta2020}
S.~Dutta, A.~H. Gandomi, Design of experiments for uncertainty quantification
  based on polynomial chaos expansion metamodels, Elsevier, 2020, pp. 369--381.
\newblock \href {https://doi.org/10.1016/B978-0-12-816514-0.00015-1}
  {\path{doi:10.1016/B978-0-12-816514-0.00015-1}}.

\bibitem{Chiu2022}
P.~H. Chiu, J.~C. Wong, C.~Ooi, M.~H. Dao, Y.~S. Ong, Can-pinn: A fast
  physics-informed neural network based on coupled-automatic–numerical
  differentiation method, Computer Methods in Applied Mechanics and Engineering
  395 (2022) 114909.
\newblock \href {https://doi.org/10.1016/J.CMA.2022.114909}
  {\path{doi:10.1016/J.CMA.2022.114909}}.

\bibitem{Wong2024}
J.~C. Wong, C.~C. Ooi, A.~Gupta, Y.~S. Ong, Learning in sinusoidal spaces with
  physics-informed neural networks, IEEE Transactions on Artificial
  Intelligence 5 (2024) 985--1000.
\newblock \href {https://doi.org/10.1109/TAI.2022.3192362}
  {\path{doi:10.1109/TAI.2022.3192362}}.

\bibitem{Chen2021}
X.~Chen, R.~Chen, Q.~Wan, R.~Xu, J.~Liu,
  \href{https://www.nature.com/articles/s41598-021-99037-x}{An improved
  data-free surrogate model for solving partial differential equations using
  deep neural networks}, Scientific Reports 2021 11:1 11 (2021) 1--17.
\newblock \href {https://doi.org/10.1038/s41598-021-99037-x}
  {\path{doi:10.1038/s41598-021-99037-x}}.
\newline\urlprefix\url{https://www.nature.com/articles/s41598-021-99037-x}

\bibitem{Li2022}
S.~Li, X.~Feng, \href{https://www.mdpi.com/1099-4300/24/9/1254/htm
  https://www.mdpi.com/1099-4300/24/9/1254}{Dynamic weight strategy of
  physics-informed neural networks for the 2d navier–stokes equations},
  Entropy 2022, Vol. 24, Page 1254 24 (2022) 1254.
\newblock \href {https://doi.org/10.3390/E24091254}
  {\path{doi:10.3390/E24091254}}.
\newline\urlprefix\url{https://www.mdpi.com/1099-4300/24/9/1254/htm
  https://www.mdpi.com/1099-4300/24/9/1254}

\bibitem{Lu2021}
L.~Lu, R.~Pestourie, W.~Yao, Z.~Wang, F.~Verdugo, S.~G. Johnson,
  \href{https://arxiv.org/abs/2102.04626v1}{Physics-informed neural networks
  with hard constraints for inverse design}, SIAM Journal on Scientific
  Computing 43 (2021) B1105--B1132.
\newblock \href {https://doi.org/10.1137/21M1397908}
  {\path{doi:10.1137/21M1397908}}.
\newline\urlprefix\url{https://arxiv.org/abs/2102.04626v1}

\bibitem{Kingma2014}
D.~P. Kingma, J.~L. Ba, \href{https://arxiv.org/abs/1412.6980v9}{Adam: A method
  for stochastic optimization}, 3rd International Conference on Learning
  Representations, ICLR 2015 - Conference Track Proceedings (12 2014).
\newline\urlprefix\url{https://arxiv.org/abs/1412.6980v9}

\bibitem{Adam}
\href{https://www.tensorflow.org/api_docs/python/tf/keras/optimizers/Adam}{tf.keras.optimizers.adam
   |  tensorflow v2.16.1}.
\newline\urlprefix\url{https://www.tensorflow.org/api_docs/python/tf/keras/optimizers/Adam}

\bibitem{tf}
\href{https://www.tensorflow.org/}{Tensorflow}.
\newline\urlprefix\url{https://www.tensorflow.org/}

\bibitem{Sukumar2022}
N.~Sukumar, A.~Srivastava,
  \href{https://www.sciencedirect.com/science/article/pii/S0045782521006186}{Exact
  imposition of boundary conditions with distance functions in physics-informed
  deep neural networks}, Computer Methods in Applied Mechanics and Engineering
  389 (2022) 114333.
\newblock \href {https://doi.org/10.1016/J.CMA.2021.114333}
  {\path{doi:10.1016/J.CMA.2021.114333}}.
\newline\urlprefix\url{https://www.sciencedirect.com/science/article/pii/S0045782521006186}

\bibitem{Ghia1982}
U.~Ghia, K.~N. Ghia, C.~T. Shin, \href{https://doi.org/10.2514/3.61116}{High-re
  solutions for incompressible flow using the navier-stokes equations and a
  multigrid method}, Journal of Computational Physics 48 (1982) 387--411.
\newblock \href {https://doi.org/10.1016/0021-9991(82)90058-4}
  {\path{doi:10.1016/0021-9991(82)90058-4}}.
\newline\urlprefix\url{https://doi.org/10.2514/3.61116}

\bibitem{Ramabathiran2021}
A.~A. Ramabathiran, P.~Ramachandran, Spinn: Sparse, physics-based, and
  partially interpretable neural networks for pdes, Journal of Computational
  Physics 445 (2021) 110600.
\newblock \href {https://doi.org/10.1016/J.JCP.2021.110600}
  {\path{doi:10.1016/J.JCP.2021.110600}}.

\bibitem{Shepherd1994}
R.~Shepherd, R.~J. Wiltshire,
  \href{https://link.springer.com/article/10.1007/BF00625515}{A periodic
  solution to a nonlinear diffusion equation}, Transport in Porous Media 15
  (1994) 175--182.
\newblock \href {https://doi.org/10.1007/BF00625515/METRICS}
  {\path{doi:10.1007/BF00625515/METRICS}}.
\newline\urlprefix\url{https://link.springer.com/article/10.1007/BF00625515}

\bibitem{Diez1992}
J.~A. Diez, J.~Gratton, F.~Minotti, Self-similar solutions of the second kind
  of nonlinear diffusion-type equations, Quarterly of Applied Mathematics 50
  (1992) 401--414.
\newblock \href {https://doi.org/10.1090/QAM/1178424}
  {\path{doi:10.1090/QAM/1178424}}.

\bibitem{Putra2022}
C.~A. Putra, P.~S. Palar, R.~Stevenson, L.~R. Zuhal, On physics-informed deep
  learning for solving navier-stokes equations, AIAA Science and Technology
  Forum and Exposition, AIAA SciTech Forum 2022 (2022).
\newblock \href {https://doi.org/10.2514/6.2022-1436}
  {\path{doi:10.2514/6.2022-1436}}.

\bibitem{Cai2021}
S.~Cai, Z.~Wang, S.~Wang, P.~Perdikaris, G.~E. Karniadakis,
  \href{https://dx.doi.org/10.1115/1.4050542}{Physics-informed neural networks
  for heat transfer problems}, Journal of Heat Transfer 143 (6 2021).
\newblock \href {https://doi.org/10.1115/1.4050542/1104439}
  {\path{doi:10.1115/1.4050542/1104439}}.
\newline\urlprefix\url{https://dx.doi.org/10.1115/1.4050542}

\bibitem{Goraya2023}
S.~Goraya, N.~Sobh, A.~Masud,
  \href{https://link.springer.com/article/10.1007/s00466-023-02334-7}{Error
  estimates and physics informed augmentation of neural networks for thermally
  coupled incompressible navier stokes equations}, Computational Mechanics 72
  (2023) 267--289.
\newblock \href {https://doi.org/10.1007/S00466-023-02334-7/FIGURES/29}
  {\path{doi:10.1007/S00466-023-02334-7/FIGURES/29}}.
\newline\urlprefix\url{https://link.springer.com/article/10.1007/s00466-023-02334-7}

\bibitem{Xing2023}
Z.~Xing, H.~Cheng, J.~Cheng,
  \href{https://www.mdpi.com/2227-7390/11/19/4049/htm
  https://www.mdpi.com/2227-7390/11/19/4049}{Deep learning method based on
  physics-informed neural network for 3d anisotropic steady-state heat
  conduction problems}, Mathematics 2023, Vol. 11, Page 4049 11 (2023) 4049.
\newblock \href {https://doi.org/10.3390/MATH11194049}
  {\path{doi:10.3390/MATH11194049}}.
\newline\urlprefix\url{https://www.mdpi.com/2227-7390/11/19/4049/htm
  https://www.mdpi.com/2227-7390/11/19/4049}

\bibitem{Carslaw1986}
H.~S. Carslaw, J.~J.~C. Jaeger,
  \href{https://books.google.com/books/about/Conduction_of_Heat_in_Solids.html?id=y20sAAAAYAAJ}{Conduction
  of heat in solids. 2nde edition, edition anglaise} (1986) 510.
\newline\urlprefix\url{https://books.google.com/books/about/Conduction_of_Heat_in_Solids.html?id=y20sAAAAYAAJ}

\end{thebibliography}

	\end{document}